\documentclass[aps,prd,showpacs,showkeys,final,11pt]{revtex4}
\usepackage{graphicx,amssymb,url}
\usepackage{amsmath}
\usepackage{showkeys}
\usepackage{float}

\begin{document}
\title{Neutral thin shell immersed into the Reissner-Nordstr\"om space-time}
\author{V. A. Berezin\email{berezin@inr.ac.ru}}
\author{V. I. Dokuchaev\email{dokuchaev@inr.ac.ru}}
\affiliation{Institute for Nuclear Research of the Russian Academy of
Sciences \\ 60th October Anniversary Prospect 7a, 117312 Moscow, Russia}

\begin{abstract}
Starting from  Israel equations for the spherically symmetric thin shells we introduce the effective potential and show how it can be used in constructing, without further thorough investigation, the corresponding Carter-Penrose diagrams describing clearly the global geometry of the composite space-time manifolds. We demonstrate, how this new method works, by considering all possible configurations for the neutral thin dust shell immersed into different types of Reissner-Nordstr\"om electro-vacuum manifolds.

\end{abstract}


\maketitle
\tableofcontents
\newpage
\section*{Notations}

{}\indent

$\Delta$ --- invariant function in (\ref{Delta})

$G$ --- Newton's constant

$e$ --- electrical charge

$i_0$ --- spatial infinities on the Carter-Penrose diagrams

$i_{\pm}$ --- past and future temporal infinities

$\cal J^{\pm}$ --- past and future null infinities

$R_\pm$--regions --- 4D regions with signature $(+,-,-,-)$

$T_\pm$--regions --- 4D regions with signature \, $\!(-,+,-,-)$

$M > 0$ --- "bare" mass of the shell

$m_{\rm in}>0$ --- mass parameter of the inner metrics

$m_{\rm out}>0$ --- mass parameter of the outer metrics

$\Delta m=m_{\rm out}-m_{\rm in}$ --- defined in (\ref{Delta})

$r_g=2Gm_{\rm out}>0$ --- gravitational radius in the outer metrics

$r_\pm=Gm_{\rm in}\pm\sqrt{G^2m_{\rm in}^2-Ge^2}$ --- horizon's radii in the inner metrics

$\rho(\tau)$ --- shell's radius as a function of its proper time $\tau$

$\rho_0$ --- turning point at the shell trajectory

$\sigma_{\rm in}(\rho)=\pm1$ è $\sigma_{\rm out}(\rho)=\pm1$ --- sign functions in (\ref{sigmainout}) and (\ref{sigmas})

\newpage
\section{Introduction}

The complexity of the Einstein equations dictates consideration of very simple and highly symmetric models. Apart from the well known examples of cosmological models, the general relativistic effects and therefore, the most interesting devi\-ations from Newtonian gravitation should be expected when considering the very concentrated massive bodies. The simplest of them, of course, is the point-like particle. The space-time around the point-like particle appeared to be the first exact solution to the Einstein vacuum equations found by Karl Schwarzschild in 1916, just few months after presentation of the new gravitational theory in the Berliner Academy of Sciences by Albert Einstein. This solution was spherically symmetric and looked, at the first sight, as a simple generalization of the Newton's gravity law.

Only in 1960's it became clear that the Schwarzschild space-time has rather nontrivial global geometry: it possesses an event horizon (which serves as a black hole boundary), two asymptotically flat regions and two singularities at zero radii, where the gravitating sources are concentrated. The appearance of event horizons separating the space-time region from where the light rays can escape to spatial infinity and those from where it is impossible to do this, is just the consequence of Special Relativity, namely, it follows from the fact that photons can not be stopped, and this exhibits the inconsistency between Special Relativity and Newtonian gravitation. In the Schwarzschild space-times there are two branches of the horizon (the past and the future ones called the particle and event horizons, respectively), and beyond them there lie two unusual regions: of inevitable expansion (in the past) and inevitable contraction (in the future. It is in such regions where the singularities appeared. These singularities are also unusual, they are space-like. Thus, the source in the Schwarzschild space-time exists only one moment in at zero radius in the past, then disappears and resurrects in the future singularity, again at zero radius. Thus, such a source is clearly nonphysical and can not be considered (even theoretically) as a limit of any extended static body.

 In a more general case, outside the point-like electrically charged massive source, the solution to the electro-vacuum Einstein equations was found by Reissner and Nordstr\"om. The global geometry of a Reissner-Nordstr\"om (R-N) manifold depends on the relation between the electrical charge and the total mass (energy) of the system (the latter includes both the energy of Coulomb field and the binding gravitational energy). For small enough charge the manifold (which is called in this case the R-N black hole) consists of the same elements as the Schwarzschild ones. But the number of these parts is infinite and they form an infinitely long ladder from the past to the future. In addition to the event (and particle) horizons, called now the outer horizons (infinitely many of them), there exist the inner horizons that serve as Cauchy horizons beyond which the trajectories of test particles, whose initial data are specified in the one of the asymptotically flat regions, cannot be unambiguously continued. What concerns the singularities at zero radii with sources (of course, there are also infinitely many of them), they are time-like, but hidden beyond the horizons of both types and, thus, can not be thought of as the limits of any usual extended static charged body. When the charge grows while the total mass remains constant, the outer and inner horizons become closer and closer to each other and they merge eventually for some critical value of charge (which is proportional to the total mass). Such a manifold with the double horizon is called the extremal R-N black hole. The singularity with the source is still hidden beyond the horizon in this case, Finally, if the charge exceeds the critical value, the horizons disappear, and the naked singularity reveals itself. In this case the global geometry is no more an infinite ladder, but it rather resembles that of the flat Minkowskian space-time with only one, but important difference: the zero radius time-like world-line is now singular and contains the charged massive source. But, physically, it is not a limit of any extended static charged body because the latter is unstable and would expand infinitely without some additional force (which, in turn, would add some energy-mass to the whole system).

 The simplest generalization of the point-like particle is a spherically symmet\-ric thin dust shell. Though it is also singular (the finite amount of mass=energy is concentrated in an infinitesimal volume), but the singularity is now spread on a sphere of finite radius, there are both interior and exterior regions where the space-time metrics are, in principle, known. These are parts of Schwarzschild and R-N manifolds, whose parameters are related to that ones describing the immersed shell and its initial state. The evolution of thin shell is governed by the so called Israel equations. The latter are nothing but the matching conditions between the inner and outer metrics. The mathematical structure of these equations reflects that of Einstein equations (of course, in three dimensions instead of four): there are both constraints and dynamical equations. Due to the spherical symmetry we will have only one constraint and one dynamical radial equation plus their differential consequence --- analog of Bianchi identity, which is nothing more but the continuity equation relating the surface energy density of the shell and its surface tension. In principle, given the shell's equation of state, ir is possible to solve this continuity equation and, thus we are left with only one equation, the constraint. And it is this equation that will be the subject of our investigation. The final goal is the construction of the so called Carter-Penrose conformal diagrams that describe quite clearly the global geometry of the composite manifold for every combination of the parameters involved.

 We have already mentioned that the global geometry of the vacuum and electro-vacuum spherically symmetric solutions to the Einstein equations is by no means trivial. For the composite manifolds with the shells and different solutions inside and outside it, the number of different combinations becomes very large. To simplify their investigation we need some method in order to recognize easily which of them is realized for any particular choice of the para\-meters of, say, an inner metrics and a shell. In this paper we propose the effective potential method. Drawing together the effective potential an two more very simple curves we are able to construct immediately the Carter-Penrose diagram for any allowed values of the shell's parameters.

 The paper is organized as follows. In the Section "Preliminaries" we describe shortly the spherical gravity, the construction of Carter-Penrose conformal dia\-grams, the thin shell formalism and the method of the effective potential. The subsequent Sections are devoted to the application of the proposed method to the construction of the global geometries for neutral spherically symmetric thin dust shells immersed into different R-N manifolds.

 Throughout the paper we use units with $\hbar = c = 1$, where $\hbar$ is the Planckian constant, and $c$ is the speed of light.

\section{Preliminaries}

\subsection{Spherical gravity}
The structure of any spherically symmetric space-time is completely
determined by two invariant functions of two variables. Indeed,
locally, the general spherically symmetric metric can be written as
\begin{equation}
ds^2 = A^2 dt^2 + 2 H dt dq - B^2 dq^2 - R^2 d\sigma^2 \, ,
\end{equation}
where $A(t,q),\, H(t,q)$ and $B(t,q)$ are functions of the time
coordinate, $t$, and some radial coordinate, $q$, $d\sigma ^2$ is
the line element of a $2- dim $ unit sphere, and $R(t,q)$ is the
radius of this sphere in the sense that its area equals $4 \pi \,
R^2$. Therefore, we are, actually, dealing with the invariant
function $R(t,q)$ and the two-dimensional metric, which by suitable
coordinate transformation can always be put in the conformally flat
form
\begin{equation}
ds^2_2 = \gamma _{ik} dx^i dx^k = \omega^2 (t,q) (dt^2 - dq^2)
\, , \;\;\; i,k = 0,1 \, .
\end{equation}
This proves the above statement about two functions of two
variables.

The first invariant function is, of course, the radius $R(t,q)$. By
geometrical reasons, we choose for the second function the invariant
(notations are obvious)
\begin{equation}
\Delta = \gamma ^{ik} \frac{\partial R}{\partial x^i} \frac{\partial
R}{\partial x^k} = \frac{1}{\omega ^2} \left( \dot R^2 - R^{\prime
2}\right) \, .
\label{Delta}
\end{equation}
This is nothing more but the square of the normal vector to the
surfaces of constant radii, $R (t,q) = const$. The invariant
function $\Delta$ brings a very important geometrical information.
If $\Delta < 0$, the surfaces $R = const$ are time-like, such
regions are called the $R_{\pm}$-regions, the signs $"\pm"$ being
denote the sign of a spatial derivative of the radial function $R$.
If $\Delta > 0$, the regions are called the $T_{\pm}$-regions,
depending on the sign of the corresponding time derivative
(inevitable expansion or inevitable contraction), and the surfaces
$R = const$ are space-like. The $R_{\pm}-$ and $T_{\pm}-$ regions
are separated by the apparent horizons with $\Delta = 0$. It is the
set of these regions and horizons together with the boundaries
(infinities and that determines the global geometry. The boundaries
are to be chosen in such a way that the space-time becomes
geodesically complete, namely, all the time-like and null geodesics
should start and end either at infinities or at singularities.

\subsection{Carter-Penrose diagrams for the Schwarzschild
and Reissner-Nordstr\"om space-times.}

The causal structure of geodesically complete spherically symmetric
space-times can be best seen on the conformal Carter-Penrose
diagrams where each point represents a sphere, and infinities are
brought to the final distances. Since every 2-dimensional space-time
is (locally) conformally flat, its Carter-Penrose diagram is the set
of that for the $2$-dimensional Minkowski manifold. To see how the latter
looks like, let us, first, transform the Minkowski  metric $ds^2 =
dt^2 - dx^2 $ to the double-null coordinates $u = t -x$ (retarded
time) and $v = t + x$ (advanced time), then $ds^2 = du\, dv$. We will
use the convention that on the diagram the time coordinate increases
from down to up, the spatial coordinate --- from left to right, and
the null curves $u = const$, $v = const$ are the straight lines
with the slope $\pm 45^{\circ}$. Making one more transformation
\begin{eqnarray}
u^{\prime} &=& \arctan {u} \, , \;\;\; - \frac{\pi}{2} \le u^{\prime}
\le \frac{\pi }{2} \nonumber \\
v^{\prime} &=& \arctan {v} \, , \;\;\; - \frac{\pi}{2} \le v^{\prime}
\le \frac{\pi }{2}
\end{eqnarray}
one gets
\begin{eqnarray}
ds^2 &=& \Omega ^2 ds^{\prime 2} \, , \;\;\; \Omega =
\frac{1}{\cos {u^{\prime}} \cos {v^{\prime}}} \nonumber \\
ds^{\prime 2} &=& du^{\prime} dv^{\prime} = dt^{\prime 2} -
dx^{\prime 2} \, .
\end{eqnarray}
Formally, the metric $ds^{\prime 2}$ looks exactly as the starting
one, but now coordinates $(u^{\prime}, \, v^{\prime})$ and
$(t^{\prime}, \, x^{\prime})$ run the finite intervals.

\begin{figure}[t]
\begin{center}
\includegraphics[angle=0,width=0.94\textwidth]{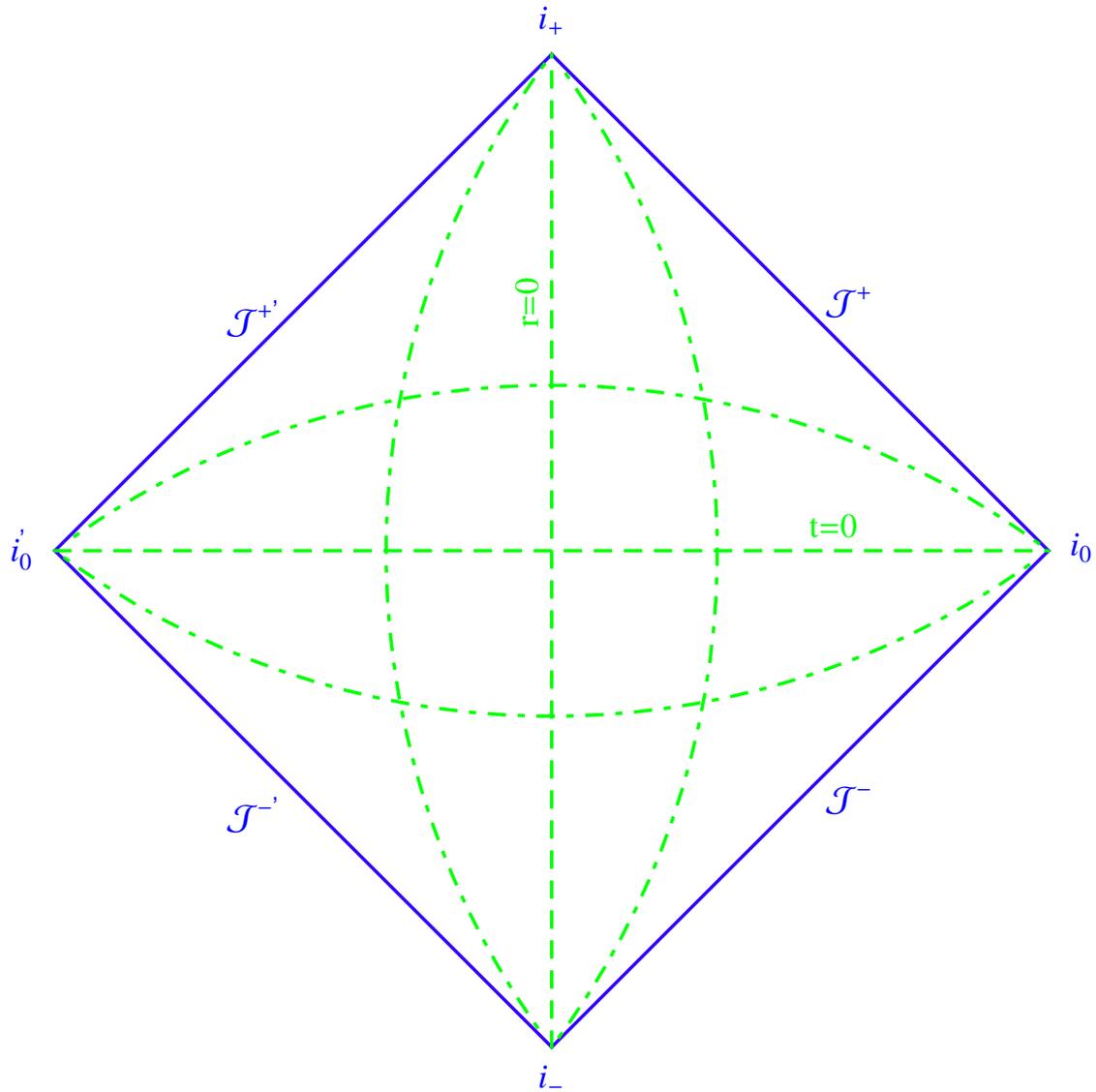}
\end{center}
\caption{The Carter-Penrose diagram for the complete $2$-dimensional
Minkowski space-time $( - \infty < t < \infty, \; - \infty < x <
\infty)$. The horizontal dashed curves represent $t = const$
lines, while the vertical ones are for $x = const$.}
\label{2DimMinkowski}
\end{figure}

The Carter-Penrose diagram for the complete $2$-dimensional Minkowski
space-time $( - \infty < t < \infty, \; - \infty < x < \infty)$ is
shown in Fig.~\ref{2DimMinkowski}.
Here $J^{\pm} (J^{\prime \pm})$ are null future $(v^{\prime}
(u^{\prime}) = \pi/2, \, v (u) = \infty)$ and past
$(u^{\prime} (v^{\prime}) = -\pi/2, \, u (v) = - \infty)$
infinities, $i_{\pm}$ are future and past $(t^{\prime} = \pm
\pi/2)$ temporal infinities, and $i_0 (i^{\prime}_0)$ are
spatial $(x^{\prime} = \pm\pi/2, \, x = \pm \infty)$
infinities. If the corres\-ponding conformally flat metric is not
complete in the sense that one of the coordinates starts from or
ends at the finite boundary value (like, for example, the zero
radius value in the case of spherical symmetry), then one should
cut the above square along the corresponding diagonal (in general,
along some time-like os space-like curve), and such part of the
complete Carter-Penrose diagram will be a triangle with the
vertical (left for $R_+$-regions and right for $R_-$-regions) or
horizontal (for $T_{\pm}$-regions) boundary.

Both the Schwarzschild and R-N metrics look the same
in the so-called curvature coordinates:
\begin{equation}
ds^2 = F dt^2 - \frac{1}{F} dR^2 - R^2 (d \vartheta ^2 +
\sin^2{\vartheta} d\varphi^2) \, ,
\end{equation}
where $R$ --- radius $(0 \le R < \infty), \, F = F(R)$, and
$\vartheta$ and $\varphi$ are spherical angles. The two-dimensional
part can easily be written in the conformally flat form by
introducing the "tortoise" coordinate $R^{\star}$:
\begin{eqnarray}
dR^{\star} &=& \frac{d R}{|F|} \, , \nonumber \\
ds^2_2 &=& F \left( d \xi^2 - dR^{\star 2}\right) \, .
\end{eqnarray}
In the $R_{\pm}$-regions $F = - \Delta > 0$ and $R^{\star}$ plays
the role of the spatial (radial) coordinate $q$, while $\xi$ is the
time coordinate $t$. In the $T_{\pm}$-regions, $R^{\star}$ plays the
role of the time coordinate $t$, while $\xi$ is the spatial
coordinate $q$.

\begin{figure}[t]
\begin{center}
\includegraphics[angle=0,width=0.9\textwidth]{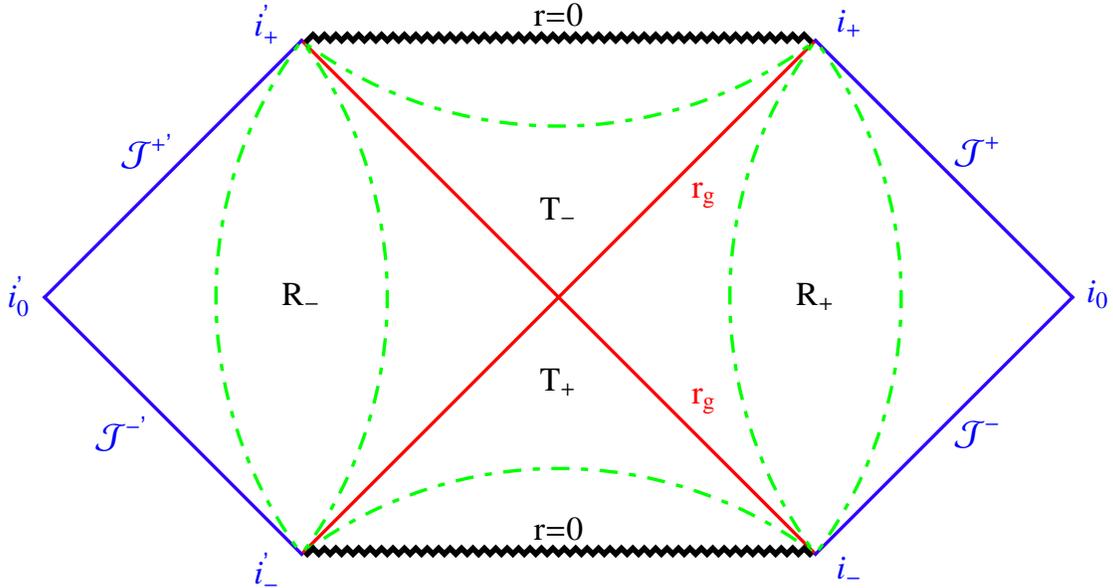}
\end{center}
\caption{The complete Carter-Penrose diagram of the Schwarzschild
metric.} \label{SchwTotal}
\end{figure}

Consider, first, the  Schwarzschild metric. In this case
\begin{equation}
F = 1 - \frac{2 \, G \, m}{R} \, ,
\end{equation}
where $G$ is the Newton's gravitational constant $m$ is the total
mass of the gravitating system measured by distant observers (at
infinity), and we put the speed of light $c = 1$. For $R > r_g = 2
\, G \, m$ we have the the $R$-region, and for $R < r_g$ --- the
$T$-region. The event horizon coincides with the apparent horizon
at $R = r_g$ (gravitational, or Schwarzschild, radius). At $R = 0$
we encounter the (space-like) curvature singularity. The complete
Carter-Penrose diagram looks as follows in Fig.~\ref{SchwTotal}.
There are two isometric $R_{\pm}$-regions bounded by two apparent
(past and future) horizons at $R = r_g$ and two asymptotically flat
regions with corresponding future and past temporal $(i_{\pm}, \,
i^{\prime}_{\pm})$, future and past null $(J_{\pm}, \,
J^{\prime}_{\pm})$ and spatial $(i_0, \, i^{\prime}_0)$ infinities.
Also we have two $T$-regions ($T_+$ and $T_-$) bounded by the
apparent horizons at $R = r_g$ and future and past space-like
singularities at $R = 0$. This is called the eternal Schwarzschild
black hole. The gravitational source is concentrated on these two
space-like singularities, i.e., it exists only for one moment in the
past and reappears again for one moment in the future.

The causal structure of the R-N space-time is much
more complex. The function $F$ equals now
\begin{equation}
F = 1 - \frac{2 \, G \, m}{R} + \frac{G \, e^2}{R^2},
\end{equation}
$e$ is the electric charge. There are three different cases

\begin{figure}[h]
\begin{center}
\includegraphics[angle=0,width=0.59\textwidth]{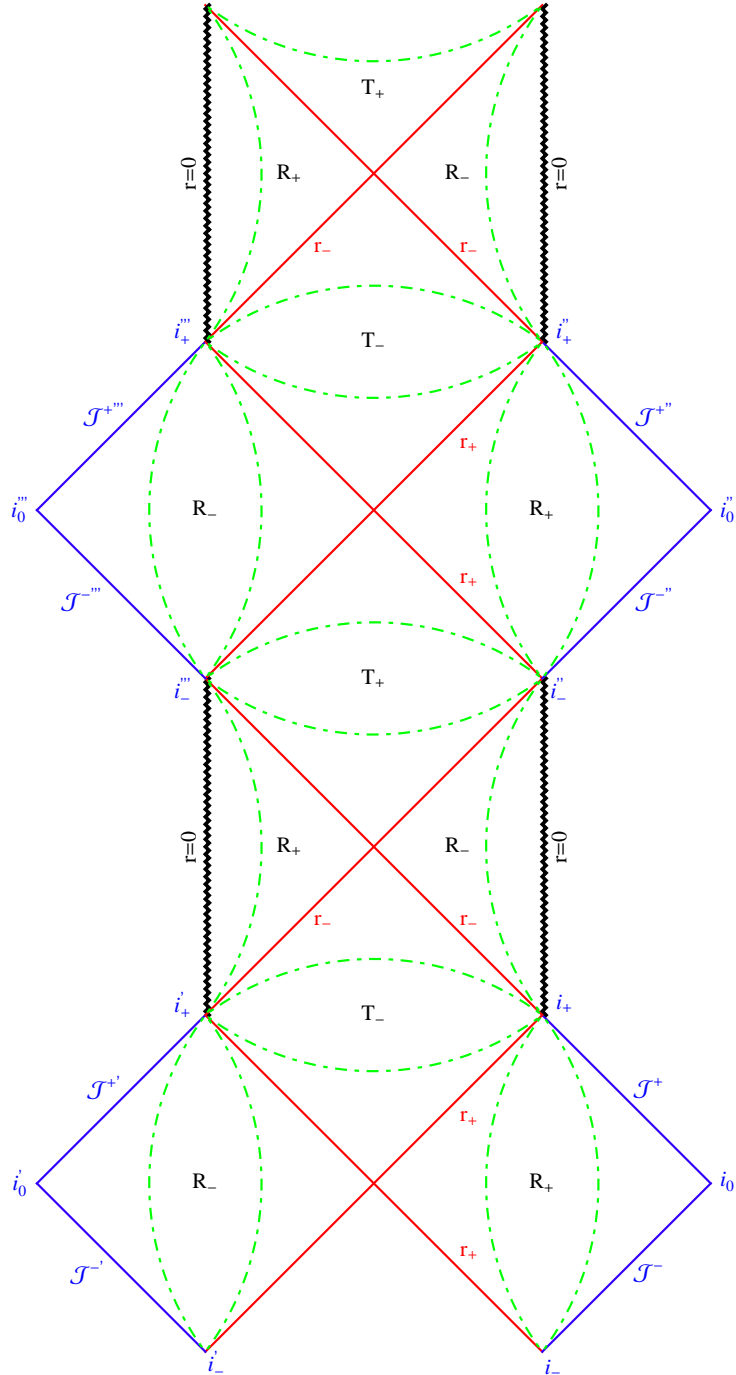}
\end{center}
\caption{The complete Carter-Penrose diagram of the Reissner-Nordstr\"om (R-N) black hole, $G \, m^2 > e^2$.} \label{RNregionsRTrconstNoFrame}
\end{figure}

(1) $G \, m^2 > e^2$ --- R-N black hole, equation $F =
0$ has two nonequal real roots $r_{\pm}$,
\begin{equation}
r_{\pm} = G \, m \pm \sqrt{G^2 \, m^2 - G \, e^2} \, .
\end{equation}
According to the signs of $F$, we have the $R$-regions for $r_+ <
R < \infty$ and $0 \le R < r_-$, $T$-regions in-between, $r_- < R
< r_+$, and two apparent horizons at $R = r_{\pm}$, the external
one, $r_+$, playing the role of the event horizon, and the inner,
$r_-$, --- the Cauchy horizon. The geodesically complete
Carter-Penrose diagram is the ladder extended infinitely to the
past and to the future as is shown in Fig.~\ref{RNregionsRTrconstNoFrame}.

In the complete (eternal) R-N black hole space-time
both the the gravitational source and the electric charge(s) are
concentrated on two (for each part of the ladder) time-like
singularities $R = 0$ (left and right on the diagram), the signs
of the electric charges on them being opposite..

\begin{figure}[h]
\begin{center}
\includegraphics[angle=0,width=0.4\textwidth]{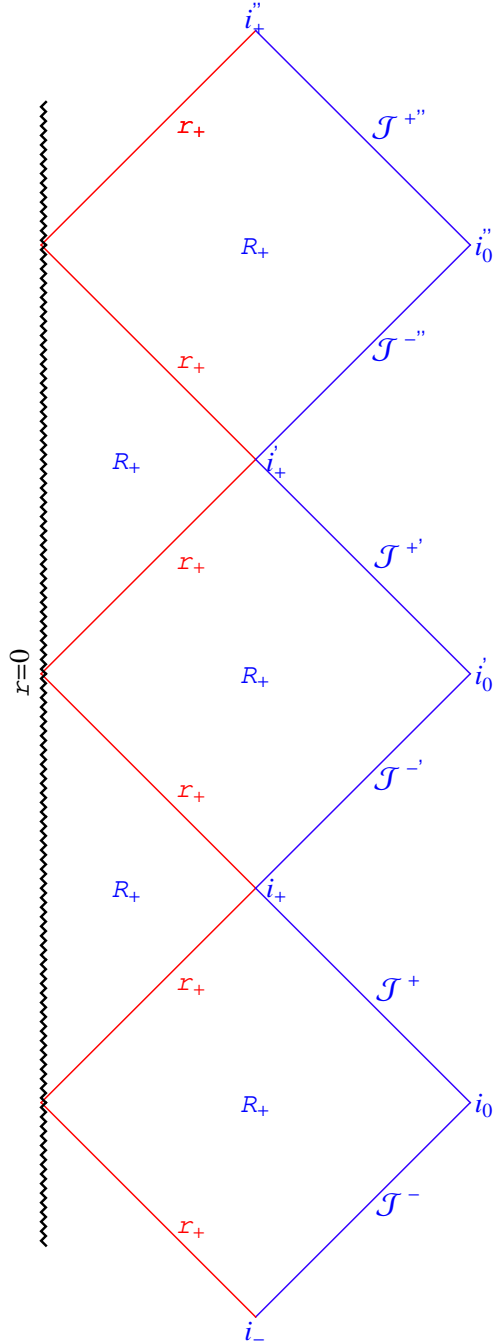}
\end{center}
\caption{Extremal Reissner-Nordstr\"om black hole, $G \, m^2 = e^2$.}
 \label{RNextremal}
\end{figure}

(2) $G \, m^2 = e^2$ --- extremal R-N black hole.
Equation $F = 0 $ has the double root $r_+ = r_- = G \, m =
\sqrt{G} |e|$.  We have $R$-regions everywhere except the apparent
(event) horizon at $R = r_+ = r_-$, as is shown in Fig.~\ref{RNextremal}.

\begin{figure}[h]
\begin{center}
\includegraphics[angle=0,width=0.55\textwidth]{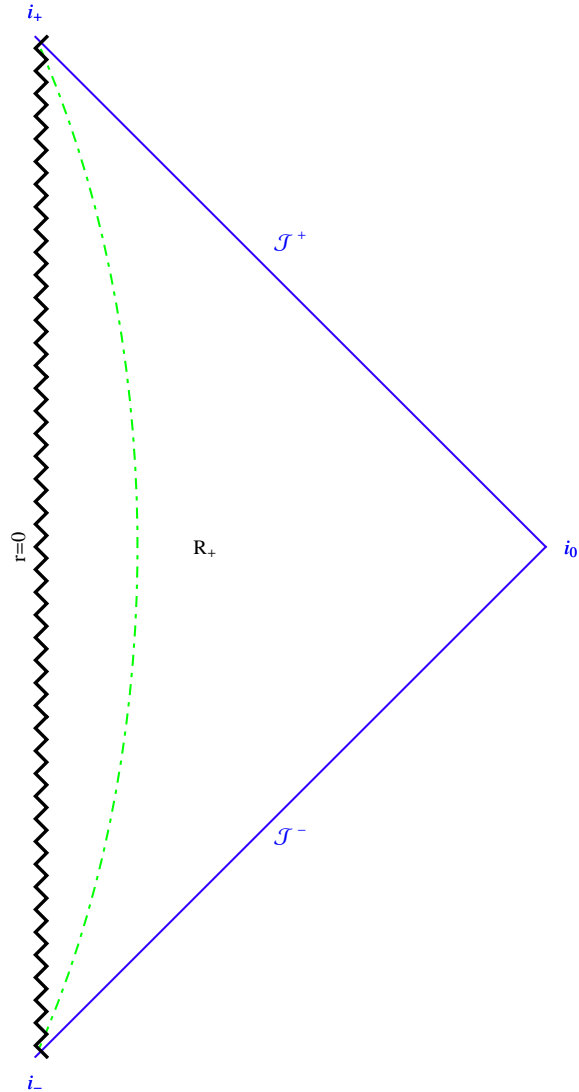}
\end{center}
\caption{The complete Carter-Penrose diagram of the Reissner-Nordstr\"om naked singularity,
$G \, m^2 < e^2$.}
\label{NakedsingularityFig5}
\end{figure}

(3) $G \, m^2 < e^2$ --- no black hole, the naked singularity at
$R=0$. The Carter-Penrose diagram is very simple (see
Fig.~\ref{NakedsingularityFig5}).

\subsection{Thin shells.}

The thin shell is a hyper-surface in the space-time on which the
energy-momen\-tum tensor is singular. If such a hyper-surface is time-
or space-like, one can introduce in its vicinity the so-called
Gaussian normal coordinates, and the line element can be written as
\begin{equation}
ds^2 = \varepsilon dn^2 + \gamma_{ij}(n,x) dx^i dx^j \, ,
\end{equation}
$n$ is the coordinate in the normal direction to the shell, and
$x^i$ --- coordinates on the shell, $\varepsilon = + 1$ in the
space-like case and $\varepsilon = - 1$ in the time-like case. The
surface is supposed to be located at $n = 0$. The energy-momentum
tensor $T^{\mu}_{\nu}$ is proportional to $\delta$-function,
\begin{equation}
T^{\mu}_{\nu} = S^{\mu}_{\nu} \, \delta (n) \, ,
\end{equation}
$S^{\mu}_{\nu}$ is called the surface energy-momentum tensor. The
dynamics of the thin shell is governed by the Israel equations
obtained by integrating the Einstein equations across the shell.
First of all, one gets $S^n_n = S^i_n = 0$, this can be considered
as the definition of the thin shell. The Israel equations are
\begin{equation}
\varepsilon \left( \left[K_{ij} \right] - \gamma_{ij} \left[ K\right] \right)
 = 8 \pi \, G S_{ij} \, ,
\end{equation}
supplemented by the Bianchi identity for the shell
\begin{equation}
S^j_{i|j} + \left[ T^n_i\right] = 0\, .
\end{equation}
Here $K_{ij} = -(1/2)\partial \gamma_{ij}/\partial n$ is the extrinsic
curvature tensor, $K$ is its trace, brackets
$[\;] = (out) - (in)$ is the jump across the shell, the vertical
line denotes the covariant derivative with respect to the metric
$\gamma_{ij}$. In what follows we will be dealing with the time-like
shells only, so, $\varepsilon = - 1$.

In the case of spherical symmetry everything is simplified
drastically. The metric becomes
\begin{equation}
ds^2 = - dn^2 + \gamma_{00} (n, \tau) d \tau^2 - \rho^2 (n,\tau) d \sigma^2,
\end{equation}
$\rho (0,\tau)$ is the shell radius as a function of the proper time
of the observer sitting on this shell, $n < 0$ inside and $n > 0$
outside. The mixed components of the surface energy momentum tensors
are $S^0_0$ (surface energy density) and $S^2_2 = S^3_3$ (surface
tension), and the Israel equations reduced to one constraint and one
dynamical equations, namely,
\begin{eqnarray}
\left[ K^2_2 \right] &=& 4 \pi \, G \, S^0_0 \nonumber \\
\left[ K^0_0 \right] + \left[ K^2_2 \right] &=& 8 \pi \, G \, S^2_2 \, .
\end{eqnarray}
The supplement equation is now
\begin{equation}
\dot S^0_0 + \frac{2 \,\dot \rho}{\rho} \left( S^0_0 - S^2_2 \right) +
\left[ T^n_0\right] = 0 \, .
\end{equation}
We are interested in the situation when both inside and outside the
shell the space-time is (electro)-vacuum one, hence, $T^n_0 = 0$.
For the sake of simplicity we will consider the dust shell, for
which $S^2_2 = 0$. Then,
\begin{equation}
S^0_0 = \frac{M}{4 \pi \, \rho^2} \, ,
\end{equation}
where $M = const$ is the bare mass of the shell (without the
gravitational mass defect). Thus, we need only the first,
constraint, equation. In order to go further we have to calculate
$K^2_2 = -(1/\rho^2)K_{22} = -1/(2 \rho^2)
\partial \rho^2/\partial n = - \rho_{,n}/\rho$.
But, from definition of the invariant $\Delta$ it follows
\begin{eqnarray}
\Delta &=& \dot \rho^2 - \rho^2_{,n} \nonumber\\
\rho_{,n} &=& \sigma \sqrt{\dot \rho^2 - \Delta} \nonumber \\
K^2_2 &=& - \frac{\sigma}{\rho} \sqrt{\dot \rho^2 - \Delta} \, .
\end{eqnarray}
Here $\sigma = \pm 1$ depending on whether radii increasee $(\sigma
= + 1)$ in the normal outward direction or decrease $(\sigma = -
1)$. Thus, the sign of $\sigma$ coincides with that of the
$R$-region, and it can change only in the $T$-regions. Finally, the
only equation we will need in our analysis is
\begin{equation}
\sigma_{\rm in} \sqrt{\dot \rho^2 - \Delta_{\rm in}} -
\sigma_{\rm out} \sqrt{\dot \rho^2 - \Delta_{\rm out}} = \frac{G \, M }{\rho} \, .
\label{sigmainout}
\end{equation}
Since in our case $\Delta = - F$, we have
\begin{equation}
\sigma_{\rm in} \sqrt{\dot \rho^2 + 1 - \frac{2 \, G \, m_{\rm in}}{\rho} +
\frac{G \, e^2_{\rm in}}{\rho^2}} - \sigma_{\rm out} \sqrt{\dot \rho^2 + 1 -
\frac{2 \, G \, m_{\rm out}}{\rho} + \frac{G \, e^2_{\rm out}}{\rho^2}} = \frac{G \, M }{\rho} \, .
\end{equation}
We will not consider exotic matter shells, so $M > 0$. From the
above constraint equation (that is nothing more but the energy
conservation law) it follows that for the qualitative analysis one
needs to investigate the behavior of the function $\rho (\tau)$ only
at several special points: $\rho \to \infty, \; \dot \rho = 0, \;
\rho = 0$ and $\rho = \rho_{\sigma}$ where $\sigma_{\rm out} (\sigma
_{\rm in})$ changes its sign.

\section{The effective potential method}

Based on the Israel equations for dynamics of thin shells, in this Section we introduce the so called effective potential which enables us to construct the conformal Carter-Penrose diagrams, describing clearly the global geometry of the corresponding composite space-time manifold, without thorough analytical investigations.

In what follows we will consider the spherically symmetric neutral thin dust shells immersed into different types of R-N space-times. So, both inside and outside the shell of bare mass $M$ we have the same value of electric charge $e$ but different mass parameters $m_{\rm in}$ and $m_{\rm out}$. We already discussed in the preceding Section that of all Israel equations the only one we need in our case is the following constraint equation
\begin{equation}
\label{constreq}
\sigma_{\rm in} \sqrt{\dot\rho^2 + 1 - \frac{2 G m_{\rm in}}{\rho} + \frac{G e^2}{\rho^2}} - \sigma_{\rm out} \sqrt{\dot\rho^2 + 1 - \frac{2 G m_{\rm out}}{\rho} + \frac{G e^2}{\rho^2}} = \frac{G M}{\rho} \, .
\end{equation}
Here $\rho$ is the shell's radius as a function of the proper time $\tau$ (the over dot means its first derivative). Let us remind that in chosen units its square root has dimension of length, time and inverse mass simultaneously. The sign functions $\sigma_{in,out} = \pm 1$ in the interior and exterior parts of the complete manifold show us whether the radii $r$ increase (+) or decrease (-) in the outward normal direction to the shell. It is the values of $\sigma's$ that define, essentially, the global geometry of the composite manifold. To specify the solution to this differential equation we need the initial data. In addition to the usual initial values of the shell's radius $\rho_0$ and rapidity $\dot \rho_0$ we should also know the initial value of $\sigma_{\rm in}$. Given, then, the parameters of our system, namely, the mass $m_{\rm in}$ and the electric charge $e$ of the inner metrics and the bare mass $M$ of the shell itself, we are able to calculate both $\sigma_{\rm out}$ of the outer metrics and its mass parameter, $m_{\rm out}$, which is the total mass (= energy) of the whole system (for this very reason the constraint equation is often called the equation of initial conditions). It is convenient to choose as initial data either the value of radius at (one of) the turning points where $\dot \rho_0 = 0$, whenever they exist, or the rapidity at infinity. Then the initial values of $\sigma_{\rm in}$ and $\sigma_{\rm out}$ show us, in which of the $R_{\pm}$-regions of the inner and outer parts of the complete manifold the shell starts to move. The signs of $\sigma's$ may change their initial values dynamically, but only in $T_{\pm}$-regions (provided they exist) when the expressions under the corresponding square roots become equal zero.

The experience in classical dynamics show us that the most convenient method to visualize particle motion under influence of different forces is the construction (whenever it is possible) the corresponding potentials. In such cases the first integral of dynamical equations (i.e. those containing the second time derivatives --- accelerations) equals the total energy of a particle which is just the sum of the kinetic energy and the potential energy. When the Newtonian (non-relativistic) gravitational force is acting, the potential energy is equal to the inertial (= gravitational) mass of the particle times the so called gravitational potential. Drawing together the potential energy graph and the horizontal line, corresponding to the particle energy, one can obtain very useful qualitative information. Namely, the allowed interval for the particle motion is that where the energy line lies above the potential curve, the minima of the potential being correspond to the stable equilibrium, while its maxima --- to the unstable ones. Also it is easy to see when the motion is bound or unbound.

In our case of the spherical symmetric thin dust shell the situation is more tricky. First of all, though the constraint, Eqn.~(\ref{constreq}), is actually the first integral of the corresponding dynamical equation (after solving the shell's continuity equation), it does not have the familiar "energy" form. The latter is recovered in the non-relativistic limit $\dot\rho^2 \ll 1$, the "particle" energy being identified with
\begin{equation}
\Delta m = m_{\rm out} - m_{\rm in}.
\label{Deltam}
\end{equation}
Fortunately, it is possible to put the constraint equation into (almost) such a form. Let us rewrite it in the following way
\begin{equation}
\label{constreq1}
\sigma_{\rm in} \sqrt{\dot\rho^2 + 1 - \frac{2 G m_{\rm in}}{\rho} + \frac{G e^2}{\rho^2}} - \frac{G M}{\rho} = \sigma_{\rm out} \sqrt{\dot\rho^2 + 1 - \frac{2 G m_{\rm out}}{\rho} + \frac{G e^2}{\rho^2}} \, .
\end{equation}
By squaring this relation one gets
\begin{equation}
\label{constreq2}
\Delta m = \sigma_{\rm in} \sqrt{\dot\rho^2 + 1 - \frac{2 G m_{\rm in}}{\rho} + \frac{G e^2}{\rho^2}} - \frac{G M^2}{2 \rho} = \sigma_{\rm in} \sqrt{\dot\rho^2 + 1 - \frac{2 G m_{\rho}}{\rho}} - \frac{G M^2}{2 \rho} \, .
\end{equation}
Here $m(\rho) = m_{\rm in} - e^2/(2 \rho)$ can be called "the running mass", because $m_{\rm in}$ is the total mass measured at spatial infinity of the complete inner space-time manifold, and $e^2/(2 \rho)$ is the Coulomb energy outside the sphere of radius $\rho$, so, $m(\rho)$ is just the total mass (= energy) confined inside this sphere. The last term in Eqn.~(\ref{constreq2}) is the self-interaction energy of the shell. Though in our relativistic expression the kinetic and (gravitational) potential energies can not be separated, we may use it in the same way as in non-relativistic mechanics. Namely, we will consider $\Delta m (\dot\rho = 0)$ as the (now effective) potential. It is convenient to deal with dimensionless entities, so we adopt as definition the following
\begin{eqnarray}
\label{effpot}
 \frac{\Delta m}{M}(\dot\rho=0)&\equiv&V_{\rm eff}
 =\sigma_{\rm in}\sqrt{1-\frac{2 G m_{\rm in}}{\rho}+\frac{G e^2}{\rho^2}}-\frac{G M}{2 \rho}\,.
\end{eqnarray}
This was use previously (with $\sigma_{\rm in} = + 1$ only) in \cite{berokhr} for studying dynamics of thin shells with orbiting constituents. Important note: the value of $\Delta m$ may be both positive and negative (at least it is negative when $\sigma_{\rm in} = - 1$), i.e., the total mass (energy) of the whole system is less then that of the inner part of the space-time. But this should not confuse us. Let us remind that, by definition, the bare mass (energy) inside the spherical layer equals
\begin{equation}
\label{M}
M = 4 \pi \int^{q_2}_{q_1} T^0_0 r^2(q) e^{\lambda/2} dq = 4 \pi \int^{n_2}_{n_1} T^0_0 r^2(n)\, dn,
\end{equation}
where the radial coordinate $q$ (or the Gaussian normal coordinate $n$) runs from inside to outside (the shell is situated at $n = 0$), while the total mass (energy), which includes the gravitational mass defect is expressed by the Landau formula \cite{LL2}
\begin{equation}
\label{constreq2b}
m = 4 \pi \int^{r_2}_{r_1} T^0_0 r^2 dr = 4 \pi \int^{n_2}_{n_1} T^0_0 r^2 r_{,n} dn = 4 \pi \int^{n_2}_{n_1} T^0_0 r^2 \sigma |r_{n}|\, dn,
\end{equation}
and for $\sigma_{\rm in} = -1$ it is negative. One more thing: of course, we can get rid of the square root and obtain the expression more like the non-relativistic one, but, in doing this, we are loosing the very important information about the sign of $\sigma_{\rm in}$, which is responsible for the global geometry of the inner part of the composite manifold. Surely, we already lost an information about $\sigma_{\rm out}$ due to first squaring, but it can easily be restored. Indeed, knowing $\sigma_{\rm in}$, one obtains from Eqs.~(\ref{constreq}) and (\ref{constreq2b}) that
\begin{eqnarray}
\label{sigmas}
\sigma_{\rm out} &=& sign \left( \frac{\Delta m}{M} - \frac{G M}{2 \rho}\right) \\
\sigma_{\rm in} &=& sign \left( \frac{\Delta m}{M} + \frac{G M}{2 \rho}\right) \, .
\end{eqnarray}
We, thus, see that in order to reconstruct the global geometry of the composite space-time manifold and visualize qualitatively the thin shell motion, one needs to know not only the potential curve, $y_1 = V_{\rm eff} (\rho)$, but, in addition, two more curves showing the change of signs $\sigma_{\rm in}$ and $\sigma_{\rm out}$, namely, $y_2 =G M/(2 \rho)$ and $\tilde y_2 = -G M/(2 \rho) = - y_2$. In our Figures we will paint the effective potential curve $y_1$ in red color, the curves $y_2$ and $\tilde y_2$ --- in blue, and the horizontal lines $\Delta m/M = \mu = const$ --- in green. The rules are the following:
\begin{figure}[h]
\begin{center}
\includegraphics[angle=0,width=0.95\textwidth]{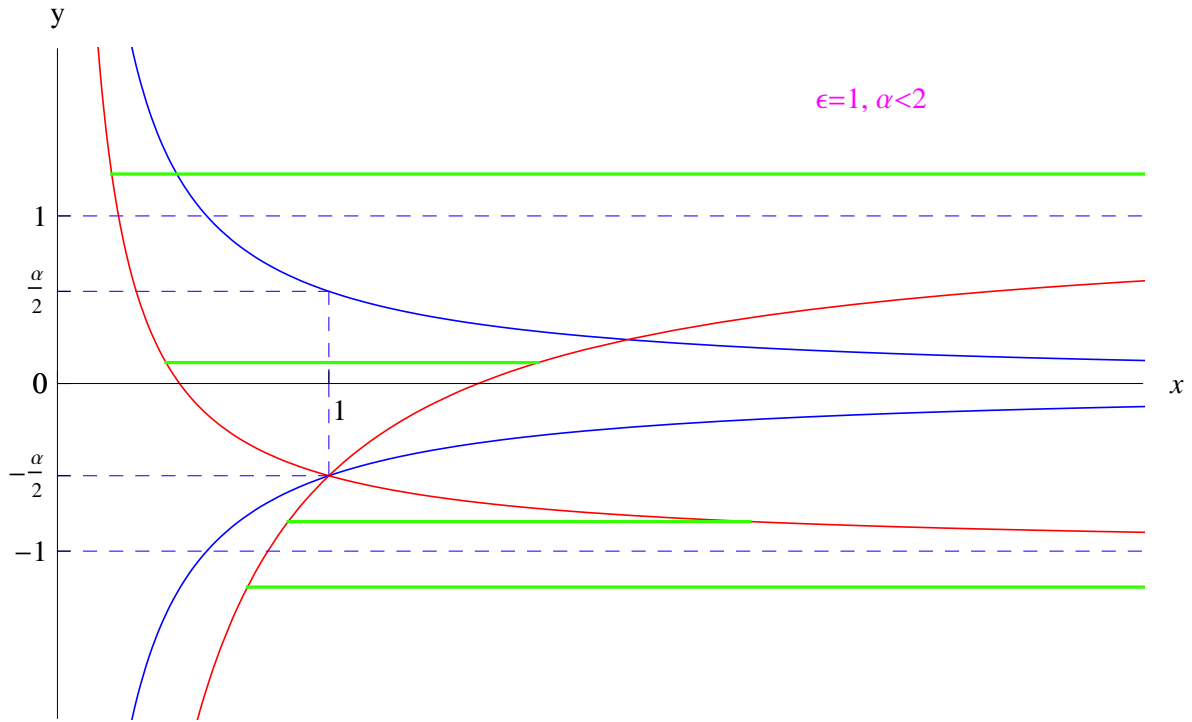}
\end{center}
\caption{The effective potential for ``light'' shells.}
\label{fig6}
\end{figure}
\begin{figure}[h]
\begin{center}
\includegraphics[angle=0,width=0.95\textwidth]{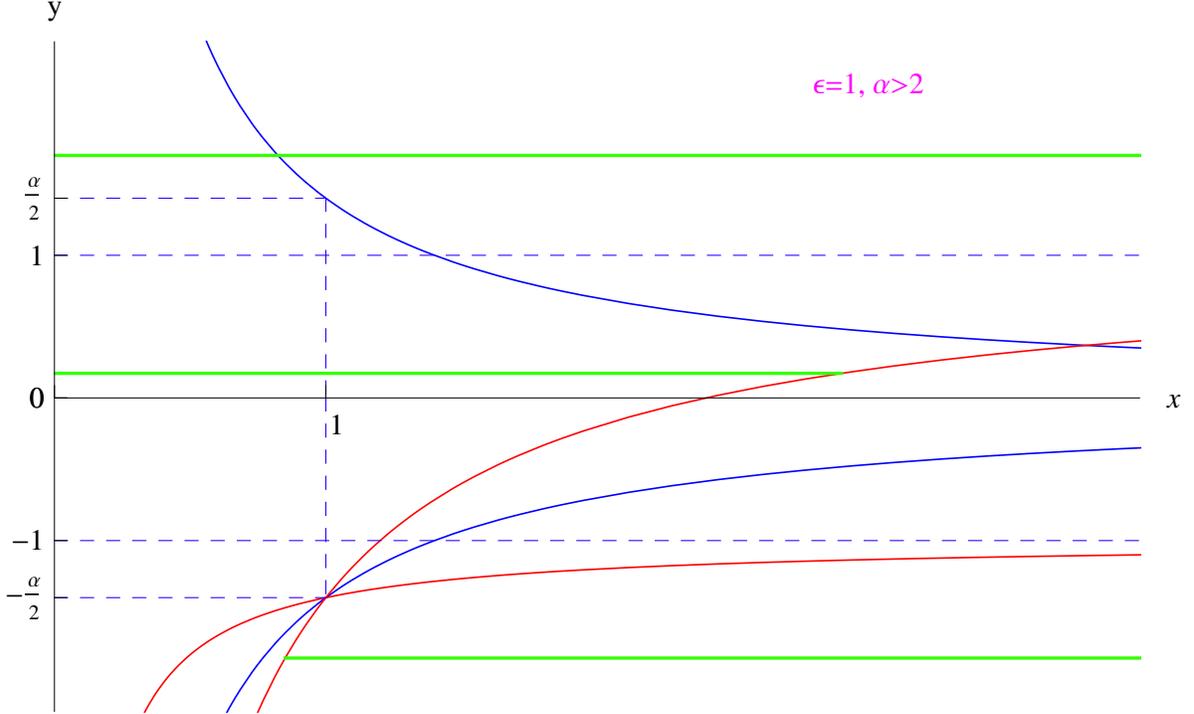}
\end{center}
\caption{The effective potential for ``heavy'' shells.}
\label{fig7}
\end{figure}
\begin{figure}[h]
\begin{center}
\includegraphics[angle=0,width=0.95\textwidth]{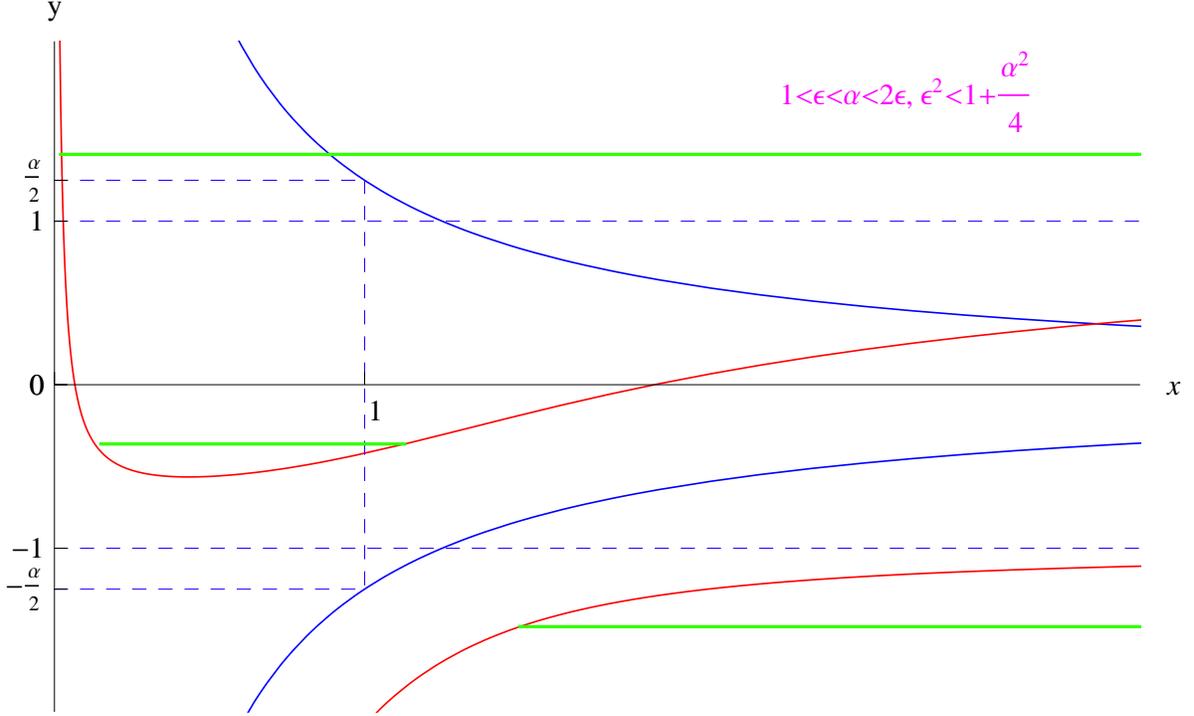}
\end{center}
\caption{The effective potential at $1<\epsilon < \alpha < 2 \epsilon$ and $\epsilon^2 < 1 +\alpha^2/4$.}
\label{fig8}
\end{figure}
\begin{figure}[t]
\begin{center}
\includegraphics[angle=0,width=0.95\textwidth]{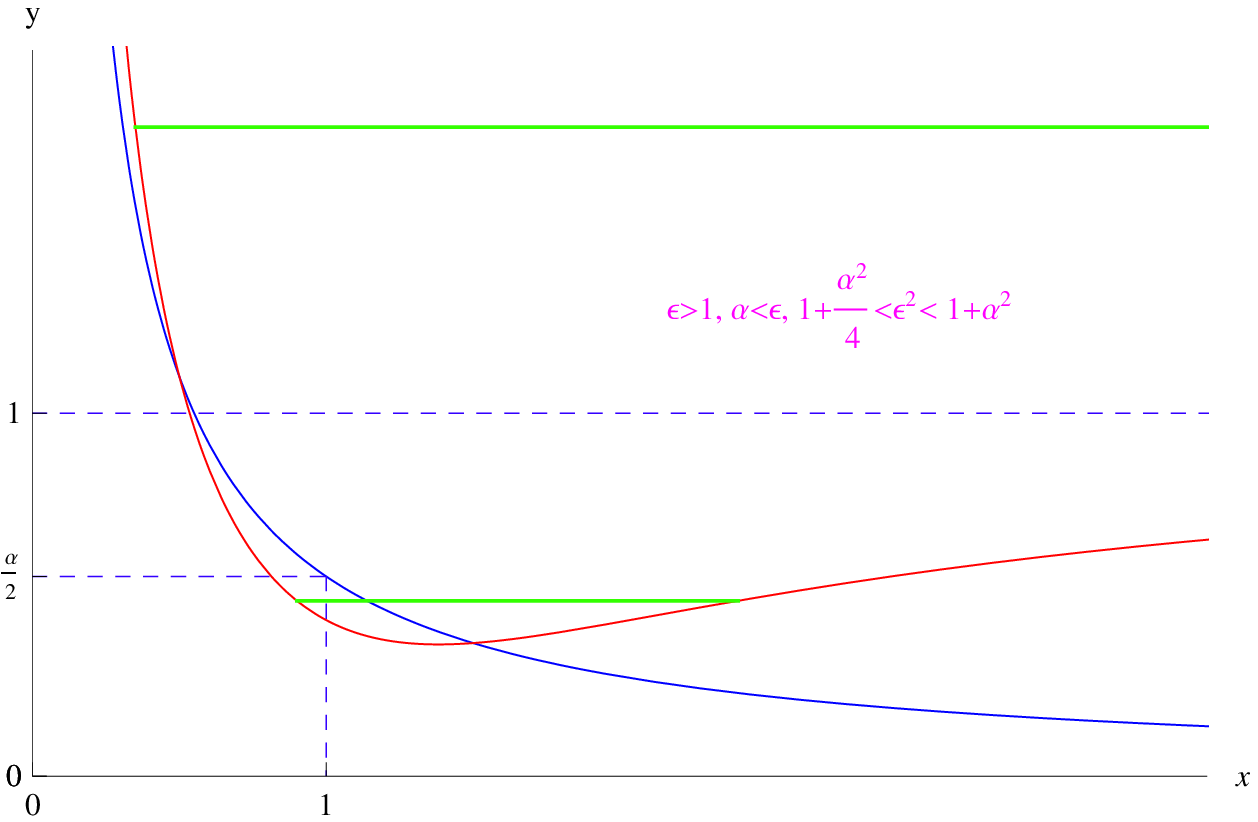}
\end{center}
\caption{The effective potentials for $\alpha^2 < \epsilon^2 < 1 + \alpha^2$.}
\label{fig9}
\end{figure}
\begin{figure}[h]
\begin{center}
\includegraphics[angle=0,width=0.95\textwidth]{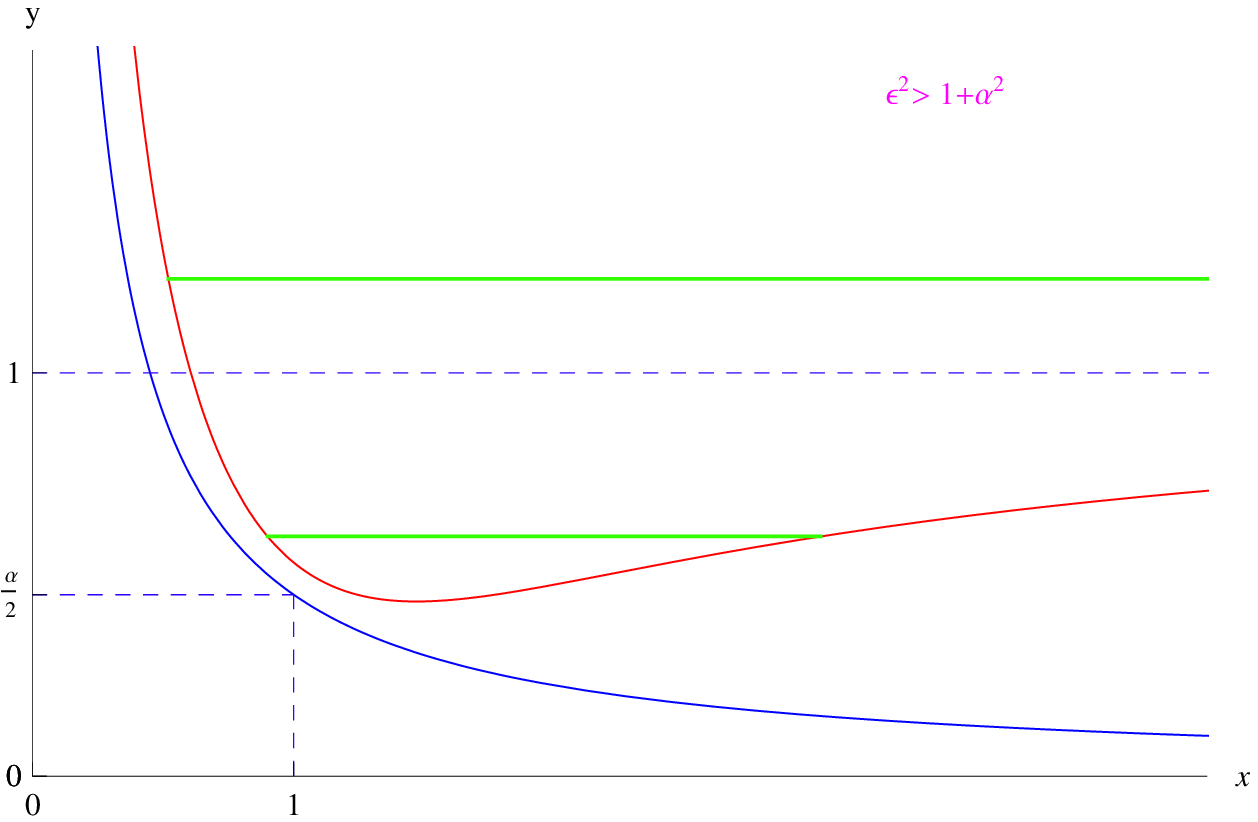}
\end{center}
\caption{The effective potentials for $\epsilon^2 > 1 + \alpha^2$.}
\label{fig10}
\end{figure}
\begin{figure}[t]
\begin{center}
\includegraphics[angle=0,width=0.95\textwidth]{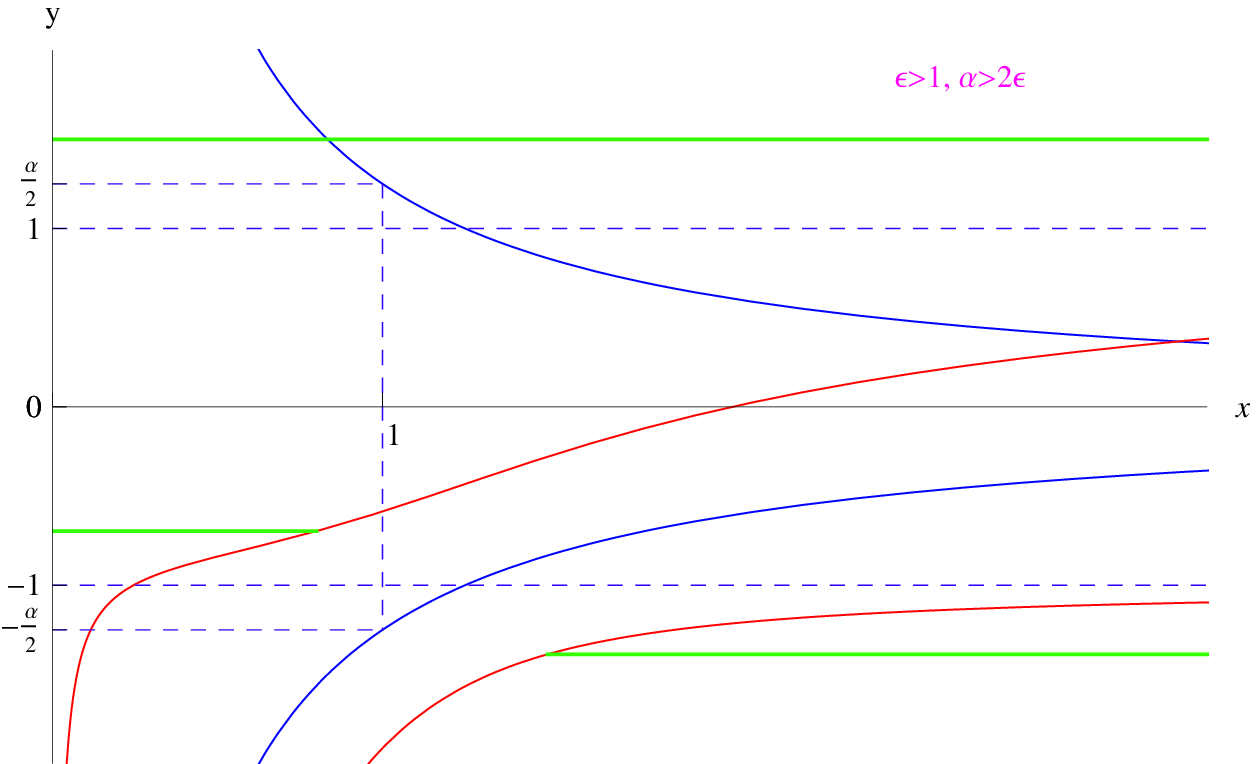}
\end{center}
\caption{The effective potentials for heavy shells at $\epsilon>1$ and $\alpha > 2 \epsilon$.}
\label{fig11}
\end{figure}
\begin{figure}[t]
\begin{center}
\includegraphics[angle=0,width=0.95\textwidth]{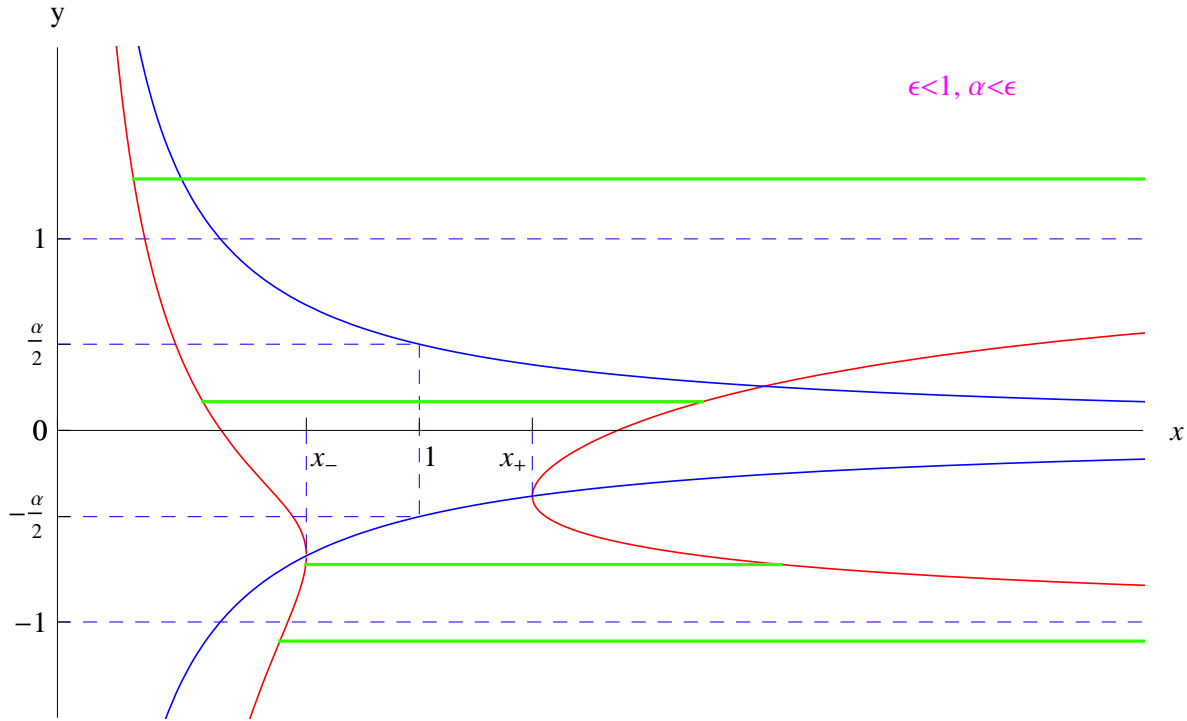}
\end{center}
\caption{The effective potential for the light shells at $\epsilon<1$ and $\alpha < \epsilon$.}
\label{fig12}
\end{figure}
\begin{figure}[t]
\begin{center}
\includegraphics[angle=0,width=0.95\textwidth]{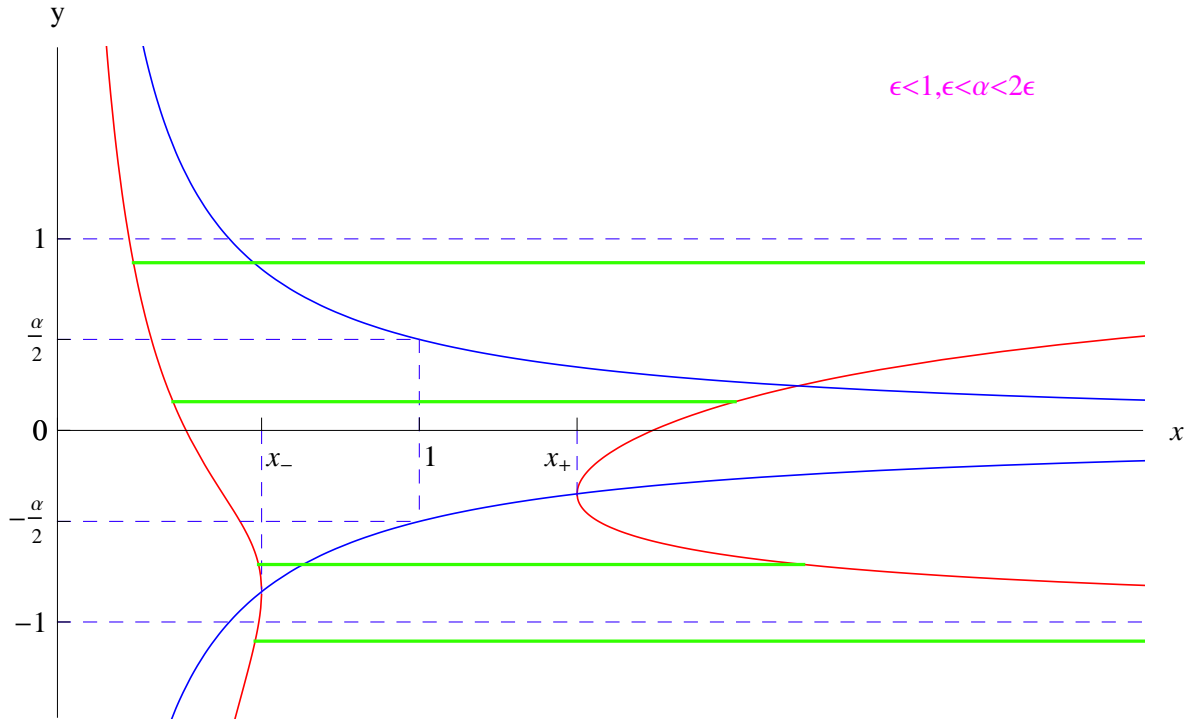}
\end{center}
\caption{The effective potential for the heavy shells at $\epsilon<1$ and
$\epsilon<\alpha <2\epsilon$.}
\label{fig13}
\end{figure}
\begin{figure}[t]
\begin{center}
\includegraphics[angle=0,width=0.95\textwidth]{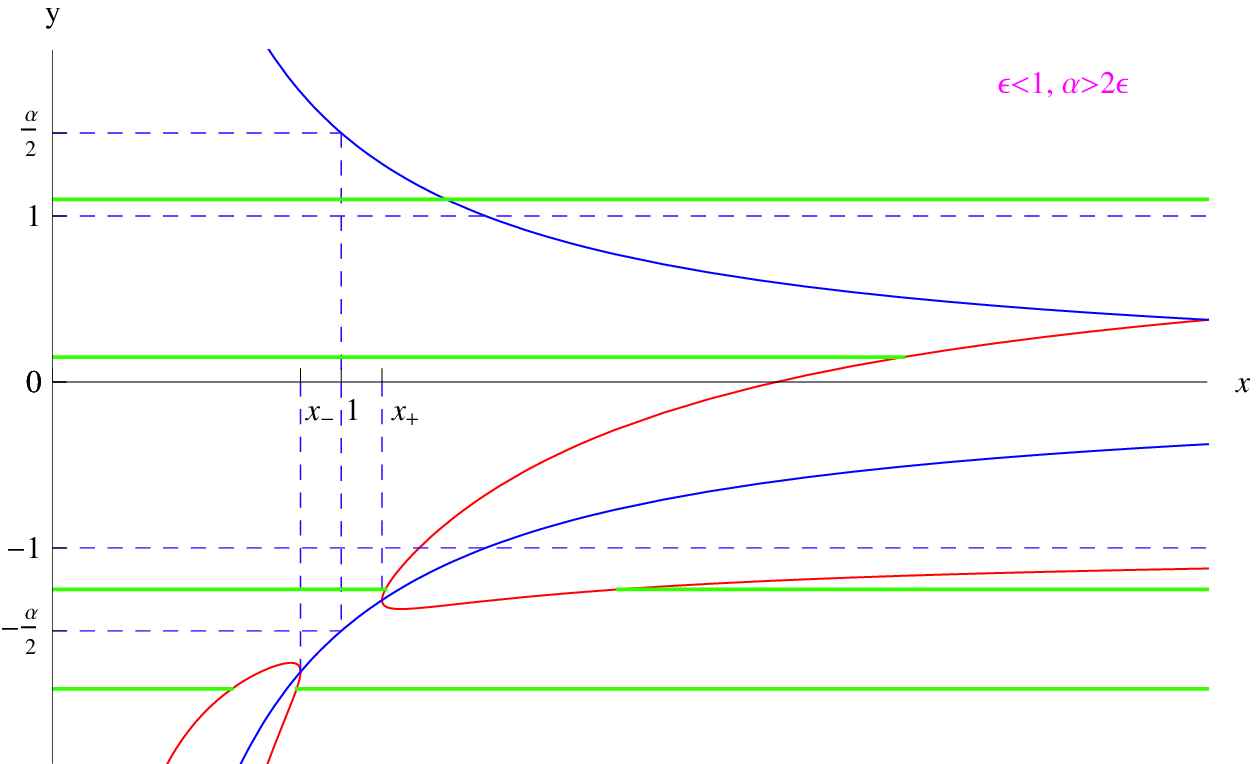}
\end{center}
\caption{The effective potentials for heavy shells at $\epsilon<1$ and $\alpha > 2 \epsilon$.}
\label{fig14}
\end{figure}

(1) the regions allowed for shells motion are:
\begin{eqnarray}
\label{rule1}
\mu > V_{\rm eff} \qquad  \mbox{if} \qquad \sigma_{\rm in} = + 1, \nonumber \\
\mu < V_{\rm eff} \qquad  \mbox{if} \qquad \sigma_{\rm in} = - 1. \nonumber
\end{eqnarray}

(2) how to determine the sign of $\sigma_{\rm in}$:
\begin{eqnarray}
\label{rule2}
\sigma_{\rm in} = + 1 \qquad \mbox{if} \qquad \mu > \tilde y_2, \nonumber \\
\sigma_{\rm in} = - 1 \qquad \mbox{if} \qquad \mu < \tilde y_2. \nonumber
\end{eqnarray}

(3) how to determine the sign of $\sigma_{\rm out}$:
\begin{eqnarray}
\label{rule3}
\sigma_{\rm out} = + 1 \qquad \mbox{if} \qquad \mu > y_2, \nonumber \\
\sigma_{\rm out} = - 1 \qquad \mbox{if} \qquad \mu < y_2. \nonumber
\end{eqnarray}
In the remaining part of the Section we will investigate the forms of the effective potential for different types of the inner R-N metrics, that depend on the values of the inner mass $m$ and electric charge $e$, and different values of the shell's bare mass $M$. As usual, it is convenient to use some dimensionless variables and parameters. So, we fix the inner total mass $m_{\rm in}$ and introduce the dimensionless radius $x = \rho/(G m_{\rm in})$ and three dimensionless parameters, $\epsilon = |e|/(\sqrt{G}m_{\rm in})$, $\alpha = M/m_{\rm in}$ and $\mu = \Delta m/M$. Then,
\begin{eqnarray}
\label{dimless}
y_1=\mu(\dot\rho=0)\equiv V_{\rm eff} = \sigma_{\rm in}\sqrt{1 - \frac{2}{x} + \frac{\epsilon^2}{x^2}} - \frac{\alpha}{2 x}, \qquad
y_2 = \frac{\alpha}{2 x}, \qquad \tilde y_2 = - \frac{\alpha}{2 x}
\end{eqnarray}
In these notations, the apparent horizons of the inner R-N black holes (at $\epsilon < 1$) are
\begin{equation}
\label{app}
x_{\pm} = 1 \pm \sqrt{1 - \epsilon^2},
\end{equation}
and we denote by $x_1$ and $x_2$ ($x_2 < x_1$) the abscissae of intersection points $y_1 = y_2$. Note, that the intersections $y_1 = \tilde y_2$ can exist only if $\sigma_{\rm in} = - 1$, and the corresponding abscissae are simply the horizon radii, $x_{\pm}$.

Let us start to study the effective potential. Consider, first, its asymptotical behavior when $x \to 0$ and $x \to \infty$. If $\sigma_{\rm in} = +1$, then
\begin{equation}
\label{xtozero}
 y_1=V_{\rm eff}\longrightarrow\frac{2\epsilon-\alpha}{2 x}-\frac{1}{\epsilon}\,, \qquad x\to0\,;
\end{equation}
so, for when $x \to 0$, $V_{\rm eff}\rightarrow+\infty$, if $2\epsilon>\alpha$, and $V_{\rm eff}\rightarrow- \infty$, if $2 \epsilon < \alpha$. For $2 \epsilon = \alpha$, $V_{\rm eff} (0) = - 1/\epsilon$. When $\sigma_{\rm in} = - 1$, then always $V_{\rm eff} \rightarrow - \infty $ for $x \to 0$. At infinity, $x \to \infty$, the behavior of the effective potential is the following. If $\sigma_{\rm in} = + 1$, then $V_{\rm eff} \rightarrow 1 - 0$. But, if $\sigma_{\rm in} = - 1$,
\begin{equation}
\label{xtoinf}
y_1 = V_{\rm eff} \longrightarrow -1 + \frac{2 - \alpha}{2 x}\,, \qquad x \to \infty \, ;
\end{equation}
so, $V_{\rm eff}$ approaches $y = -1$ from above, when $\alpha < 2$, and from below, when $\alpha > 2$.

Now, how about extrema? It is easy to see that the extremum condition is
\begin{equation}
\label{cond1}
\sigma_{\rm in} \sqrt{1 - \frac{2}{x} + \frac{\epsilon^2}{x^2}} = 2 (\frac{\epsilon^2}{x} - 1).
\end{equation}
So, the relevant solution to this equation should obey the inequality $\sigma_{\rm in}(\epsilon^2/x-1)>0$. We get
\begin{equation}
\label{xextr}
\frac{1}{x_{\rm extr}} = \frac{1}{\epsilon^2}\left( 1 + \alpha \sigma_{\rm in} \sqrt{\frac{\epsilon^2 - 1}{4 \epsilon^2 - \alpha^2}}\right)\, .
\end{equation}
It is clear that the extreme exist only if either simultaneously $\epsilon^2 > 1$ and $4 \epsilon^2 > \alpha^2$, or $\epsilon^2 < 1$ and $4 \epsilon^2 < \alpha$. We do not discuss here the nature of extrema, this will become quite evident while considering particular cases. Here we would like only to note, that for R-N black holes when $\epsilon < 1$, $x_{\rm extr} < x_-$ for $\sigma_{\rm in} = + 1$, and $x_{\rm extr} > x_+$ for $\sigma_{\rm in} = - 1$.

Finally, before coming to drawing Figures, we need to consider intersections of the effective potential curve $y_1$ with the curves $y_2$ and $\tilde y_2 (= - y_2)$. The latter is quite trivial: the intersections occur just at the horizons $x_{\pm}$ when they exist. Clearly, the intersections $y_1 = y_2$ exist only for $\sigma = + 1$. Their abscissae are
\begin{equation}
\label{x12}
x_{1,2} = 1 \pm \sqrt{1 + \alpha^2 - \epsilon^2} \,.
\end{equation}
If $\epsilon^2 < \alpha^2$, we have only one point with
\begin{equation}
\label{x1}
x_1 = 1 + \sqrt{1 + \alpha^2 - \epsilon^2} \,.
\end{equation}
For $\alpha^2 < \epsilon^2 < 1 + \alpha^2$ we have two intersections while for $\epsilon^2 > 1 + \alpha^2$ --- no intersections at all. Note, that the latter may happen only for the inner R-N metrics with naked singularity.

We now start drawing Figures for the effective potential and begin with the case when the inner metrics is a part of the extreme R-N black hole, i.e., when $\epsilon = 1$. Then,
\begin{equation}
\label{extrbh}
y_1 = V_{\rm eff} = \sigma_{\rm in} \left|1 - \frac{1}{x}\right| - \frac{\alpha}{2 x} \,.
\end{equation}
Though for such a metrics $\sigma_{\rm in}$ can not change its sign (it is either $+ 1$ everywhere, or $- 1$ everywhere) we prefer to put both curves on the same Figure. Due to the modulus sign, each curve has a jump in the first derivative at $x = 1$ (at doubled apparent horizon). Evidently, instead of drawing curves separately for $\sigma_{\rm in} = + 1$ and $\sigma_{\rm in} = - 1$, we can draw two intersecting smooth curves
\begin{equation}
\label{V1}
V_1 = 1 - \frac{1}{x} - \frac{\alpha}{2 x} \quad   \mbox{and} \quad
V_2 = - 1 + \frac{1}{x} - \frac{\alpha}{2 x},
\end{equation}
the resulting Figure being the same. The curves behave differently when $M/m_{\rm in} = \alpha < 2$, such shells we will call ``light'' shells, and when $M/m_{\rm in} = \alpha > 2$ --- these are ``heavy'' shells. The effective potentials are shown in Fig.~\ref{fig6} and Fig.~\ref{fig7}, respectively.

Each of the resulting Figures consists of four branches that begin at the vortex with the coordinates $(1, \, -\alpha/2)$. The upper branches bound the region where $\sigma_{\rm in} = + 1$, while the lower ones --- the region with $\sigma_{\rm in} = - 1$. One may observe the sharp difference between the potentials for light and heavy shells. In the case of light shells the left branch for $\sigma_{\rm in} = + 1$ goes to $+\infty$ when $x \to 0$, while in the case of heavy shells it goes down to $-\infty$. Also, the potentials for light shells form the wedges, looking down when $\sigma_{\rm in} = + 1$, and up --- when $\sigma_{\rm in} = - 1$, while for the heavy shells there are no wedges. The types of allowed motions are also different. The light shells have either bound trajectories with two turning points, or unbound ones with one turning point, and they never reach the naked singularity at $r = 0$. The heavy shells, on the contrary, start from the naked singularity, when $\sigma_{\rm in} = + 1$, and may have either bound trajectories with one turning point, or the unbound ones with no turning point at all. If $\sigma_{\rm in} = - 1$, the motion is always unbound with one turning point. Finally, one should distinguish between two cases: $0 < \alpha < 1$ and $1 < \alpha < 2$. In the first there are two intersection points of the effective potential $y_1 = V_{\rm eff}$ with $\sigma_{\rm in} = + 1$ and the curve $y_2$ where $\sigma_{\rm out}$ changes its sign, while in the second we have the single intersection point with the right hand branch of the effective potential. In our Fig.\ref{fig6} we showed only the latter case.

What will happen when we start to increase or decrease the electric charge $|e| = \epsilon\,\sqrt{G} m_{\rm in}$? The vortex disappears. For $\epsilon > 1$, the curve with $\sigma_{\rm in} = + 1$ will go up, and that with $\sigma_{\rm in} = - 1$ will go down. The wedges become the minimum and maximum of the potentials for $\sigma_{\rm in} = + 1$ and $\sigma_{\rm in} = - 1$, respectively. For $\epsilon < 1$ there forms the combined potential with two branches, left and right ones, each of them contains both the part with $\sigma_{\rm in} = + 1$ and that with $\sigma_{\rm in} = - 1$.

Let us begin with the case $\epsilon > 1$, i.e., when the space-time inside the shell is some part of R-N manifold with naked singularity. First of all, we need to generalize the notions of light and heavy shells. Now, the light shells are those with $\alpha < 2 \epsilon$. For them, the effective potential is shown in Fig.~\ref{fig8}.

In the Figure above we showed the case when there is only one intersection point of the effective potential $y_1$ for $\sigma_{\rm in} = + 1$ with the curve $y_2$ where $\sigma_{\rm out}$ changes its sign. This corresponds to the double inequality $\epsilon < \alpha < 2 \epsilon$ (surely, this requires $\alpha > 1$). In our Figure the value of the effective potential at the minimum is negative, it means that also should be $\epsilon^2 < 1 +\alpha^2/4$. Keeping $\alpha$ constant and increasing further $\epsilon$ we enter the region $\alpha^2 < \epsilon^2 < 1 + \alpha^2$ where curves $y_1$ and $y_2$ have two intersections. Note, that now it become possible to have $\alpha < 1, \; \epsilon^2 < 1 + \alpha^2/4$ as well. The effective potentials for $\alpha^2 < \epsilon^2 < 1 + \alpha^2$ and $\epsilon^2 > 1 + \alpha^2$ are shown in Figs.~\ref{fig9} and \ref{fig10}, respectively.

We did not show in these Figures the lower parts of the effective potentials corresponding to $\sigma_{\rm in} = - 1$, because they remain qualitatively the same. Note only that the lower branch (for $\sigma_{\rm in} = - 1$) has a maximum when $\alpha < 2$, and approaches  at infinity the line $y = - 1$ from above, and it is monotonically increases when $\alpha > 2$. The curves $\tilde y_2 = - y_2$ in either of the cases do not intersect the curves $y_1 = V_{\rm eff}$, since for $\epsilon > 1$ there are no horizons. Also, we do not discuss the special role played in our method by the points $y_1 = y_2$. This is postponed to the subsequent Sections where we will be dealing with the global geometries and Carter-Penrose diagrams.

And what about the heavy shells, $\alpha > 2 \epsilon$? Everything becomes much more simple: no minima, no maxima, just two increasing curves (for $\sigma = \pm 1$) and the curve $\tilde y_2 = - y_2$ in-between as is shown in Fig.~\ref{fig11}.

Let us turn to investigation of the most interesting case, $\epsilon < 1$, when the inner part of the complete manifold represents the R-N black hole. While before (for $\epsilon \ge 1$) the value of $\sigma_{\rm in}$ was given once and forever (i.e., there may be everywhere either $R_+$-region with $\sigma_{\rm in} = + 1$, or $R_-$-region with $\sigma_{\rm in} = - 1$), now it may change its sign dynamically during the shell's evolution. The global geometry of this inner space-time is much more rich, it contains $R_{\pm}$- and $T_{\pm}$-regions and also two types of apparent horizons. As we already know, when moving from $\epsilon = 1$ to $\epsilon < 1$, the vortex with four branches at the doubled horizon is divided into two horizons, $x_{\pm} = 1 \pm \sqrt{1 - \epsilon^2}$, and for the effective potential one obtains two separate curves, left hand $(0 < x \le x_-)$ and right hand $(x \ge x_+)$ ones, each of them contains two branches, the upper one with $\sigma_{\rm in} = + 1$ and the lower one with $\sigma_{\rm in} = - 1$. These two branches merge exactly at $x = x_-$ (on the left hand curve) and at $x = x_+$ (on the right hand curve, where the curve $\tilde y_2 = - y_2$ intersects the potential curves. Note that in the interval $x_- < x < x_+$ we have no potential curves at all because, in the Reissner-Nordstr\"om black holes between two horizons there lie $T_{\pm}$-regions where the existence of turning points with $\dot\rho = 0$ is impossible.

Again, one should distinguish between the light and heavy shells, with $\alpha < 2 \epsilon$ and $\alpha > 2 \epsilon$, respectively. In contrast to the case considered previously, now for the light shells the effective potential has no extrema, while for the heavy shells there is always maximum on the left hand curve and minimum on the right hand curve. The effective potentials for the light and heavy shells are shown in Fig.~\ref{fig12} and \ref{fig13}, correspondingly.

We showed in this Figure the case, when there exists a minimum in the left hand branch, but it is true only when $\alpha > 2$. For $\alpha < 2$ (and still $\alpha > 2 \epsilon$) the minimum disappears, and at infinity the curve will approach the line $y = - 1$ from above.

\section{Global geometries for the combined systems}

In the previous Section we introduced the effective potential in order to illustrate dynamics of thin shells immersed into the R-N manifold (which describes metrics outside an electrically charged point-like mass) and have drawn pictures for different types of its inner part (with naked singularity, extremal case and R-N black hole) and different values of shell's bare mass. In addition, two curves indicating the change of sign in $\sigma = \pm 1$ for inner and outer parts of the complete manifold were shown in the same pictures. It is these sign functions that determine, actually, the global geometry of the space-time under consideration. Let us remind that the allowed shell's trajectories are green horizontal lines $y = \Delta m/M = \mu = const$ lying above the effective potential curve $y = y_1$ for $\sigma_{\rm in} = + 1$ and below it for $\sigma_{\rm in} = - 1$. Moreover, if this green line goes above the corresponding curve of changing $\sigma's$ ($y = y_2$ or $y = \tilde y_2 = - y_1$), then $\sigma = + 1$, and $\sigma = - 1$ if it goes lower. Thus, the interplay between all of them allows us to construct the Carter-Penrose conformal diagram for any set of parameters of the combined system.

The aim of the present Section is to illustrate how our method works in all qualitatively different cases. For the sake of convenience, we list below the needed notations and definitions. The inner part of the R-N metrics is characterized by two parameters, the mass $m_{\rm in}$ and the electric charge $e$, while its outer part --- by the mass $m_{\rm out}$ (= the total mass of the system) and the same charge $e$ (remember that our shell is neutral). The shell has the bare mass $M$, and its total mass (which includes the gravitational mass defect) equals $\Delta m = m_{\rm out} - m_{\rm in}$. Also, we introduced the dimensionless parameters $\mu = \Delta m/M$, $\epsilon = |e|/(\sqrt {G} m_{\rm in})$, $\alpha = M/m_{\rm in}$ and the dimensionless radius of the shell $x = \rho/(G m_{\rm in})$. In these notations the effective potential $y_1$ looks as follows
\begin{equation}
y_1 = V_{\rm eff} = \sigma_{\rm in} \sqrt{1 - \frac{2}{x} + \frac{\epsilon^2}{x^2}} - \frac{\alpha}{2 x} \, , \nonumber
\end{equation}
while the curves $y_2 \,(\tilde y_2 = - y_2)$ where $\sigma_{\rm out} = \pm 1\, (\sigma_{\rm in} = \pm 1)$ change their signs are represented by
\begin{equation}
y_2 = \frac{\alpha}{2 x} \;, \;\;\;\; \tilde y_2 = - \frac{\alpha}{2 x} \, . \nonumber
\end{equation}
The apparent horizons of the inner R-N black holes lie at $x_{\pm} = 1 \pm \sqrt{1 - \epsilon^2}$.

\subsection{The case $\epsilon = 1$}
\begin{figure}[t]
\begin{center}
\includegraphics[angle=0,width=0.95\textwidth]{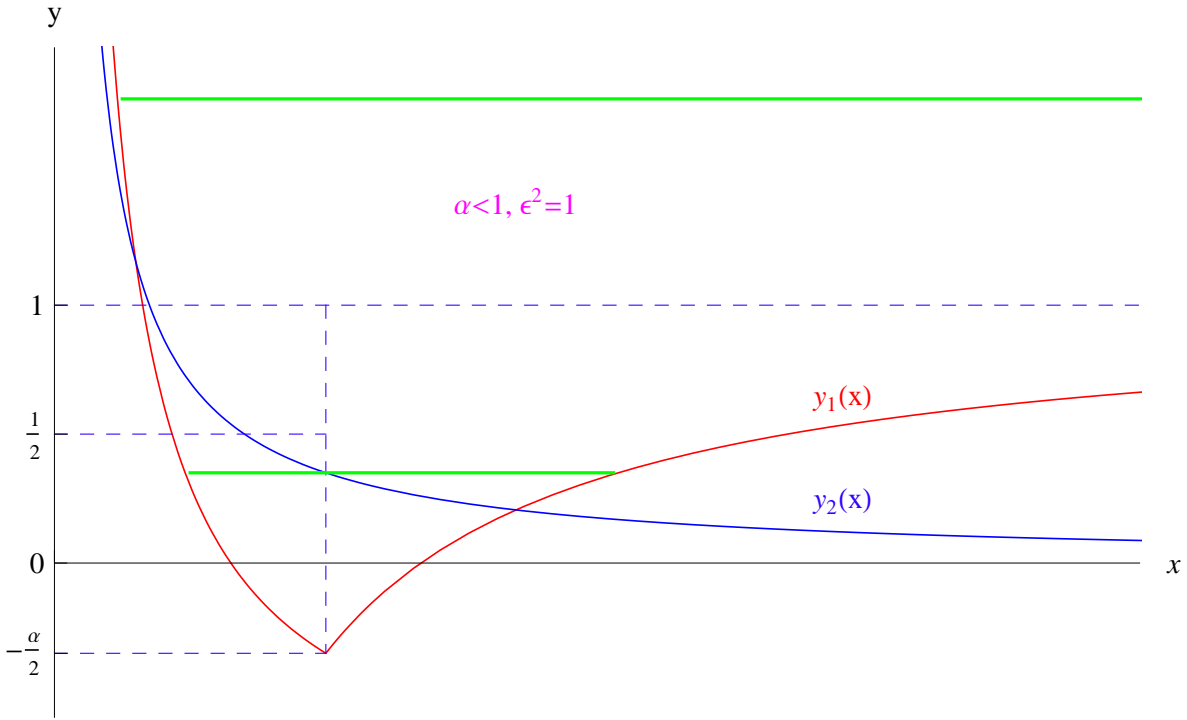}
\end{center}
\caption{The effective potentials at $\epsilon^2=1$ and $\alpha<1$.}
\label{fig15}
\end{figure}
\begin{figure}[t]
\begin{center}
\includegraphics[angle=0,width=0.53\textwidth]{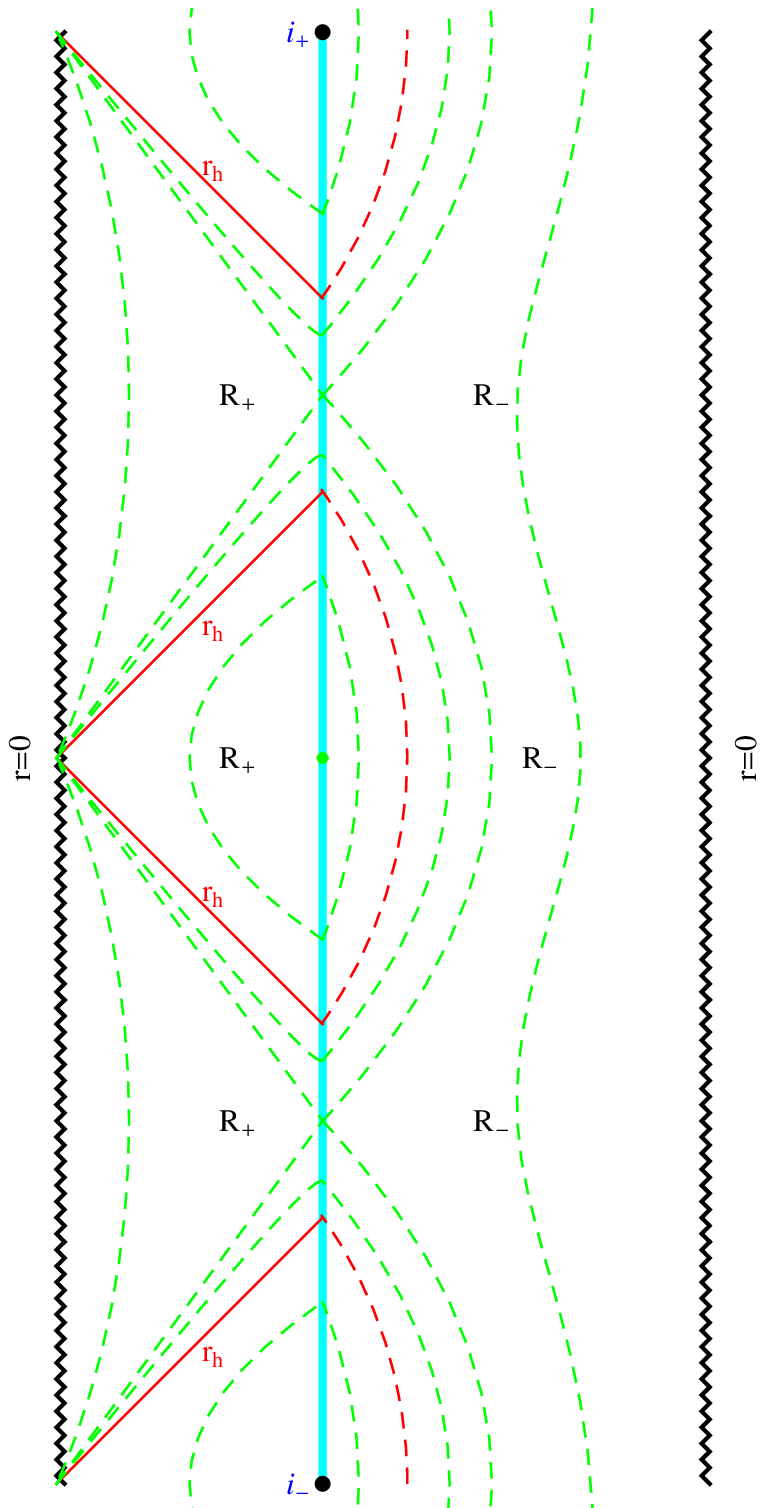}
\hfill
\includegraphics[angle=0,width=0.45\textwidth]{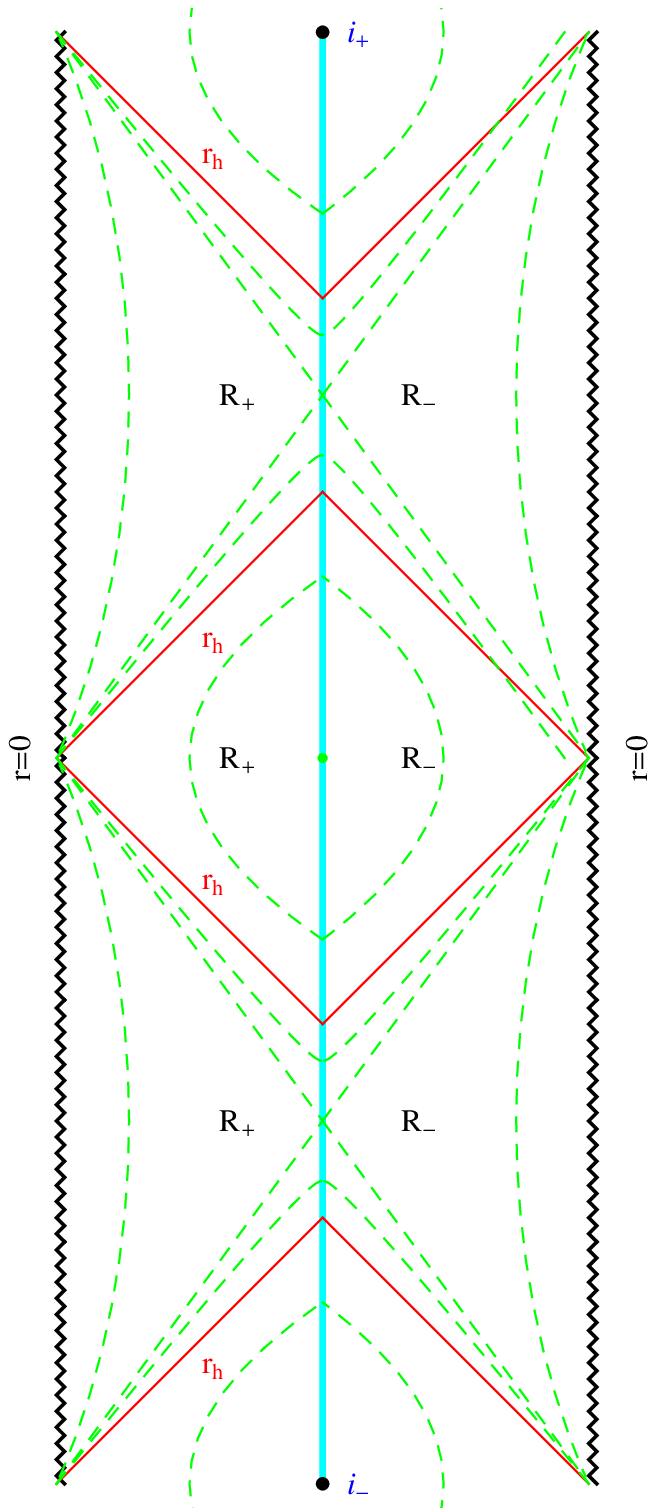}
\end{center}
\caption{The combined space-time manifolds at $\epsilon = 1$ for $\mu = \mu_1 < 0$ (left panel) and $\mu =0$ (right panel).}
\label{fig16}
\end{figure}
\begin{figure}[t]
\begin{center}
\includegraphics[angle=0,width=0.5\textwidth]{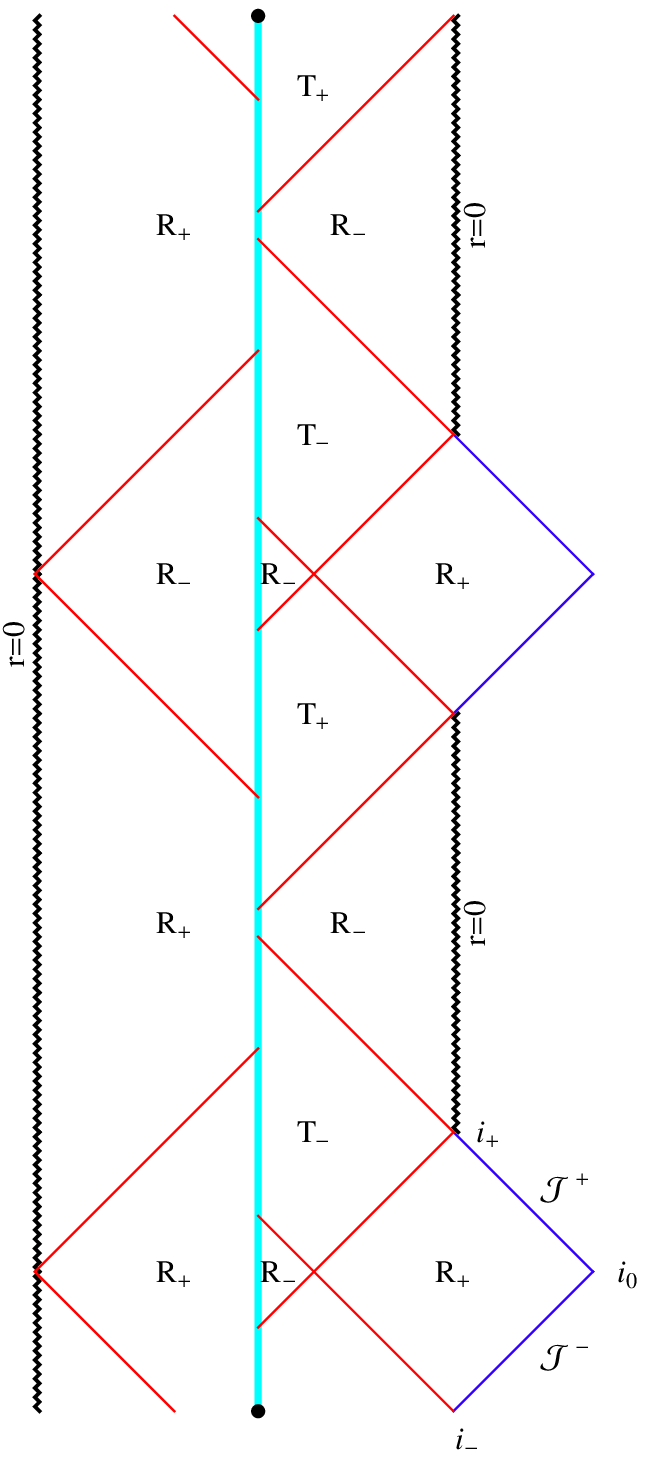}
\hfill
\includegraphics[angle=0,width=0.46\textwidth]{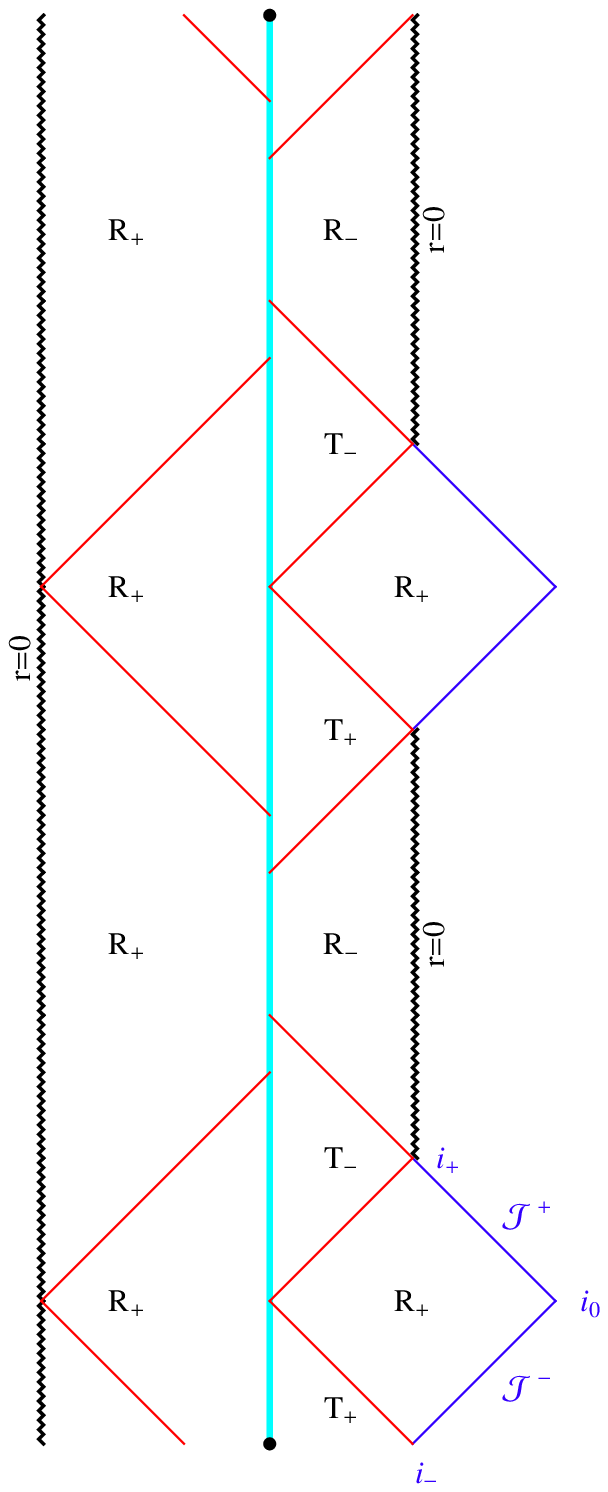}
\end{center}
\caption{The combined space-time manifolds at $\epsilon = 1$, $\alpha<1$ at $\mu =0$ (left panel) and $\mu =\mu_1=\alpha/[2 (1 - \alpha)]$ (right panel).}
\label{fig17}
\end{figure}
\begin{figure}[t]
\begin{center}
\includegraphics[angle=0,width=0.41\textwidth]{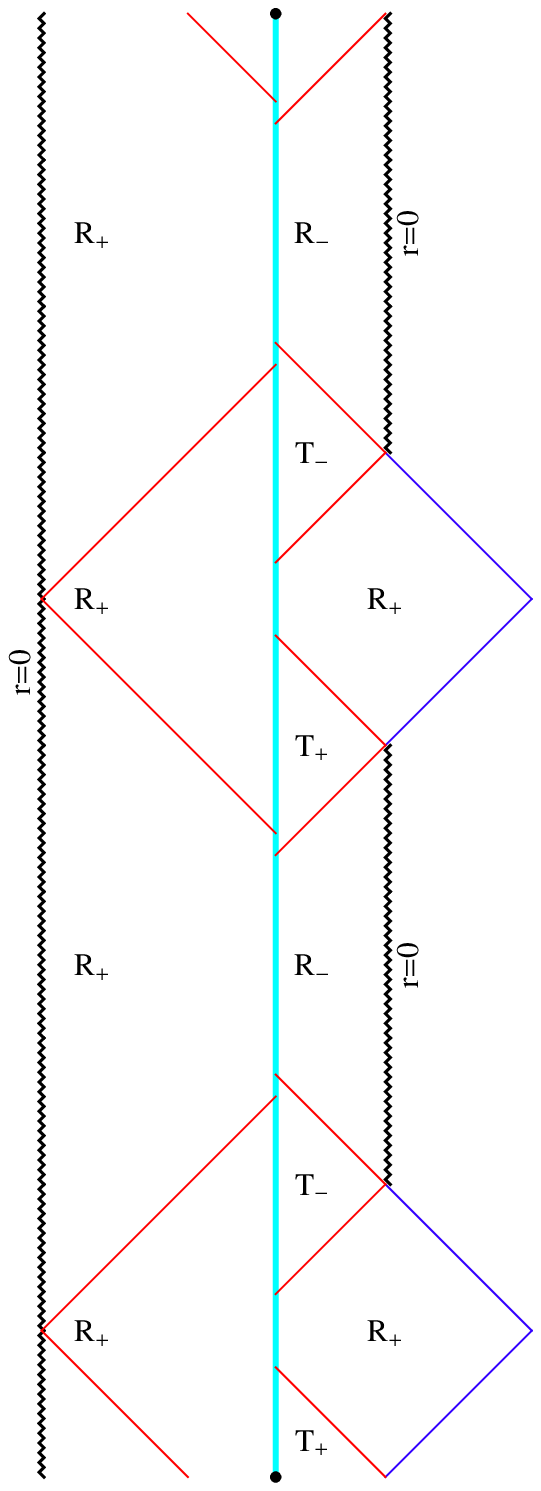}
\hfill
\includegraphics[angle=0,width=0.49\textwidth]{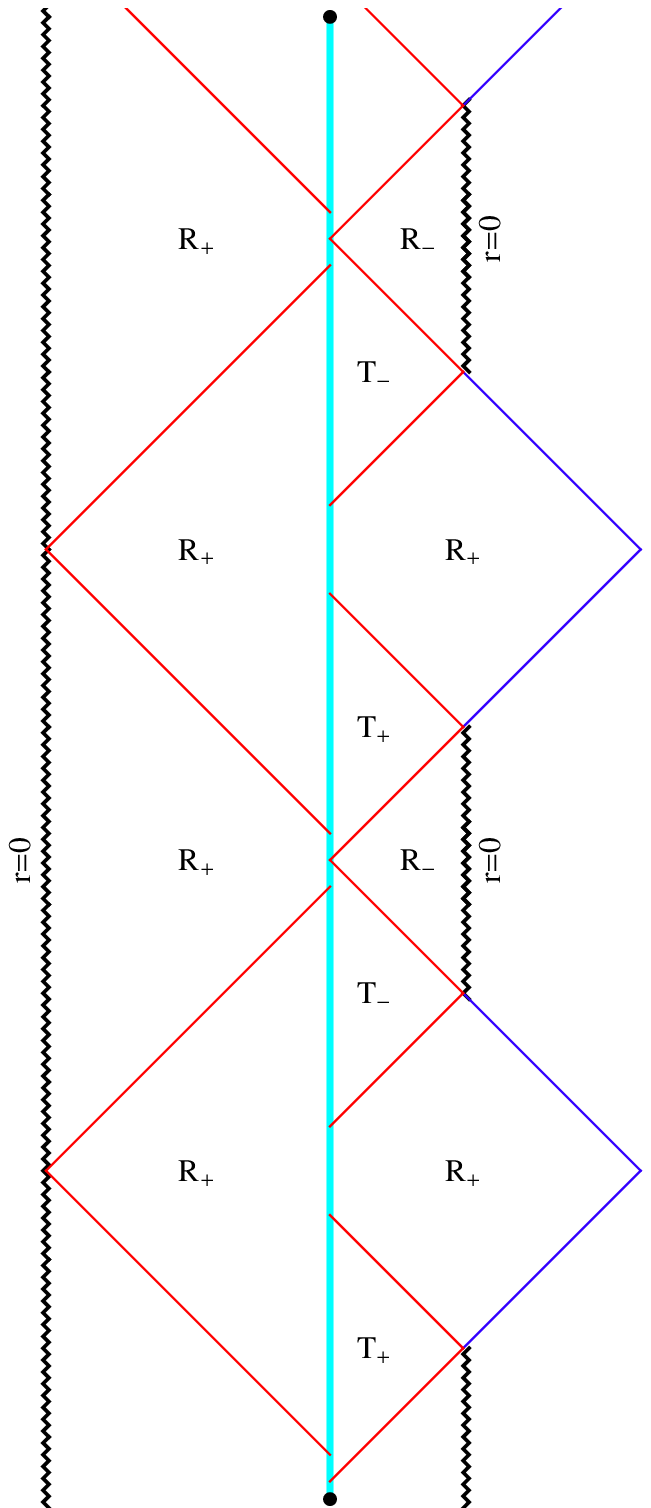}
\end{center}
\caption{The combined space-time manifolds at $\epsilon = 1$, $\alpha<1$ and
$2(1+\alpha)<\mu_3<\alpha/[2 (1-\alpha)]< 1$ (left panel) and $\mu =\mu_1=\alpha/[2 (1 - \alpha)]<1$ (right panel).}
\label{fig18}
\end{figure}
\begin{figure}[t]
\begin{center}
\includegraphics[angle=0,width=0.51\textwidth]{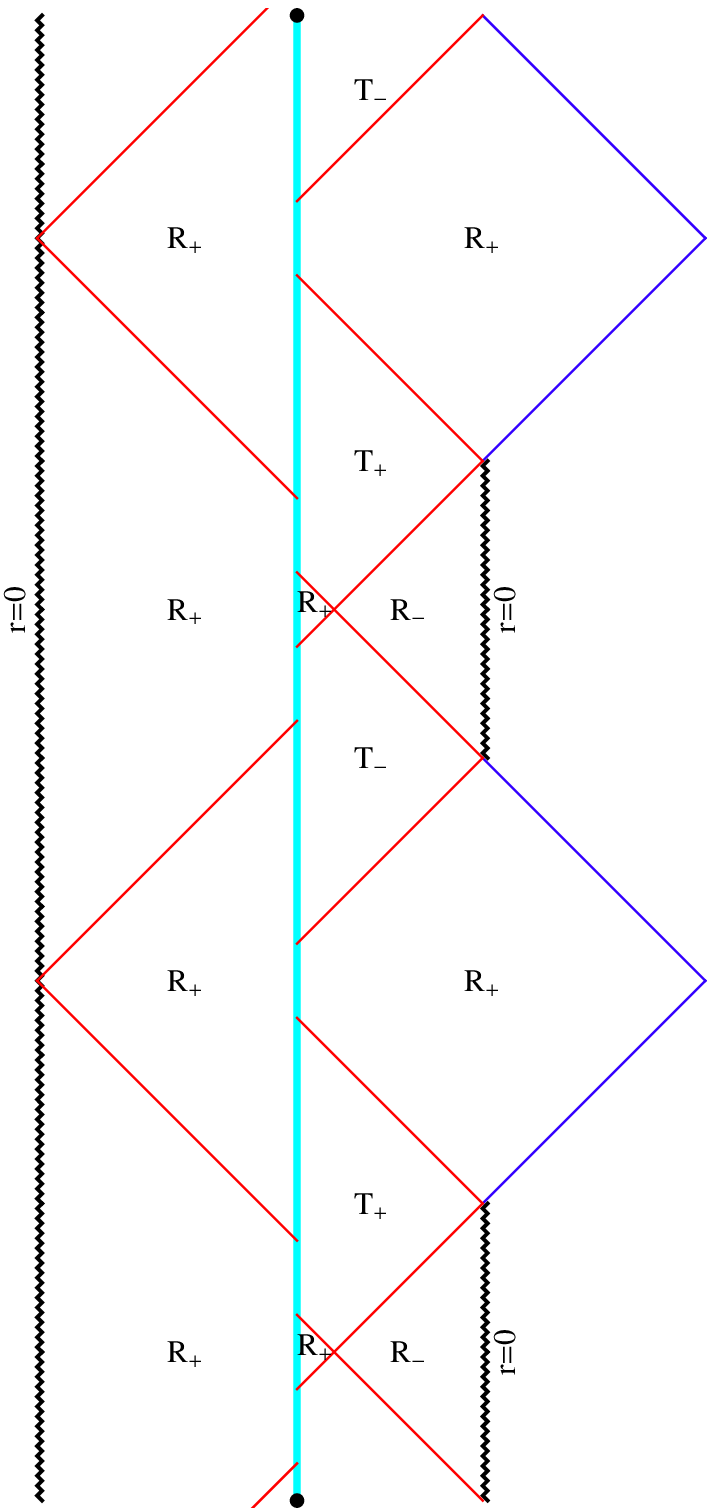}
\hfill
\includegraphics[angle=0,width=0.4\textwidth]{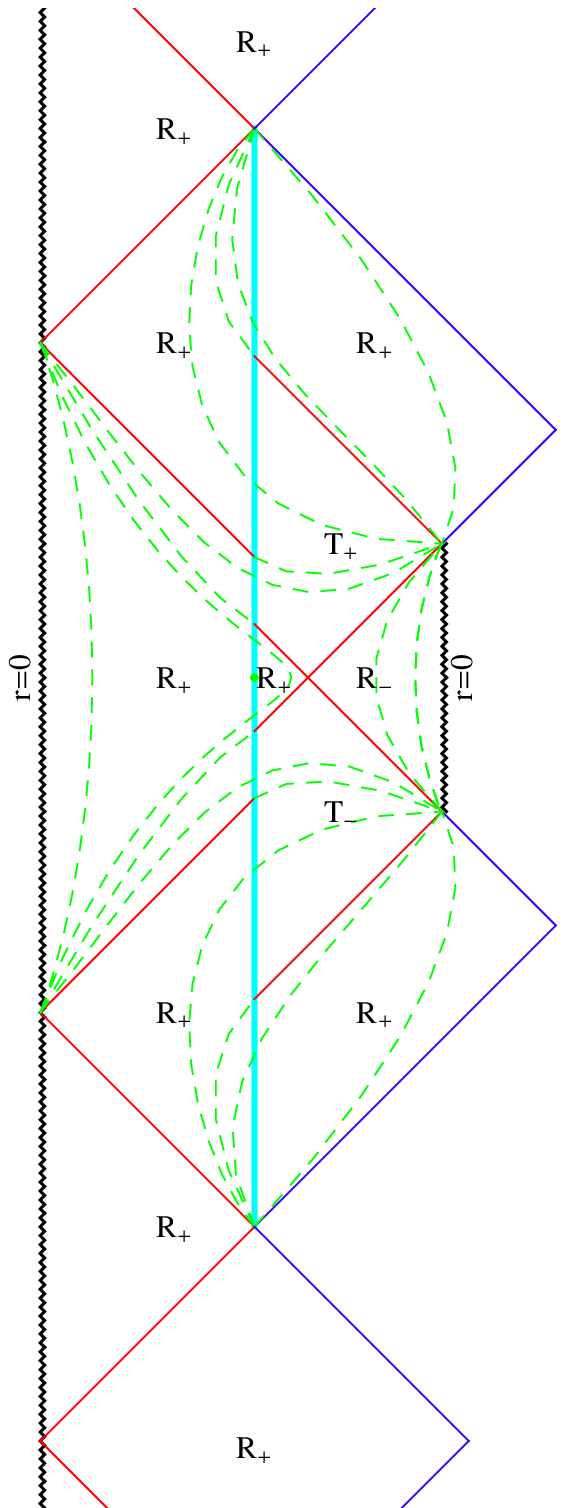}
\end{center}
\caption{The combined space-time manifolds at $\epsilon=1$, and, respectively, at $\alpha/[2 (1-\alpha)]<\mu_4<1$ and $\alpha<2/3<1$ (left panel) and at $\mu>1$
and $\alpha>2/3$ (right panel).}
\label{fig19}
\end{figure}
\begin{figure}[t]
\begin{center}
\includegraphics[angle=0,width=0.38\textwidth]{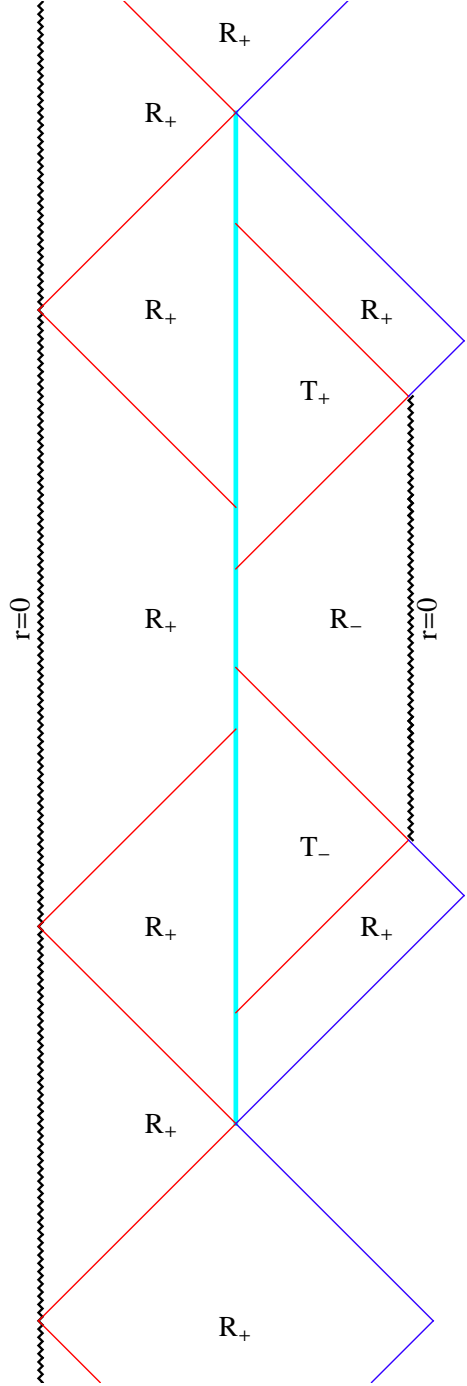}
\end{center}
\caption{The combined space-time manifolds at $\epsilon=1$, $1 < \mu < \alpha/[2 (1 - \alpha)]$
and $2/3< \alpha<1$.}
\label{fig20}
\end{figure}
\begin{figure}[t]
\begin{center}
\includegraphics[angle=0,width=0.8\textwidth]{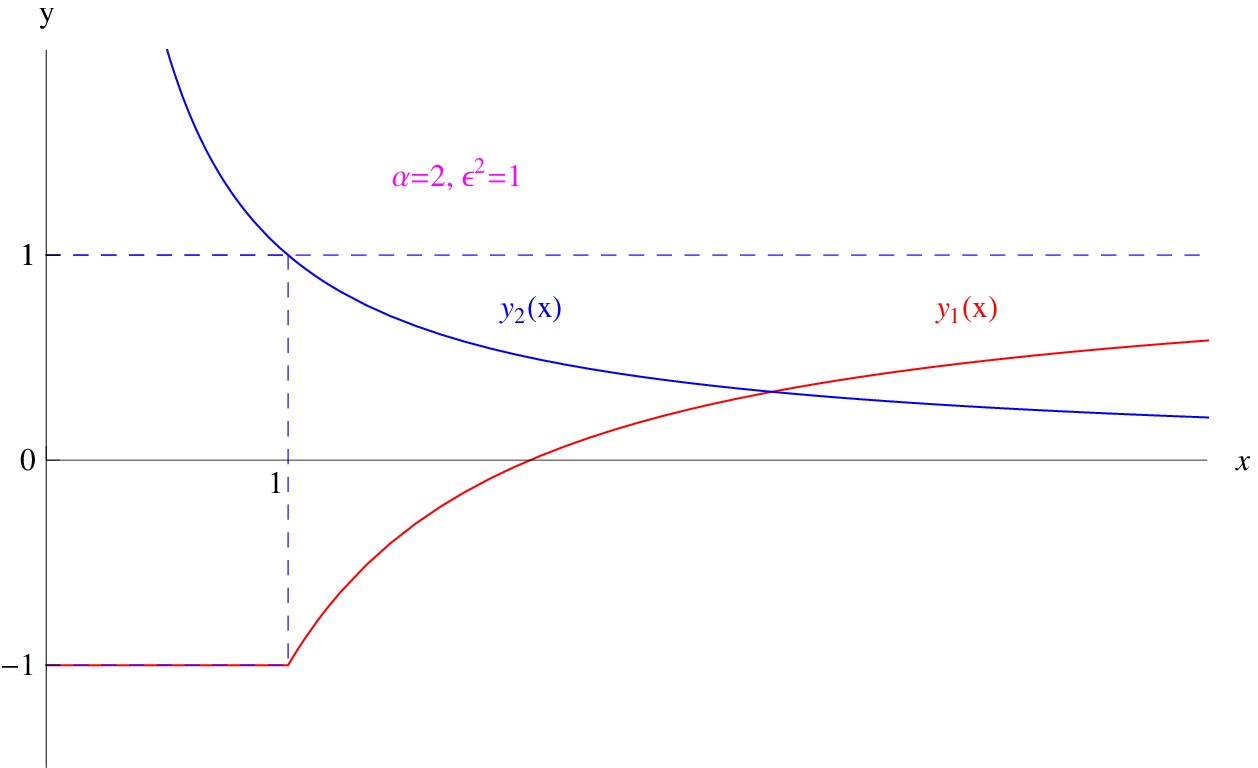}
\end{center}
\caption{The effective potentials at $\epsilon^2=1$, $\alpha=2$ and $\sigma_{\rm in}=+1$.}
\label{fig21}
\end{figure}
\begin{figure}[h]
\begin{center}
\includegraphics[angle=0,width=0.8\textwidth]{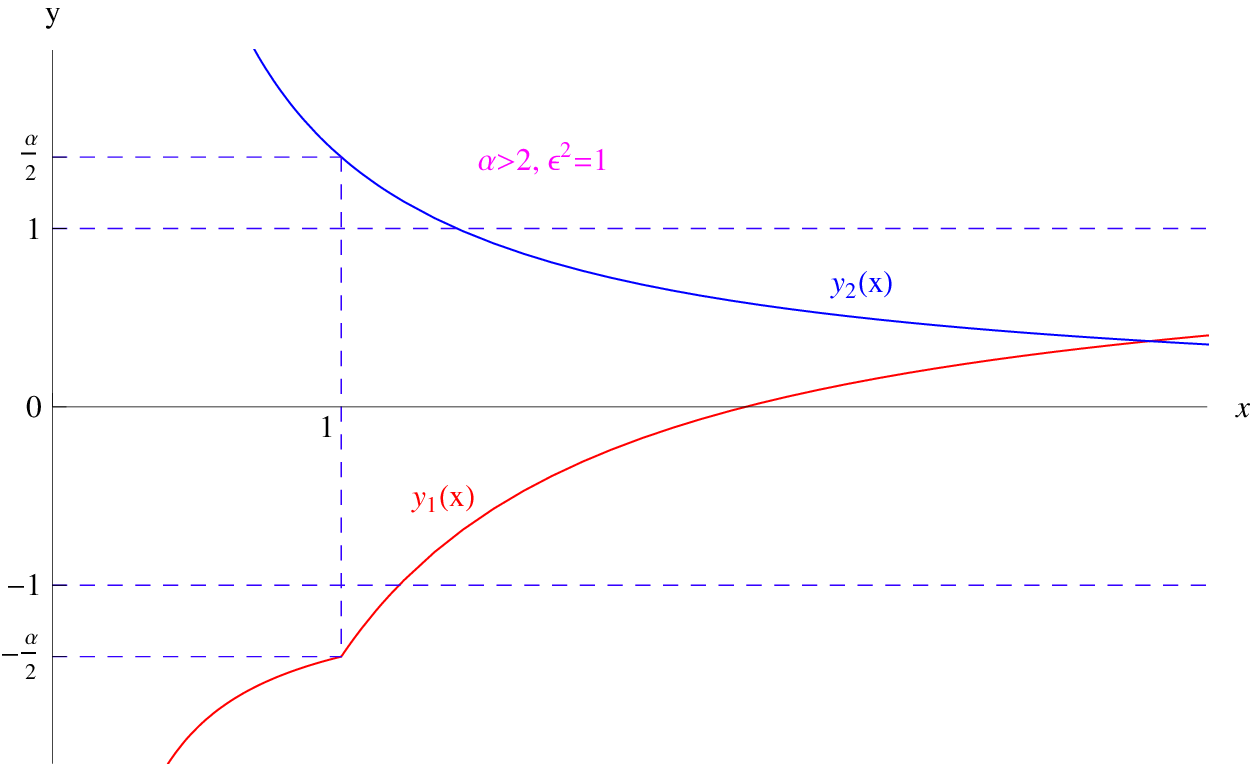}
\end{center}
\caption{The effective potentials at $\epsilon^2=1$, $\alpha > 2$ and $\sigma_{\rm in}=+1$.}
\label{fig22}
\end{figure}

We start with the case when the inner part of complete manifold is described by the metrics of extreme R-N black hole, i.e., when $\epsilon = |e|/(\sqrt{G} m_{\rm in}) = 1$. Thus, only three parameters are left free:

1. The value of $\sigma_{\rm in} = \pm 1$, which determines the part of the full extreme black hole space-time lying inside the shell, that one with the naked singularity $(\sigma_{\rm in} = + 1)$, or with the spatial infinity $(\sigma_{\rm in} = - 1)$;

2. The value of $\alpha = M/m_{\rm in}$ which enters both the effective potential $y = y_1$ and the curves of changing $\sigma's$: $y = y_2$ for $\sigma_{\rm out}$ and $y = \tilde y_2 = - y_2$ for $\sigma_{\rm in}$;

3. The value of $\mu = \Delta m/M$ that determines the position of the shell's trajectory $y = \mu$.

Consider, first, the situation with $\sigma_{\rm in} = + 1$. The effective potential curve has different profiles for light shells, $\alpha < 2$, and heavy ones, $\alpha > 2$. In the case of light shells, the pictures look differently for $\alpha < 1$, when curves $y_1$ and $y_2$ have two intersection points at $x = 1 \pm \alpha$, and for $1 < \alpha < 2$, when these curves have only one common point at $x =1+\alpha$. Moreover, if $\alpha < 2/3$, the left intersection point (at $x = 1 - \alpha$) lies below the horizontal line $y = 1$ separating the bound and unbound trajectories, while if $2/3<\alpha<1$, then $y_1 = y_2 (x = 1 - \alpha) > 1$. So, our starting point is $\epsilon=1$, $\sigma_{\rm in} = + 1$, $\alpha < 2/3$. The corresponding effective potential $y = y_1$ together with the curve $y=y_2$ is shown in Fig.~\ref{fig15}.
We enumerated different types of horizontal lines $y = \mu$ for further convenience. We did not draw the curve $y = \tilde y_2 = - y_2$ because it lies well below the present effective potential curve and, thus, does not influence the shell's trajectories. The common feature is that there are two turning points for bound and one --- for unbound motion.
\begin{figure}[h]
\begin{center}
\includegraphics[angle=0,width=0.49\textwidth]{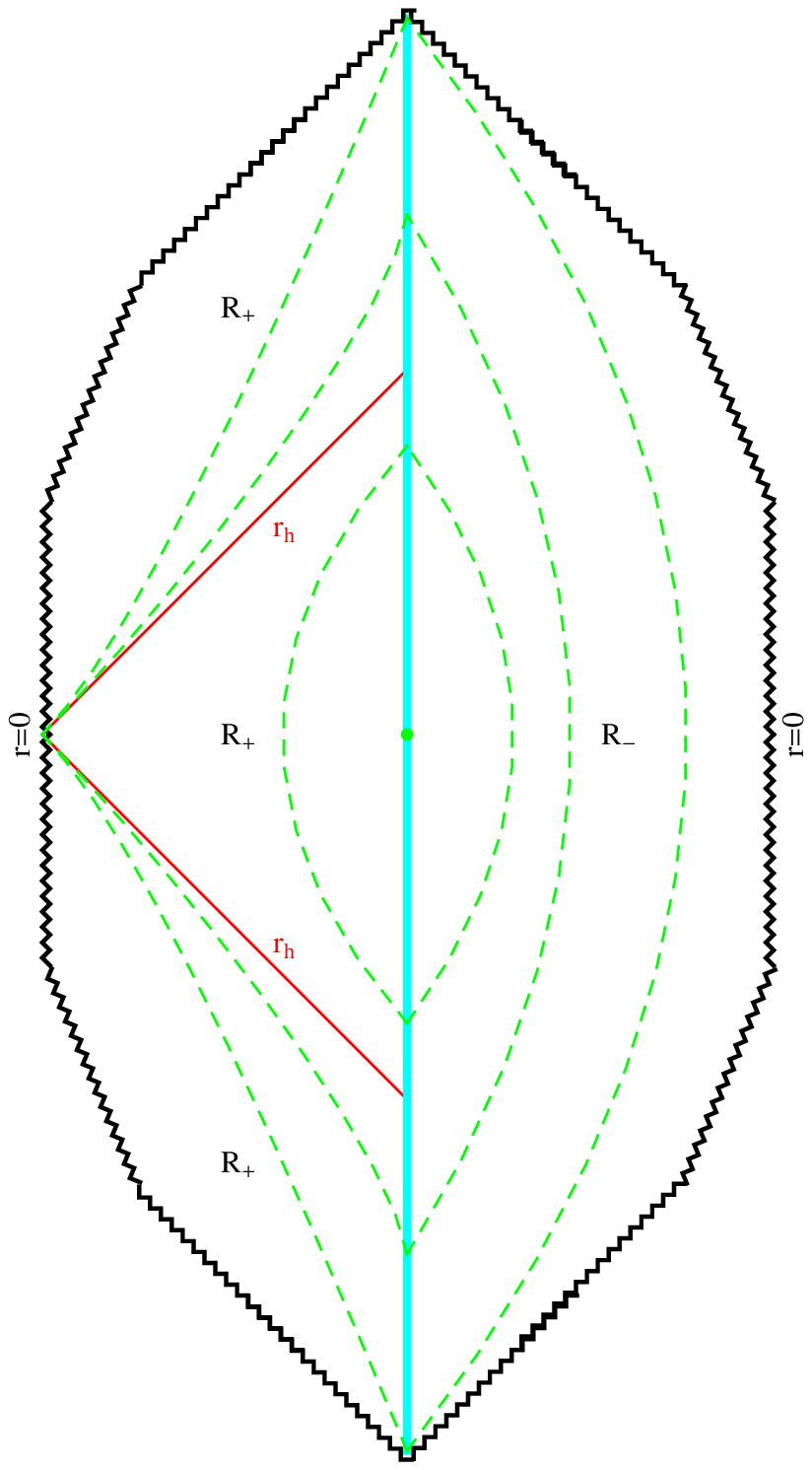}
\hfill
\includegraphics[angle=0,width=0.49\textwidth]{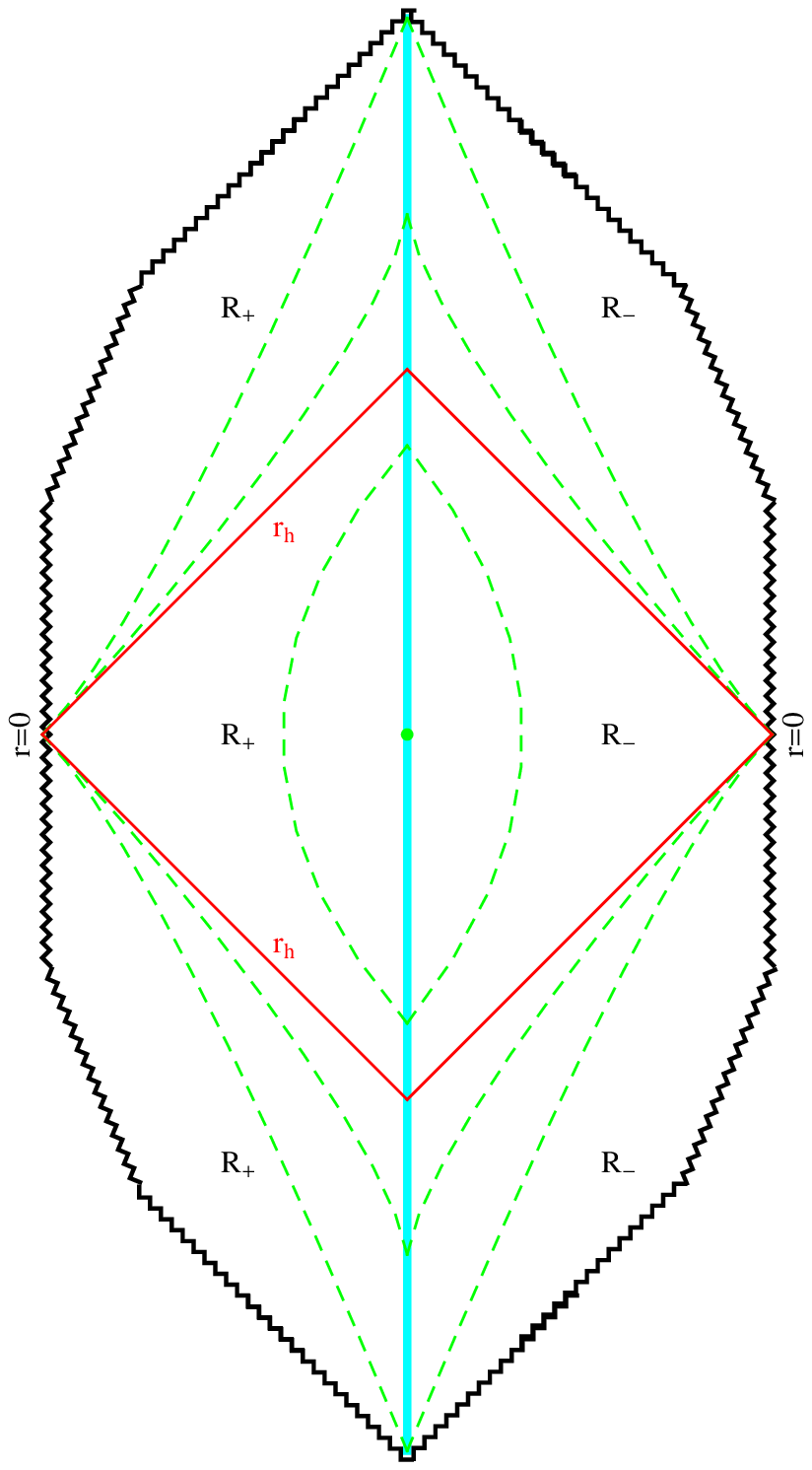}
\end{center}
\caption{The combined space-time manifolds at $\alpha>2$ and $<\mu<0$ (left panel) and $<\mu=0$ (right panel).}
\label{fig23}
\end{figure}
\begin{figure}[h]
\begin{center}
\includegraphics[angle=0,width=0.53\textwidth]{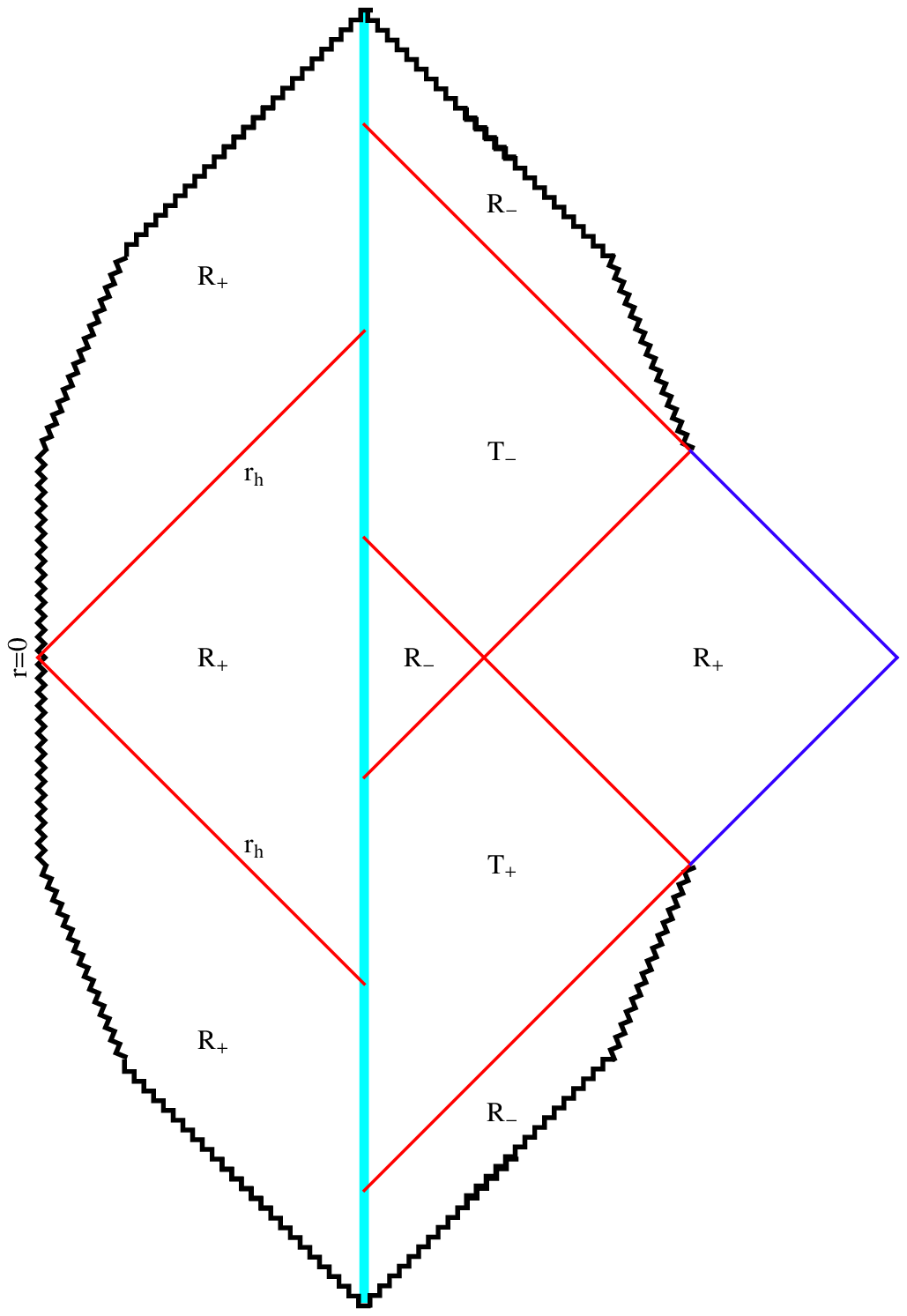}
\hfill
\includegraphics[angle=0,width=0.46\textwidth]{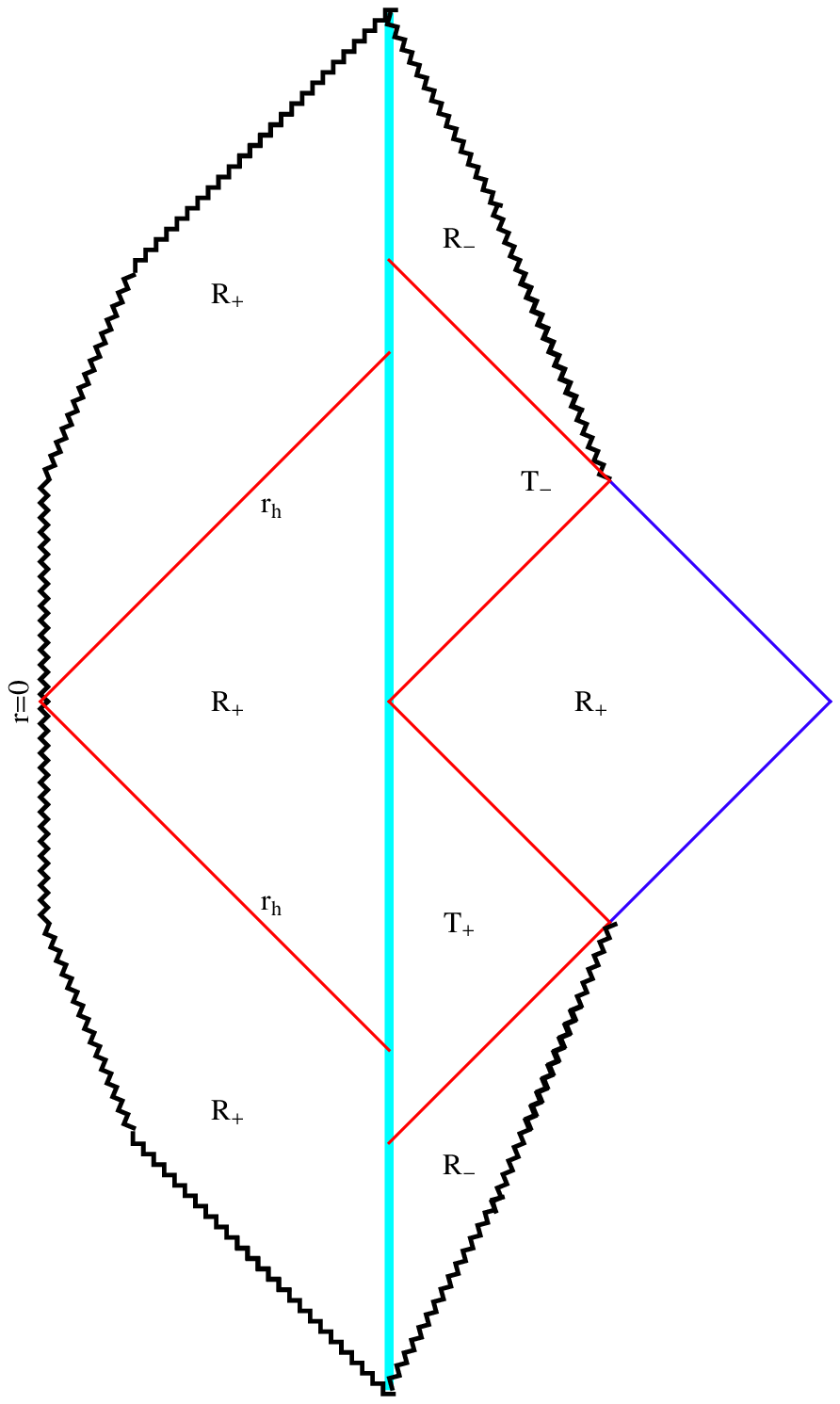}
\end{center}
\caption{The combined space-time manifolds at $\alpha>2$ and  $\mu<\alpha/[2 (1 + \alpha)]$ (left panel) and at $\alpha>2$ and  $\mu=\alpha/[2 (1 + \alpha)]$ (right panel).}
\label{fig24}
\end{figure}
\begin{figure}[h]
\begin{center}
\includegraphics[angle=0,width=0.52\textwidth]{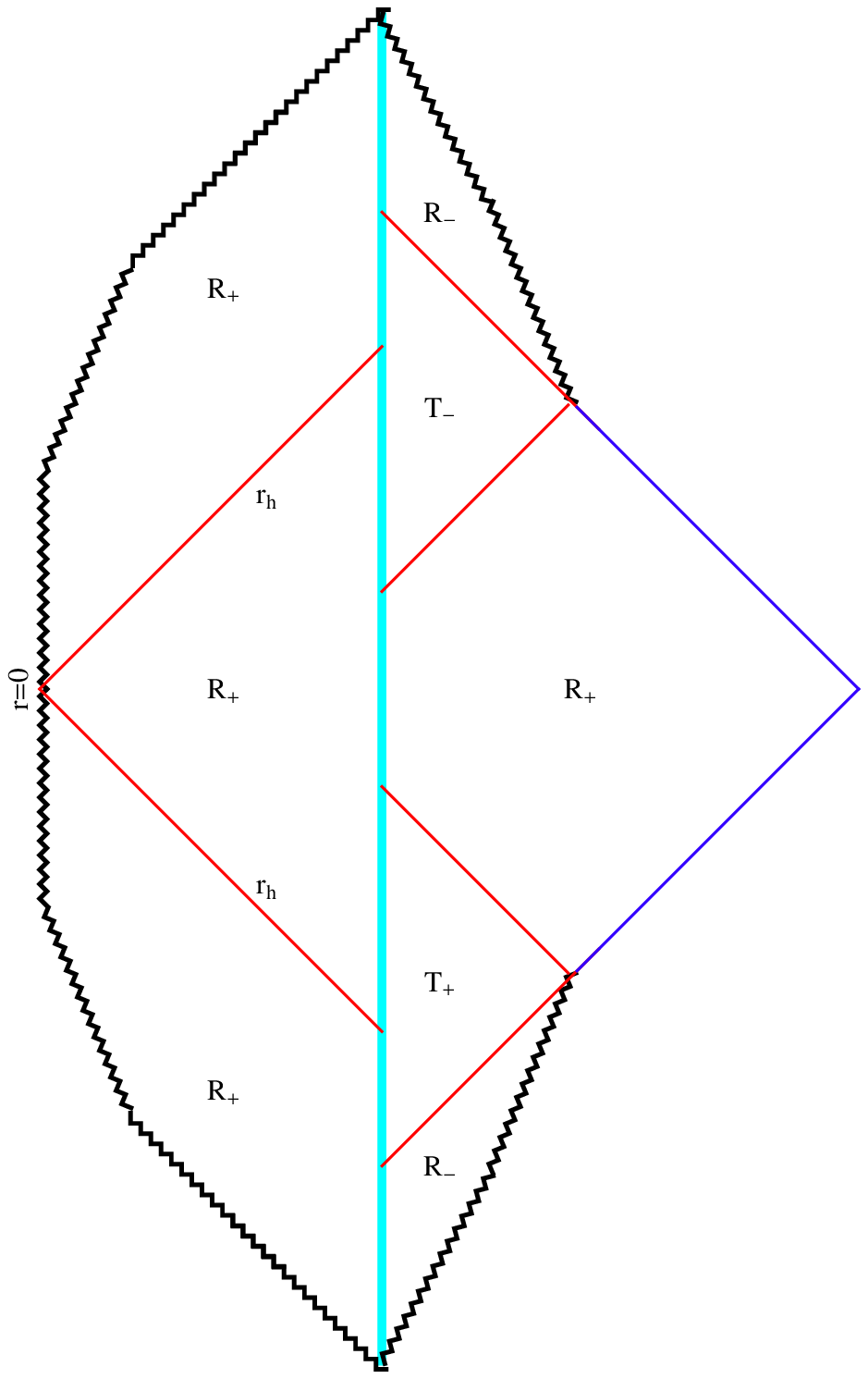}
\hfill
\includegraphics[angle=0,width=0.47\textwidth]{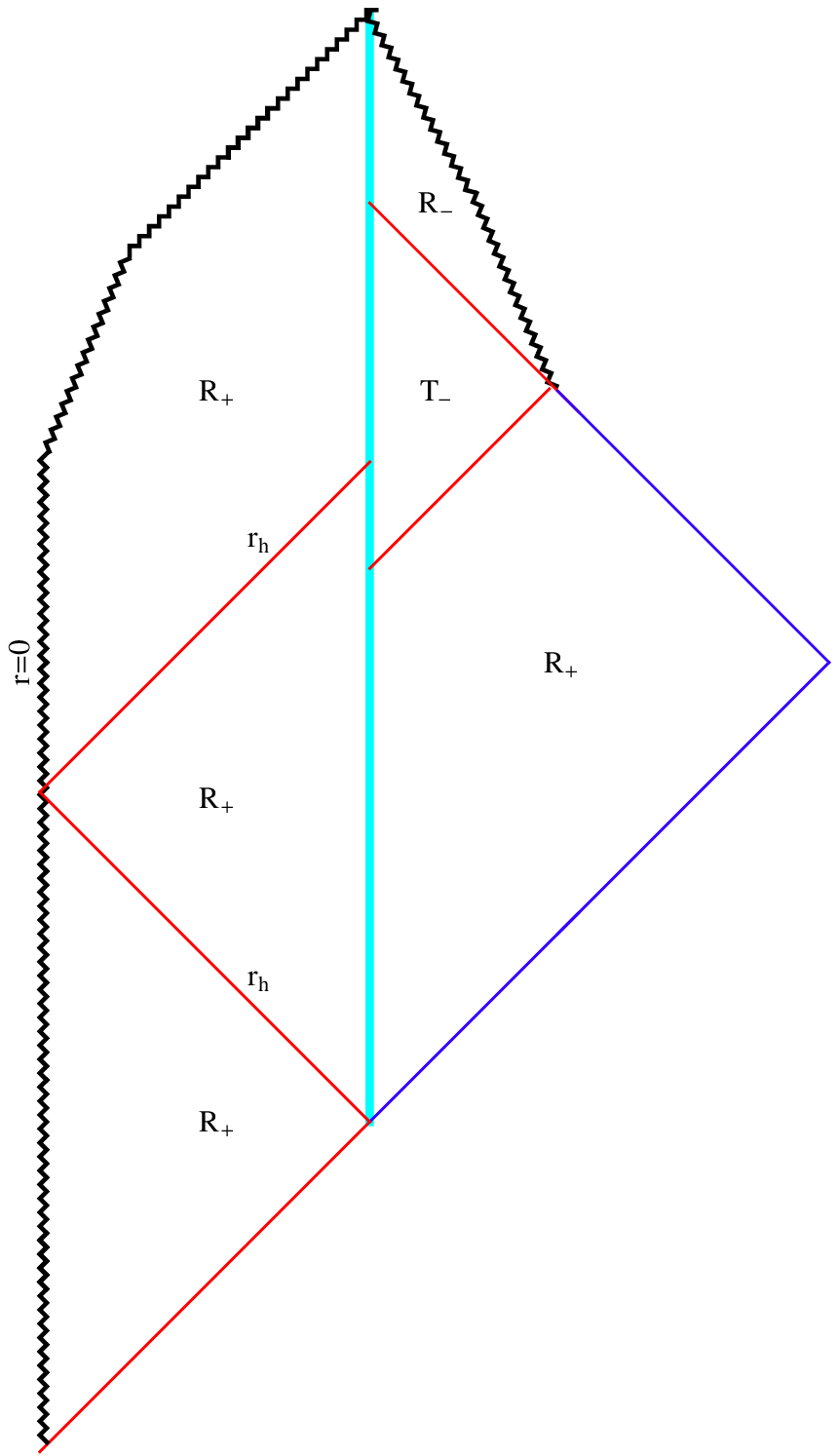}
\end{center}
\caption{The combined space-time manifolds at $\alpha>2$ and  $\alpha/[2 (1 + \alpha)]<\mu<1$ (left panel) and at $\alpha>2$ and  $\mu>1$ (right panel).}
\label{fig25}
\end{figure}

Let us begin to investigate the possible trajectories of the shell and construct the corresponding Carter-Penrose conformal diagrams that show us clearly the global geometry of the combined space-time manifolds. Consider, first, the trajectories for which the initial conditions are such that $\mu = \mu_1 < 0$. Since $\epsilon = 1$, the relation $m_{\rm out} < m_{\rm in}$ and $\Delta m < 0$ means that the outer part of the complete manifold corresponds to the case of naked singularity. Also, the horizontal line $y = \mu_1 < 0$ lies clearly below the curve $y = y_2(> 0)$, and so $\sigma_{\rm out} = - 1$. The Carter-Penrose diagram for this case is shown in Fig.~\ref{fig16} (left panel). The shell's motion is bound with two turning points. On the diagram the left hand turning point lies between the singularity at zero radius and the doubled horizon, while the right hand one is beyond the horizon. By dashed curves we indicated the surfaces of constant radii, this helps better understanding of the global geometry.

In the limiting case $\mu = 0$ the Carter-Penrose diagram is left-right symmetric (see right panel in Fig.~\ref{fig16}) . Both inside and outside the shell we have the extreme R-N black hole.

Now, let us turn to the positive values of $\mu$ and consider the case when the horizontal line $y = \mu$ lies below the right hand (lower) intersection point $y_1 = y_2$, i.\,e., $0 < \mu_2<\alpha/[2(1+\alpha)]$. In this case $\sigma_{\rm out} = - 1$ everywhere on the trajectory. Since with $\mu$ increasing from zero value, the turning points $x_1$ and $x_2$ goes, respectively, to the left and to the right off the primarily doubled horizon, the continuity properties of the relevant expressions require that $x_1 < x_-$ and $x_2 < x_+$, until we reach the level $\mu = y_1 = y_2$ at $x = x_2$ when this turning point lies exactly at the event horizon $x_+$ of the outer metrics. So, for $\mu = \mu_2$, the right hand turning point lies in the $R_-$-region of the outer metrics outside the event horizon $(x = x_+)$ (with asymptotically flat infinity), while the left hand one --- in the $R_-$-region beyond the inner (Cauchy) horizon $x =x_-$ near the singularity at zero radius. The corresponding conformal diagram (with the shell's trajectory in light-blue) for the combined manifold is shown in Fig.~\ref{fig17} (left panel).

When $\mu = \alpha/[2 (1 + \alpha)]$, the right hand turning point $(x = x_2)$ coincides with the event horizon $x =x_+$ (just at the bifurcation point) of the R-N metrics outside the shell. In the case of
$2(1+\alpha)<\mu_3<\alpha/[2 (1-\alpha)]< 1$, where the last inequality holds because of $\alpha < 2/3$, the shell starts, say, from one of the turning points $x_2$ in the $R_+$-region $(\sigma_{\rm out} = + 1)$ of the outer metrics outside the event horizon $x_+$. This shell collapses and goes through the $T_-$-region between two horizons, where $\sigma_{\rm out}$ changes sign, then crosses the inner horizon $x_-$ and enters the $R_-$-region $(\sigma_{\rm out} = - 1)$ near the singularity and reaches afterwards the another turning point $x_1$. When $\mu = \alpha/[2 (1 - \alpha)]$, this second turning point lies exactly at the inner horizon $x = x_1 = x_-$. The corresponding Carter-Penrose diagrams are shown in Figs.~\ref{fig17} and \ref{fig18}.

The trajectories with $\alpha/[2 (1 - \alpha)] < \mu_4 < 1$  differ from that of $\mu_3$-type in that the turning point $x_1 (< x_-)$ lies now in the $R_+$-region near the singularity of the outer metrics< and $\sigma_{\rm out}=+1$ everywhere. The global geometry is shown in Fig.~\ref{fig19} (left panel).

If $\mu > 1$, the motion of the shell is unbound, we denoted it as of $\mu_5$-type. The shell starts to collapse from the past temporal infinity in the $R_+$-region of the outer metrics and comes through the $T_-$-region to the $R_+$-region near the singularity at zero radius where there lies the single turning point $x_1 \, (< x_-)$, and it then expands through the $T_+$-region into the $R_+$-region outside the event horizon $(x_+)$ (which is different from the primary one) and, finally, the motion ends at the future temporal infinity. The Carter-Penrose diagram is finite in time for the outer part, but still has the infinite ladder structure for its inner part. It is shown in Fig.~\ref{fig19} (right pamel) with the corresponding level-lines $r=const$.

Let now be $2/3< \alpha<1$. What is changing? The left hand intersection point moves up, $y_1 = y_2 > 1$, so, the $\mu_4$-type trajectories for bound motion disappear. Instead, the unbound motion with the left hand turning point $x_1 \, (<x_-)$ in $R_-$-region near the singularity becomes possible for $1 < \mu < \alpha/[2 (1 - \alpha)]$. The Carter-Penrose diagram for this case is shown in Fig.~\ref{fig20}. All other types of the diagrams were already given, and it is needless to repeat drawing.

If $1 < \alpha < 2$, the situation is even more simple. The curves $y_1$ and $y_2$ have only one intersection --- with the right hand branch of the effective potential. So, for $\mu >\alpha/[2 (1 + \alpha)]$ all the trajectories will undergo the change of $\sigma_{\rm out}$, and the left hand turning point $x_1 \, (< x_-)$ lies in the $R_-$-region of the outer metrics near singularity. Again, all the corresponding diagrams were drawn before.

We are coming now to investigation of heavy shells, when $\alpha > 2$. There is no more a decreasing branch of the effective potential, so, there can be only one turning point for bound motion and no turning points at all for unbound motion. The shells are moving from the singularity at zero radius, then either reach the turning point and collapse back, or escape to infinity. In the limiting case $\alpha = 2$ the picture for the curves $y_1$ and $y_2$ is shown in Fig.~\ref{fig21}, while the case corresponding to $\alpha > 2$ is shown in Fig.~\ref{fig22}.

We confine ourselves to positive values of the system's total mass $m_{\rm out}$. In our notations this means that $\mu > -1/\alpha> - 1/2$. Thus, as it can be easily seen in the picture above, there are four different types of allowed shell's trajectories: $-1/\alpha<\mu_1<0$, $0 < \mu_2 <\alpha/[2 (1 + \alpha)]$, $2 (1 + \alpha) < \mu_3 < 1$ and $\mu_4 > 1$. We are not going to discuss the details here, it was already done before. We just show the corresponding Carter-Penrose diagrams one by one in Figs.~\ref{fig23}-\ref{fig25}. Meanwhile, two notes are in order: (1) at the singularity at zero radius (which is the starting point of all the heavy shell trajectories) we always have $\sigma_{\rm out} = - 1$; (2) in the case of unbound motion we showed only the collapsing shell, while the time reversal picture is also possible.

\begin{figure}[t]
\begin{center}
\includegraphics[angle=0,width=0.57\textwidth]{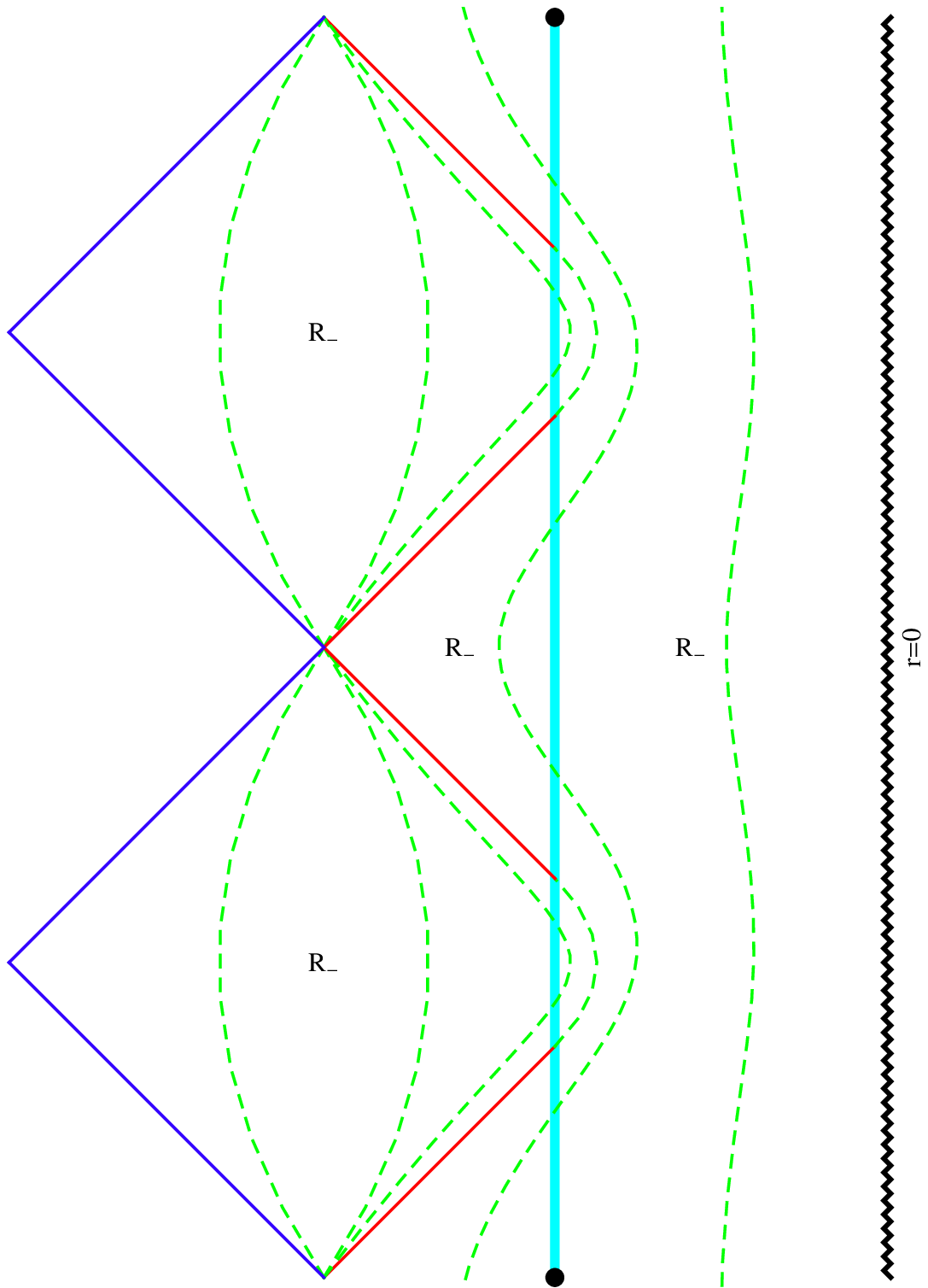}
\hfill
\includegraphics[angle=0,width=0.42\textwidth]{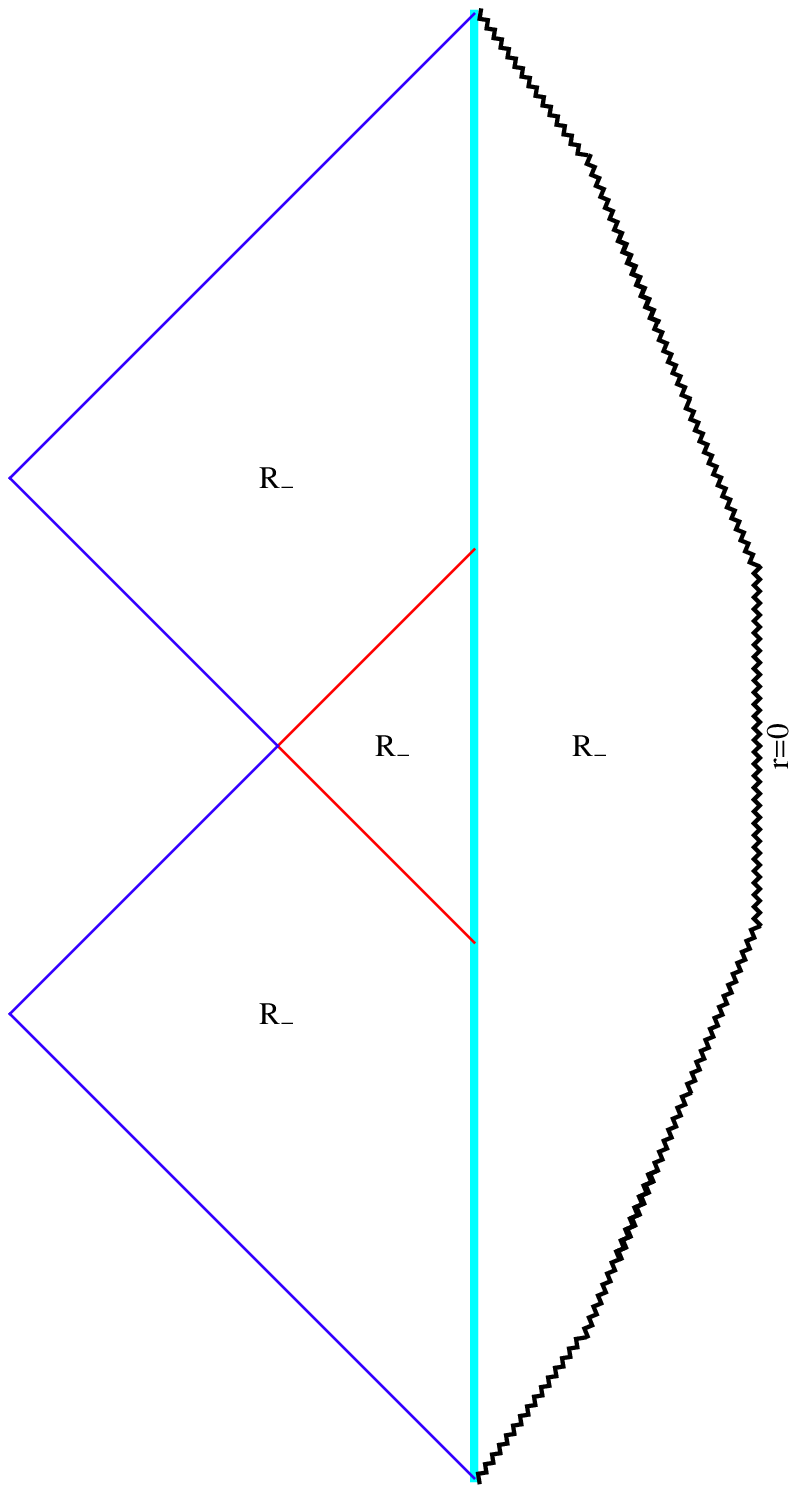}
\end{center}
\caption{The combined space-time manifolds for the light shells at $\alpha<2$ and, respectively,  $-1<\mu<-\alpha/2$ (left panel) and $\mu<-1$ (left panel).}
\label{fig26}
\end{figure}

We investigated all possible global geometries in the situation when the inner part of the complete manifold represents the extreme R-N black hole with $\sigma_{\rm in} = + 1$. And how about $\sigma_{\rm in} = - 1$? In such a case there is no center (with zero value of radius) inside the shell, but instead, there are infinities (spatial, null and temporal ones). Again, the pictures of the effective potential are different for light and heavy shells. We would like to remind that now for the allowed trajectories the corresponding parts of the horizontal lines $y = \mu$ should lie below the curves $y = y_1$. As it was done before, we begin with the light shells, $\alpha < 2$. We see that there are two types of the trajectories: for bound motion with two turning points if $- 1 < \mu < -\alpha/2$, and that ones for unbound motion if $\mu < - 1$. The Carter-Penrose diagrams are shown in Fig.~\ref{fig26}.

\begin{figure}[t]
\begin{center}
\includegraphics[angle=0,width=0.8\textwidth]{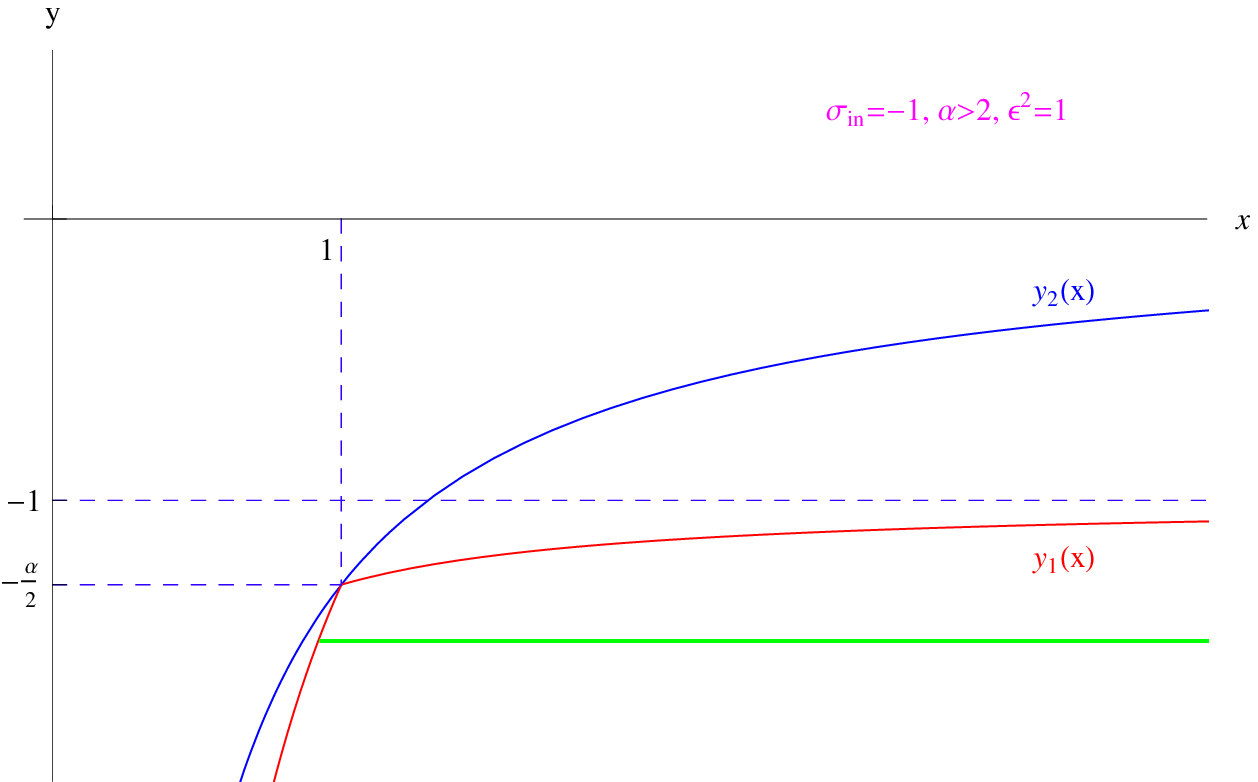}
\end{center}
\caption{The effective potential for heavy shells  at $\alpha>2$, $\epsilon^2=1$ and $\sigma_{\rm in}=-1$.}
\label{fig27}
\end{figure}

For heavy shells, $\alpha>2$, the effective potential looks as follows in Fig.~\ref{fig27}.
\begin{figure}[h]
\begin{center}
\includegraphics[angle=0,width=0.8\textwidth]{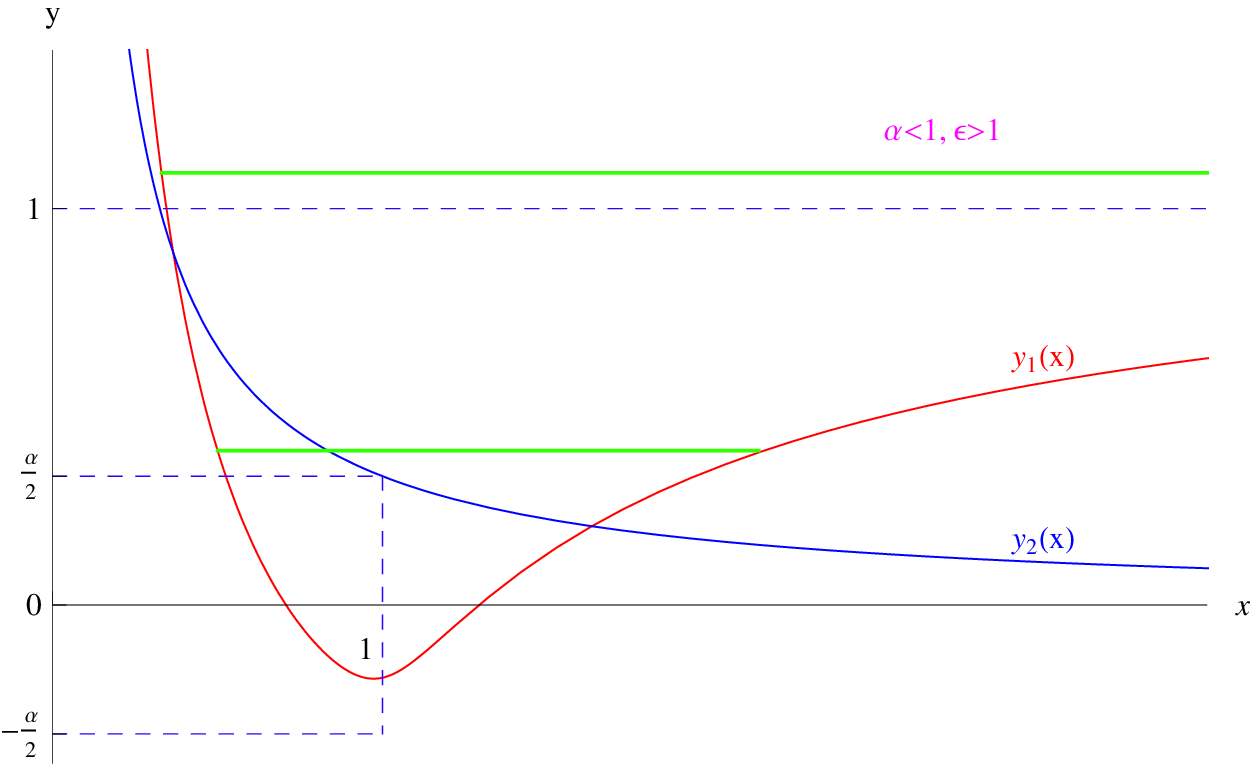}
\end{center}
\caption{The effective potential at $\alpha < 1 (< \epsilon < 2 \epsilon)$.}
\label{fig28}
\end{figure}
Evidently, there can exist only unbound trajectories with one turning point when $\mu<-\alpha/2<-1$. The conformal diagram is the same as for the light shells in the case of unbound motion.

\subsection{The case $\epsilon > 1$}

In the present Section we investigate the case when the inner part of combined space-time manifold is represented by the Reissner-Nordstrom metrics with naked singularity, i.e., when $\epsilon > 1$. Here, as before, we will consider separately the cases $\sigma_{\rm in} = + 1$ and $\sigma_{\rm in} = - 1$. We already know that if $\sigma_{\rm in} = + 1$, the effective potential curve $y = y_1 = V_{\rm eff}$ behaves differently at $x \to 0$ for ``light'' shells, $\alpha < 2 \epsilon$, and ``heavy'' ones, $\alpha > 2 \epsilon$. In the case of light shells the minimal value of the effective potential may be both negative and positive, and the interplay between the curves $y_1$ and $y_2$ ( which is of most importance in constructing the global geometries) becomes rather tricky.

Of course, we begin with the light shell, $\alpha < 2 \epsilon$. Let us fix the value of $\alpha$ and gradually increase $\epsilon$, starting from $\epsilon = 1$. In the course of doing this we meet three key points:

1. $\epsilon_1^2 = \alpha^2$, when there appears the second (left) intersection point of the curves $y_1$ and $y_2$;

2. $\epsilon_2^2 = 1 +\alpha^2/4$, when the minimum of the effective potential crosses the abscissae axis;

3. $\epsilon_3^2 = 1 + \alpha^2$, when two intersection points $y_1 = y_2$ merges. Note that the order of the first two points may be reversal ($\epsilon_1 < \epsilon_2$ if $\alpha^2 < 4/3$).

We start our consideration from the case $\alpha < 1 (< \epsilon < 2 \epsilon)$. Then, for $\sigma _{\rm in} = + 1$ and $\epsilon$ slightly greater  than $1$, the picture of the curves $y_1 = V_{\rm eff}$ and $y_2$ is shown in Fig.~\ref{fig28}.

\begin{figure}[t]
\begin{center}
\includegraphics[angle=0,width=0.525\textwidth]{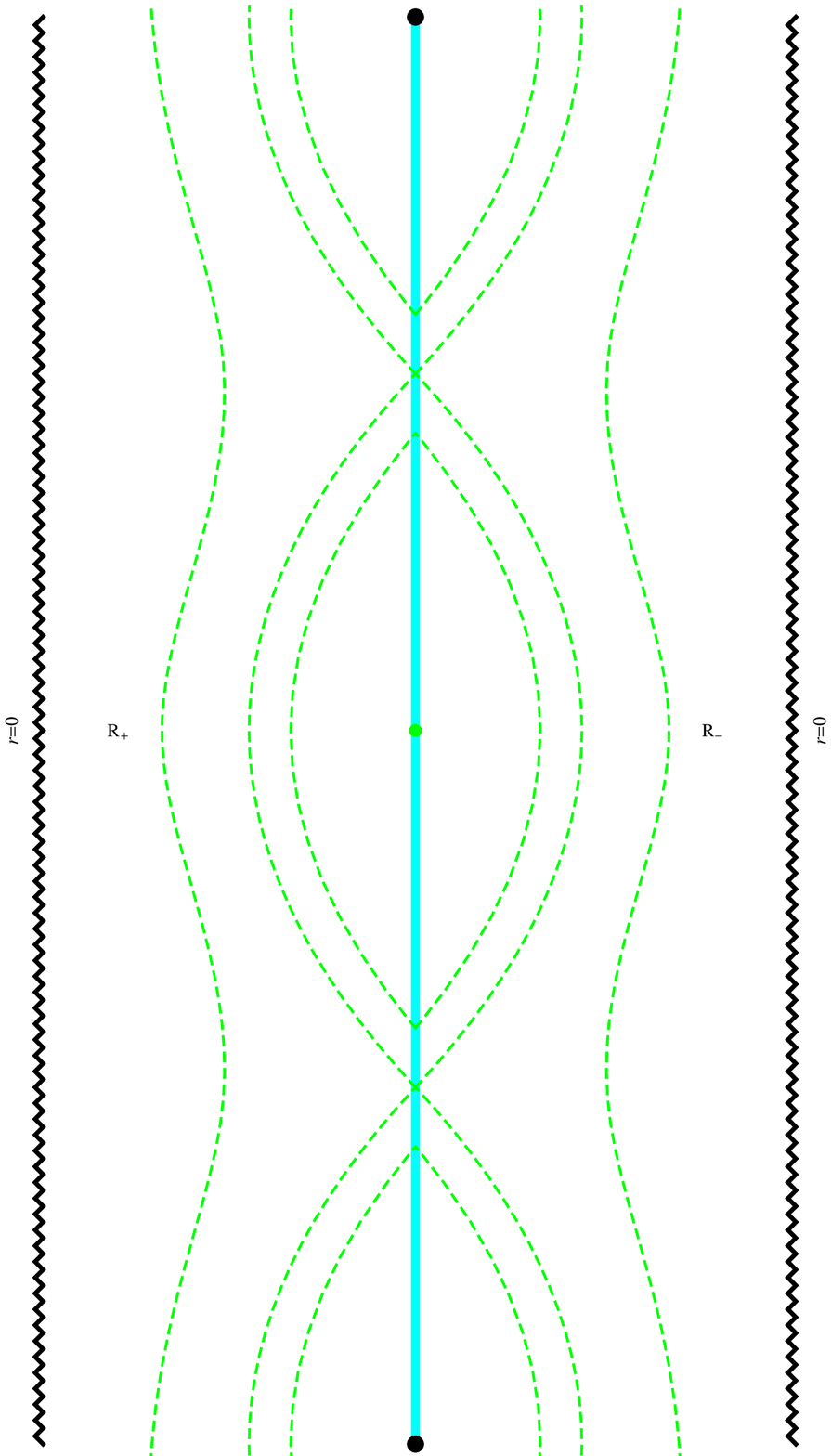}
\hfill
\includegraphics[angle=0,width=0.46\textwidth]{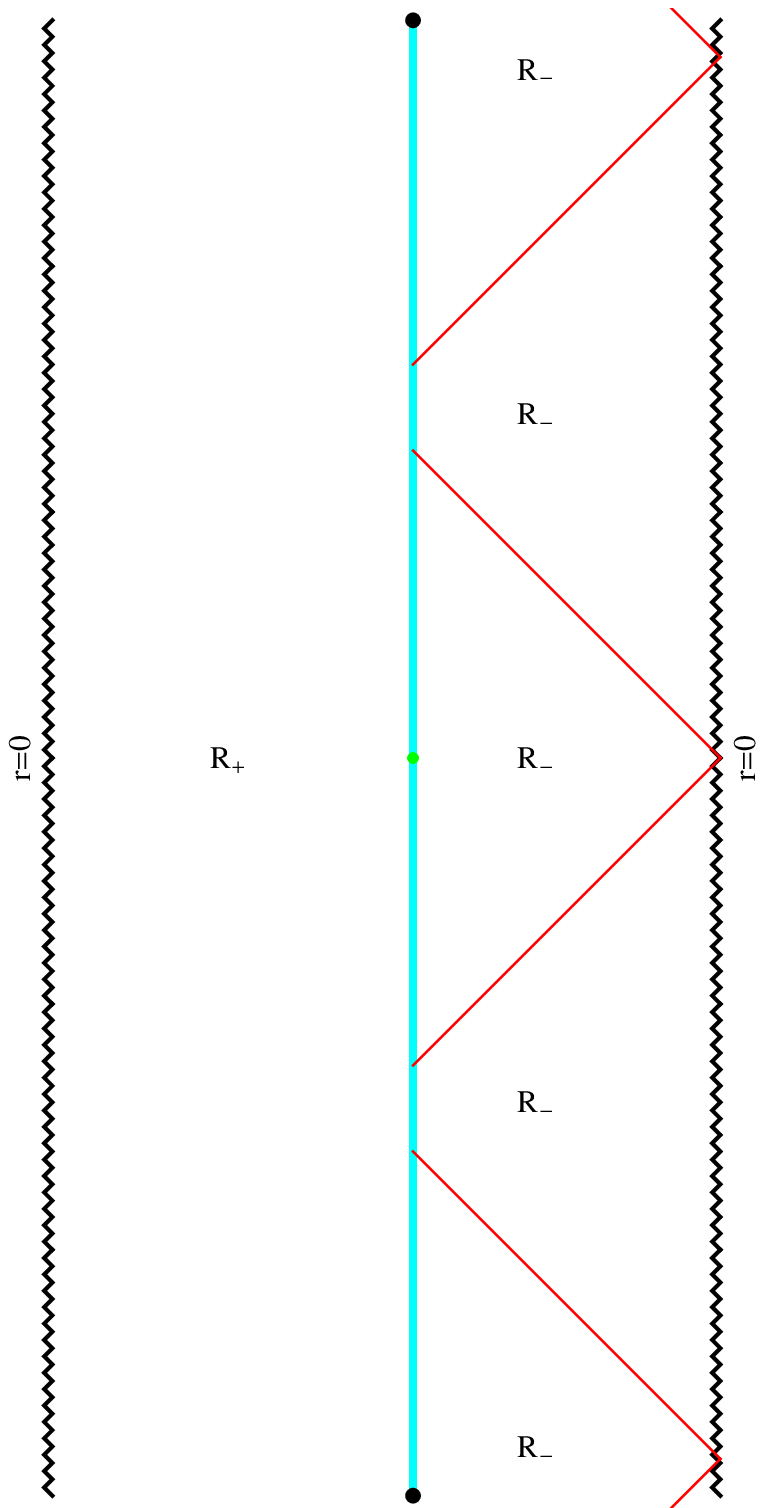}
\end{center}
\caption{The Carter-Penrose diagram at $\alpha < 1 (< \epsilon < 2 \epsilon)$, $\sigma_{\rm in} =+ 1$ and, $\sigma_{\rm out} = - 1$ at, respectively, $\mu < (\epsilon - 1)/\alpha$ (left panel) and $\mu = (\epsilon - 1)/\alpha$ (right panel).}
\label{fig29}
\end{figure}

Here we deliberately chose the case when the effective potential is negative at the minimum in order to show that such a case is, in principle, may take place. Let us investigate it in details and construct all the possible global geometries. We start with the trajectories of $\mu_1$-type. By this we denoted all the cases. when there is a naked singularity in the outer metrics. The value of $\mu_1$ can be both negative (but, of course, greater than the minimum of the potential) and positive, but less than $\mu =(\epsilon - 1)/\alpha$, when the outer metrics becomes that of the extreme Reissner-Nordstrom black hole (note, that $\sigma_{\rm out} = - 1$). The Carter-Penrose diagram looks as follows in Fig.~\ref{fig29} (left panel).

In the limiting case $\mu =(\epsilon - 1)/\alpha$ the conformal diagram is shown in Fig.~\ref{fig29} (right panel).

\begin{figure}[h]
\begin{center}
\includegraphics[angle=0,width=0.53\textwidth]{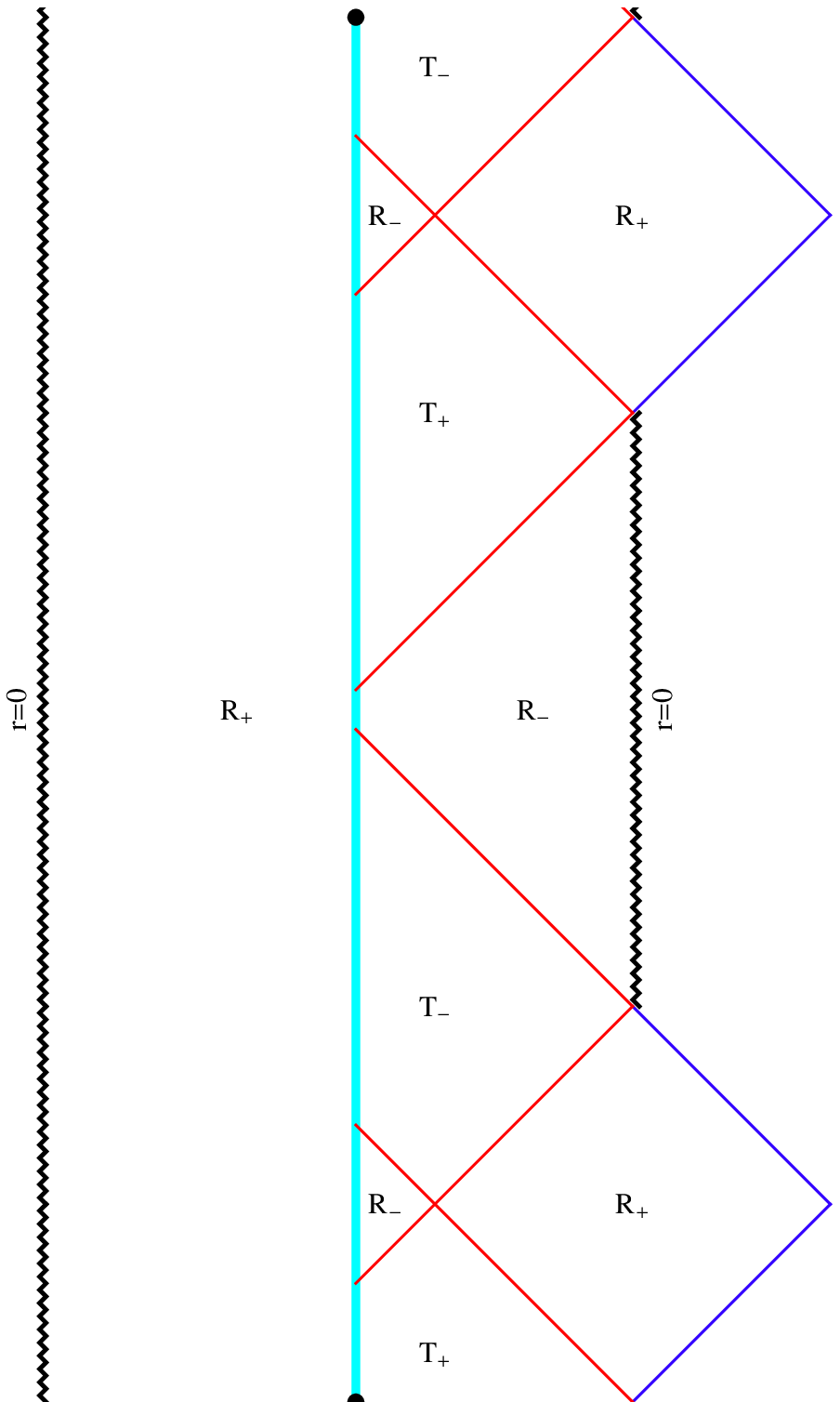}
\hfill
\includegraphics[angle=0,width=0.46\textwidth]{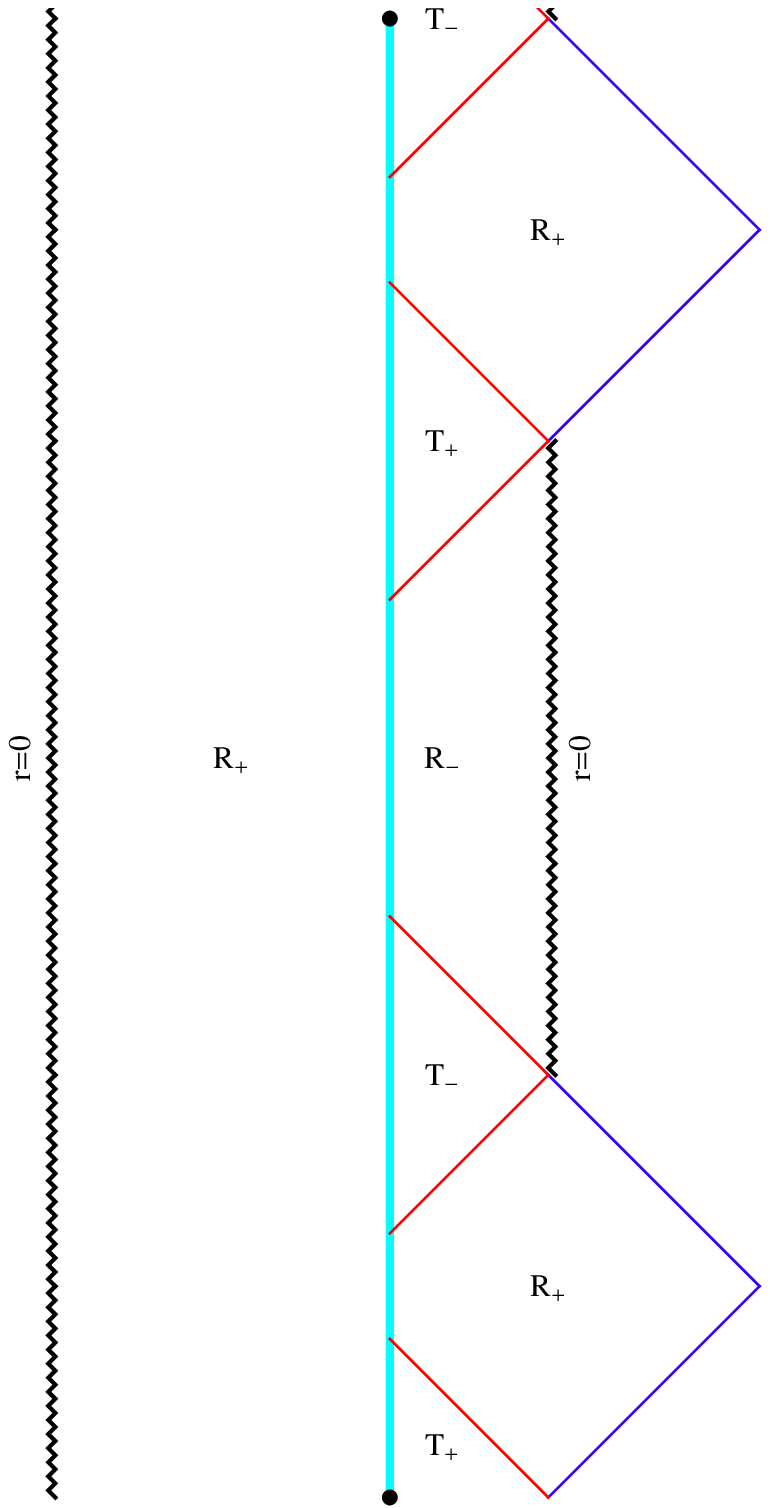}
\end{center}
\caption{The Carter-Penrose diagram at $\alpha < 1< \epsilon$ at, respectively  $(\epsilon - 1)/\alpha<\mu<\mu_2$ (left panel) and $\mu_2<\mu<\mu_1$ (right panel).}
\label{fig30}
\end{figure}

The next step is the $\mu_2$-type trajectories. Since $\mu_2 >(\epsilon - 1)/\alpha$, the outer metrics is that of the Reissner-Nordstrom black hole, but we have it still lower than the right hand intersection point $y_1 = y_2$, so everywhere along the trajectory $\sigma_{\rm out} = - 1$. The corresponding conformal diagrams are shown in Fig.~\ref{fig30}.

\begin{figure}[h]
\begin{center}
\includegraphics[angle=0,width=0.55\textwidth]{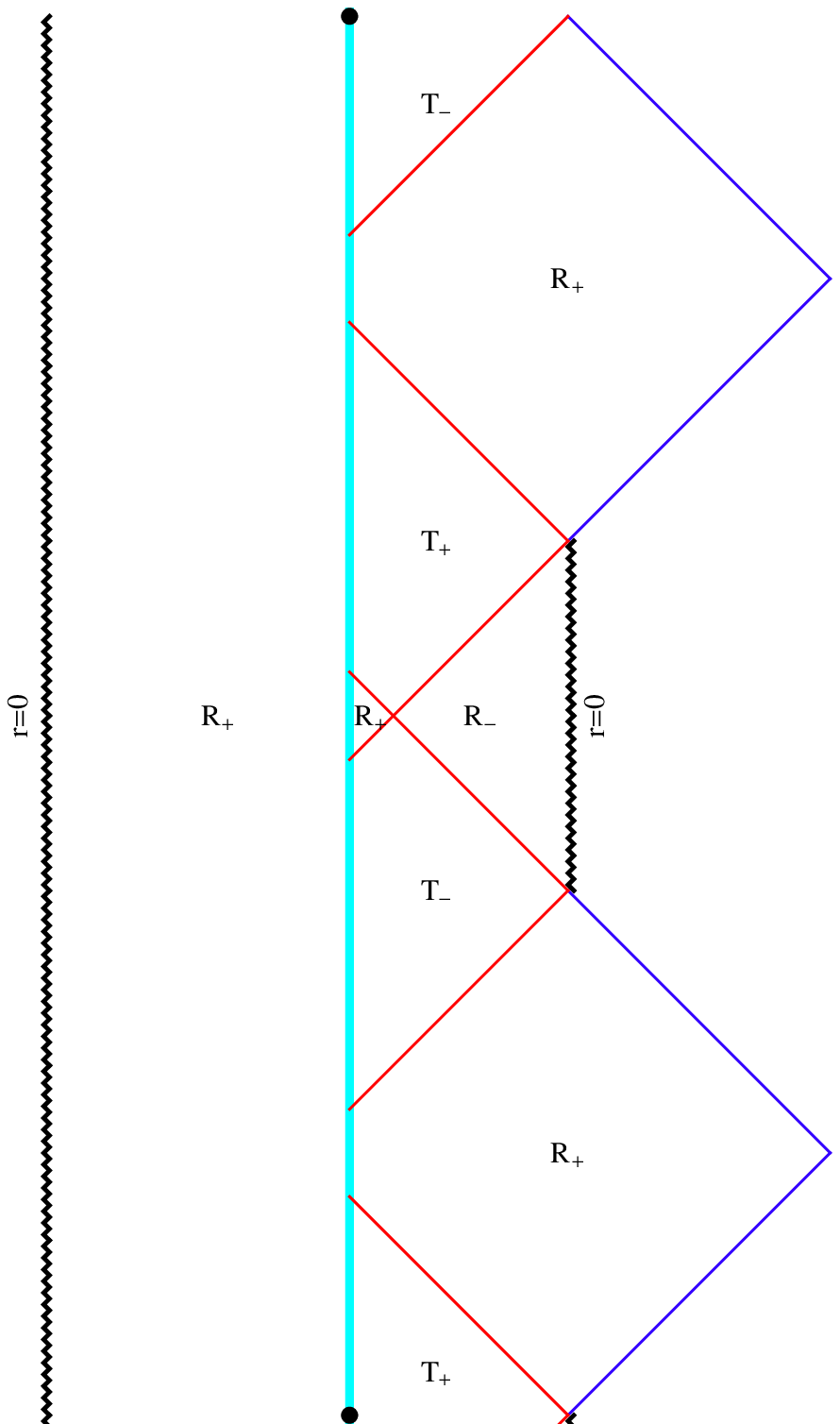}
\hfill
\includegraphics[angle=0,width=0.43\textwidth]{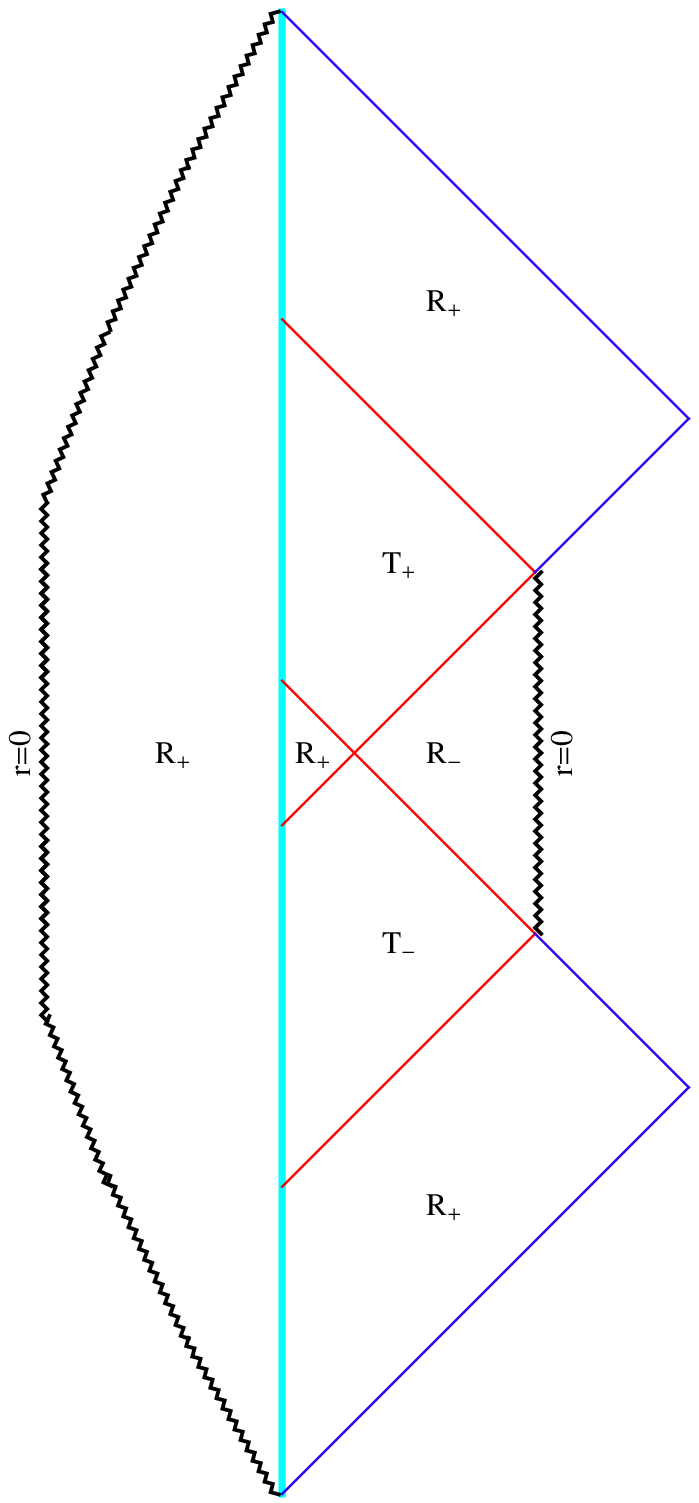}
\end{center}
\caption{The Carter-Penrose diagram at $\alpha <1<\epsilon$, respectively, for bound motion at $\mu_1<\mu<1$ (left panel) and for unbound motion at $\mu>1$ (right panel).}
\label{fig32}
\end{figure}

Increasing $\mu$ further, we come to the $\mu_3$-type trajectories when they start from the left hand turning point with $\sigma_{\rm out} = - 1$, ($R_-$-region near singularity), go through the $T_+$-region where $\sigma_{\rm out}$ changes its sign, enter the $R_+$-region outside the event horizon $(\sigma_{\rm out} = + 1)$ and then either meet the second (right) turning point, or escape to infinity. The relevant Carter-Penrose diagrams for bound and unbound motions are shown in Fig.~\ref{fig32}.

\begin{figure}[t]
\begin{center}
\includegraphics[angle=0,width=0.8\textwidth]{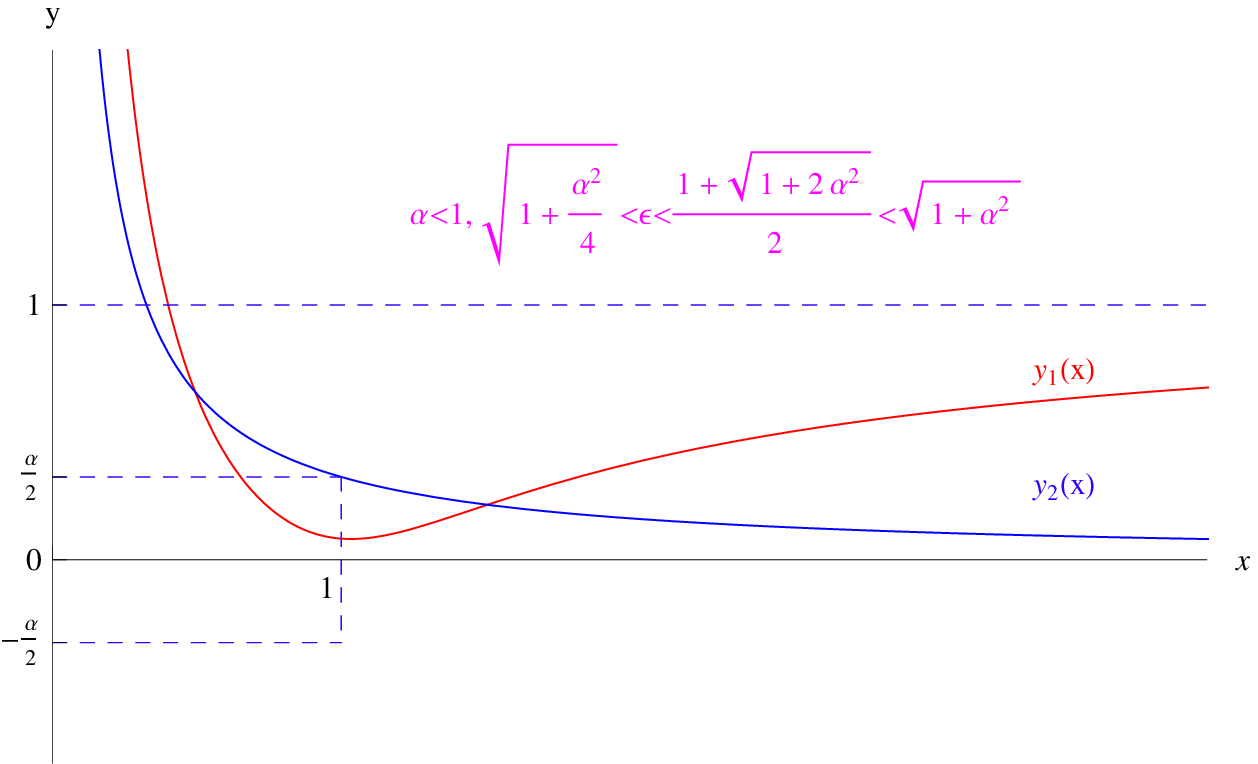}
\end{center}
\caption{The effective potential for heavy shells at $\alpha<1$, $1+\alpha^2/4 < \epsilon^2 < 1 + \alpha^2$ and $\alpha^2 > 2 \epsilon (\epsilon - 1)$.}
\label{fig33}
\end{figure}

Finally (for our specific values of $\alpha$ and $\epsilon$), the $\mu_4$-types of the trajectories, when $\sigma_{\rm out} = + 1$ everywhere. Again, there can be both bound and unbound motions. See below the corresponding conformal diagrams in Fig.~\ref{fig33}.

Keeping $\alpha < 1$, let us increase $\epsilon$ and enter the layer $1+\alpha^2/4 < \epsilon^2 < 1 + \alpha^2$, where there are two intersection points $y_1 = y_2$. But, one should distinguish between two cases: $\alpha^2 > 2 \epsilon (\epsilon - 1)$, when the intersection points lie on different (left-right, decreasing-increasing) branches of the effective potential, and $\alpha^2 < 2 \epsilon (\epsilon - 1)$, when both of them are on the same (left-decreasing) branch. The sets of possible global geometries are quite different in these two cases. To understand the problem better one should consider the triple intersection points $y = \mu = y_1 = y_2$, where the turning points of the trajectories lie exactly at the horizons of the outer metrics. The question is: which of the horizons, inner (Cauchy) or outer (event) ones? It appears that when the intersection points are on different branches of the effective potential, then the lower (right hand) one corresponds to the turning point at the event horizon and the upper (left hand) one - to the Cauchy horizon. If the intersection points are on the same (decreasing) branch, both of them correspond to the turning points at the inner horizons (surely, different ones due to different values of $\mu (m_{\rm out})$. Note, that if $\alpha^2 = 2 \epsilon (\epsilon - 1)$, the right hand intersection point lies exactly at the minimum of the effective potential. Moreover, the triple point in such a case corresponds to the extreme Reissner-Nordstrom black hole outside the shell (i.e., $m_{\rm out}/m_{\rm in} = \epsilon$). Consider, first, the smaller value of electric charge, when $\alpha^2 > 2 \epsilon (\epsilon - 1)$. The effective potential curve $y_1$ together with the curve $y = y_2$ is shown in Fig.~\ref{fig33}.
\begin{figure}[h]
\begin{center}
\includegraphics[angle=0,width=0.8\textwidth]{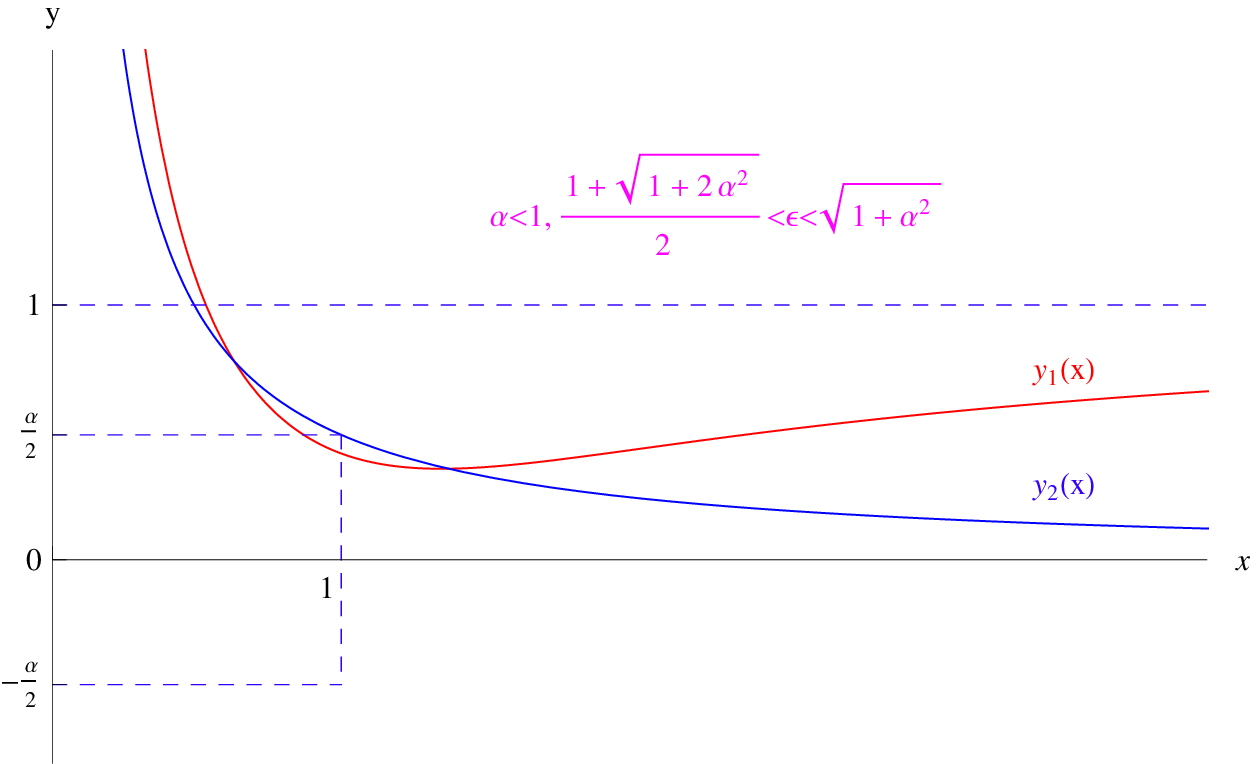}
\end{center}
\caption{The effective potential for heavy shells at $\alpha<1$, $1+\alpha^2/4 < \epsilon^2 < 1 + \alpha^2$ and $\alpha^2<2\epsilon(\epsilon - 1)$.}
\label{fig34}
\end{figure}

Evidently, in what concerns the global geometries, there is nothing new compared to the previous results. And we will not repeat drawings. Now, let us come to the case $\alpha^2 < 2 \epsilon (\epsilon - 1)$. The effective potential curve $y = y_1$ together with the curve $y = y_2$ take now the form,  shown in Fig.~\ref{fig34}.

\begin{figure}[h]
\begin{center}
\includegraphics[angle=0,width=0.55\textwidth]{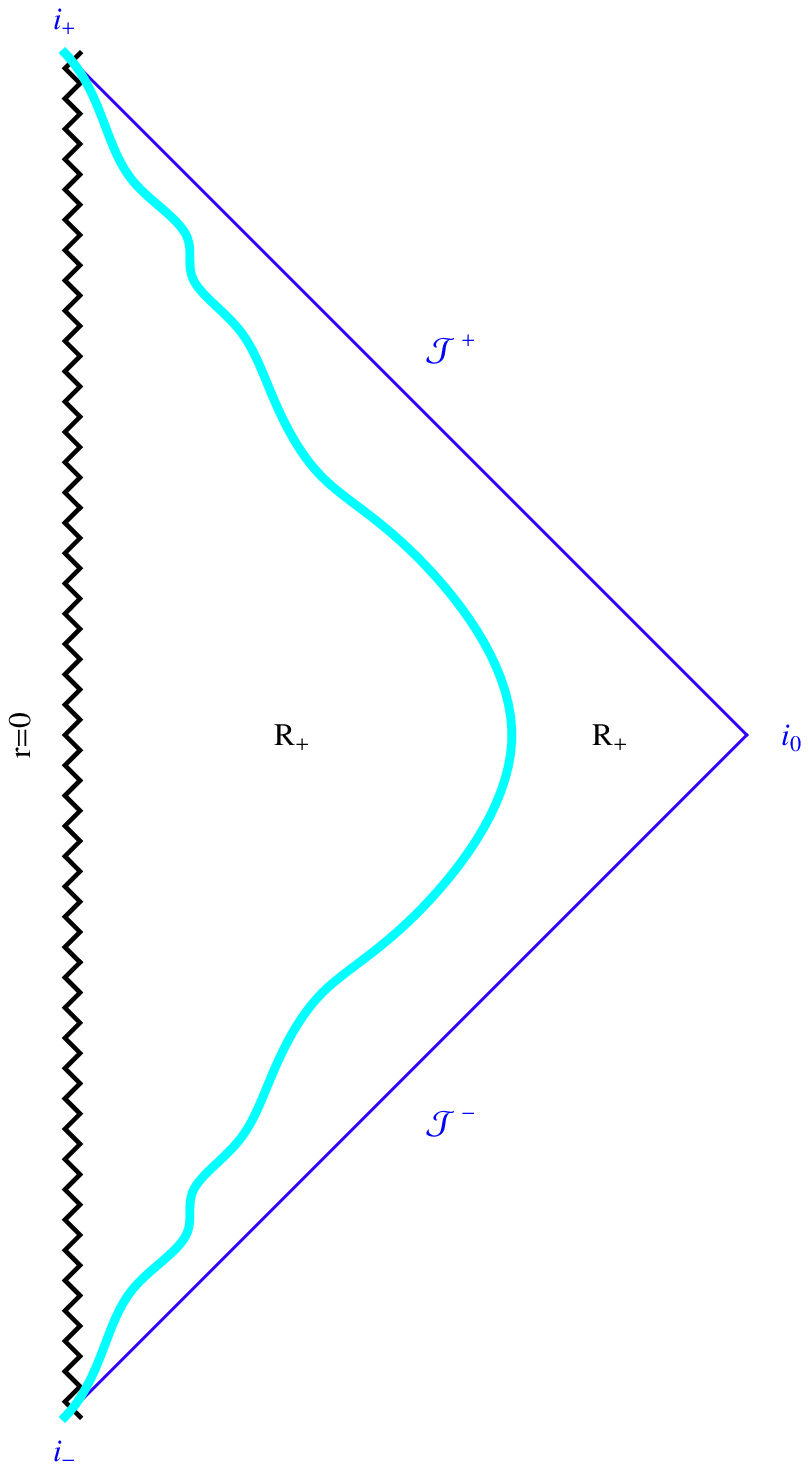}
\hfill
\includegraphics[angle=0,width=0.44\textwidth]{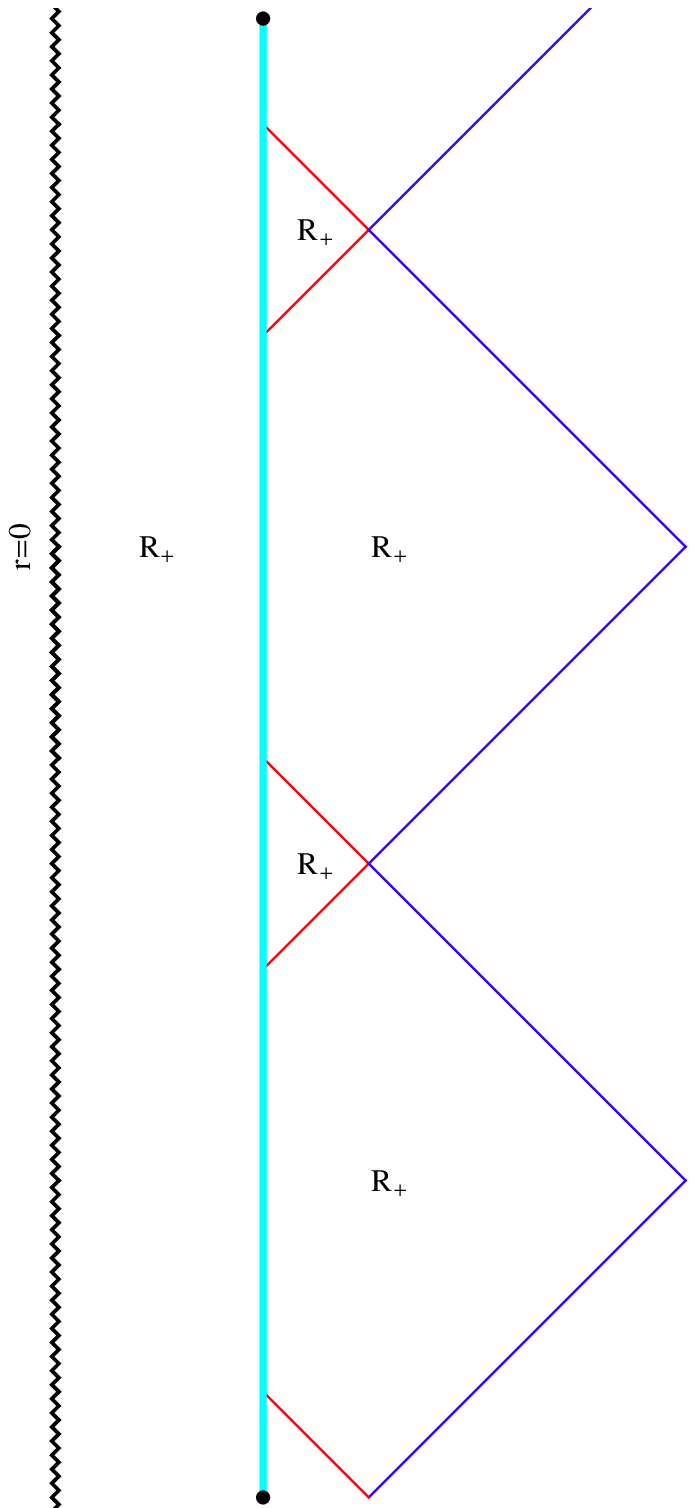}
\end{center}
\caption{The Carter-Penrose diagram at $\alpha<1$,  $\sigma_{\rm out} = + 1$ and $(1+\sqrt{1+2\alpha^2})/2<\epsilon<1+\alpha^2$ at, respectively, $\mu<(\epsilon-1)/\alpha$ (left panel) and  $\mu=(\epsilon-1)/\alpha$ (right panel).}
\label{fig35}
\end{figure}

If the horizontal line $y = \mu$ lies between the minimum of the effective potential and the lower (which is now to the left) intersection point $y_1 = y_2$, then it is possible (not, of course, always, but for certain intervals of parameters) to have the Reissner-Nordstrom naked singularity or the extreme black hole in the outer metrics and the shell's trajectories with $\sigma_{\rm out} = + 1$ everywhere. The Carter-Penrose diagrams in these two case are shown in Fig.~\ref{fig35}.
All other possible global geometries were already illustrated before.

Let now $\epsilon^2 > 1 + \alpha^2$. In this case the curve $y = y_2$ is lower than the potential curve $y = y_1$, so for all the trajectories $\sigma_{\rm out} = + 1$ everywhere. What is new? The main thing is that the horizontal line $y =(\epsilon - 1)/\alpha$ indicating the extremal black hole in the outer metrics can lie above the line $y =1$ (the threshold between the bound and unbound motions), therefore, when such situation occurs, in the case of the extreme black hole or naked singularity (in some interval of $\epsilon$) the only possible type of motion is the unbound one. These diagrams that are shown below in Fig.~\ref{fig36}.
\begin{figure}[h]
\begin{center}
\includegraphics[angle=0,width=0.53\textwidth]{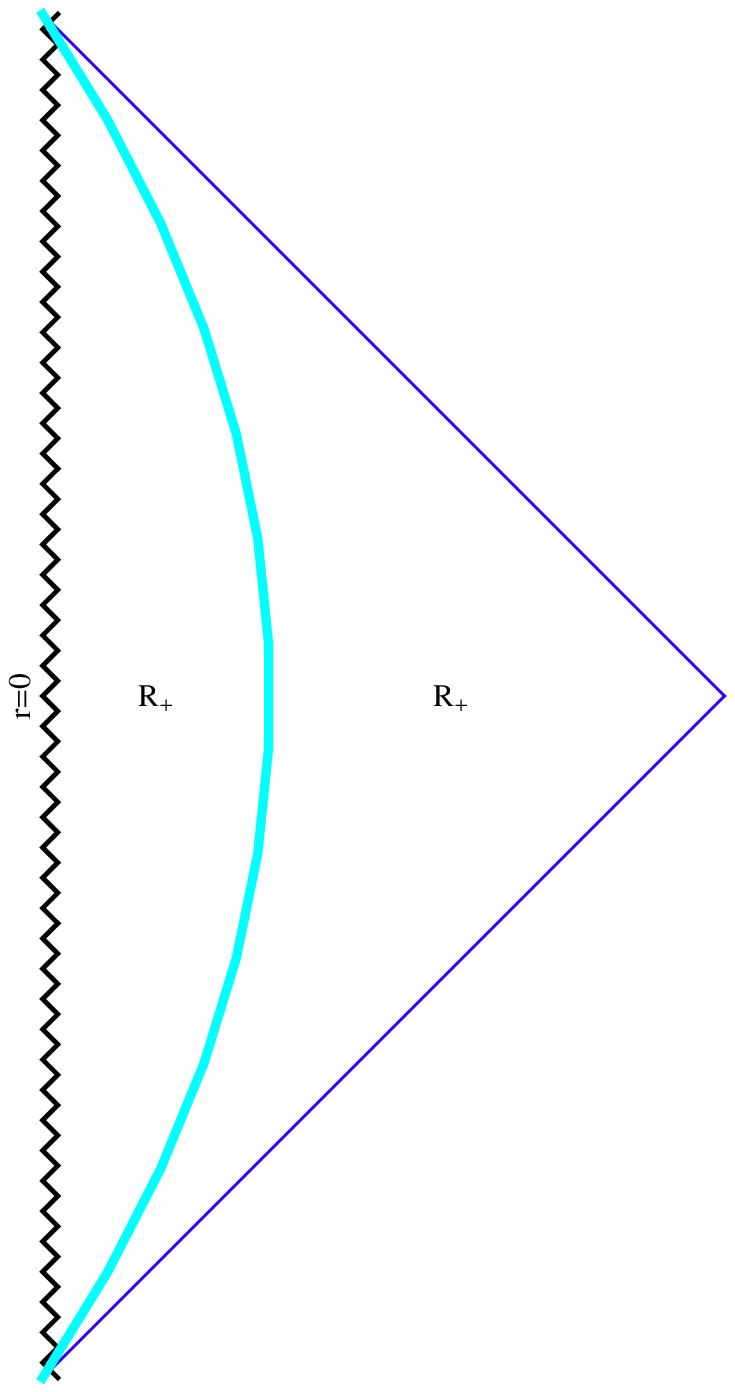}
\hfill
\includegraphics[angle=0,width=0.46\textwidth]{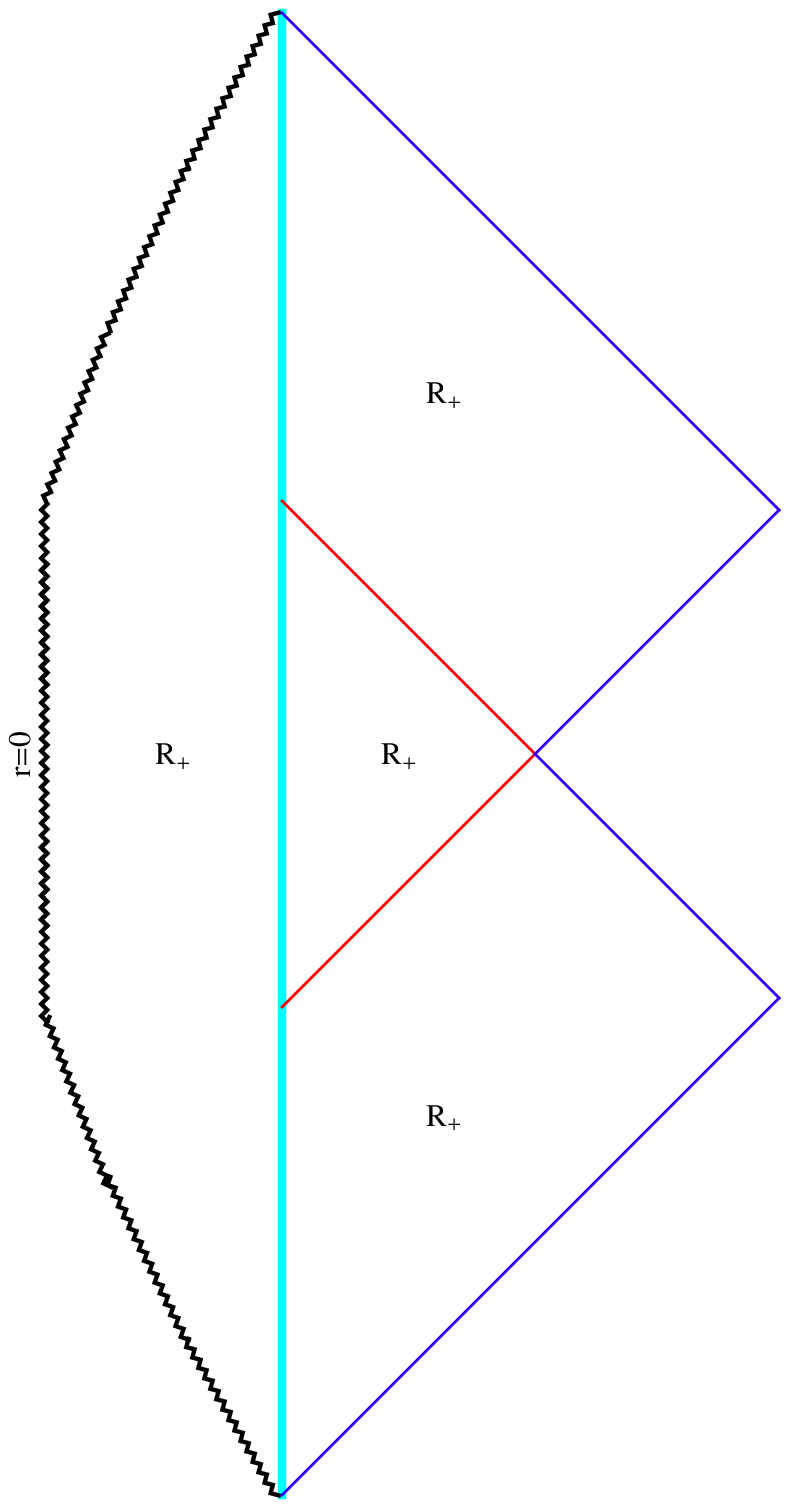}
\end{center}
\caption{The Carter-Penrose diagram at $\alpha<1$, $\epsilon>1+\alpha^2$, $1<\mu<(\epsilon-1)/\alpha$  and (left panel) and $1<\mu=(\epsilon-1)/\alpha$ (right panel).}
\label{fig36}
\end{figure}

\begin{figure}[h]
\begin{center}
\includegraphics[angle=0,width=0.95\textwidth]{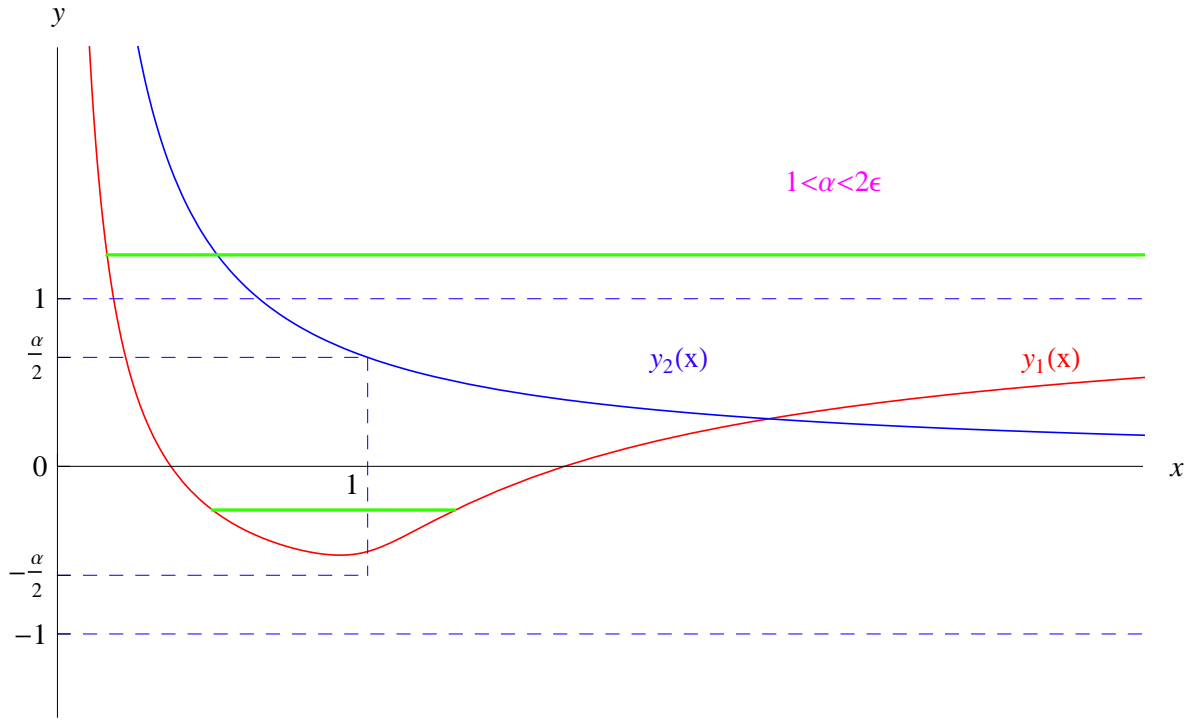}
\end{center}
\caption{The effective potential for heavy shells at $\alpha/2<\epsilon=1<\alpha$ and $\sigma_{\rm in} = + 1$.}
\label{fig37}
\end{figure}

What will happen when we shift the parameter $\alpha$ to $1 < \alpha < 2 \epsilon$? There appears the interval for $\epsilon$, namely, $1 < \alpha < 2 \epsilon$, for which $y_2 > y_1\; (x \to 0)$, and, therefore, we have only one intersection point $y_1 = y_2$. The picture for these two curves is shown in Fig.~\ref{fig37}.
There are no qualitatively new Carter-Penrose diagrams (\,i.\,e,, the global geometries) in this case.

What concerns the heavy shells, $\alpha > 2 \epsilon$, the situation is qualitatively the same as in the case $\epsilon = 1$. One should only remove the red lines indicating the doubled horizons in the inner part of the Carter-Penrose diagrams, Figs.~\ref{fig23}---\ref{fig25}.

\begin{figure}[th]
\begin{center}
\includegraphics[angle=0,width=0.8\textwidth]{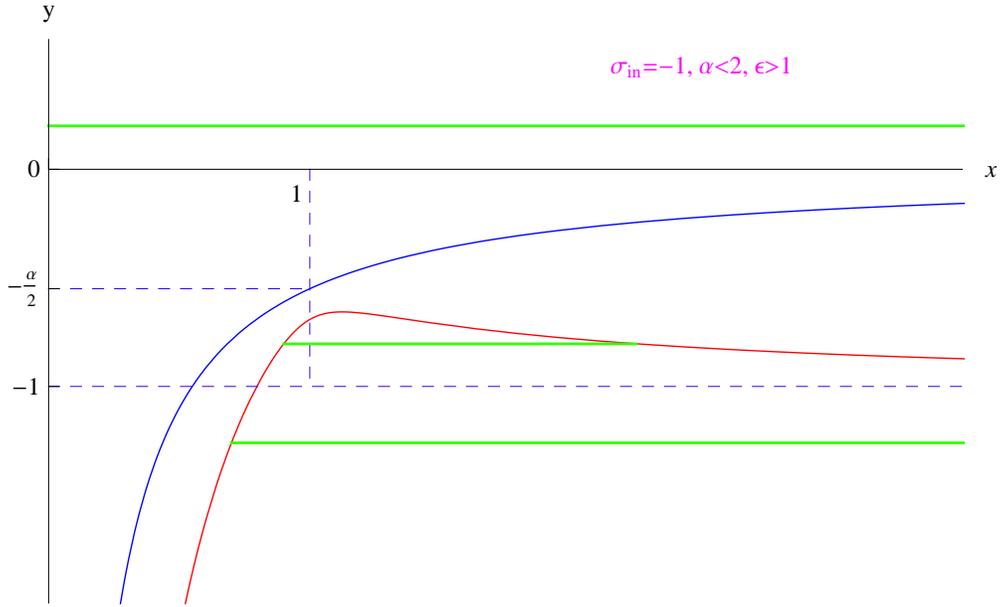}
\end{center}
\caption{The effective potential for light shells at $\alpha < 2 \epsilon$ at $\sigma_{\rm in} =-1$}
\label{fig38}
\end{figure}

\begin{figure}[h]
\begin{center}
\includegraphics[angle=0,width=0.8\textwidth]{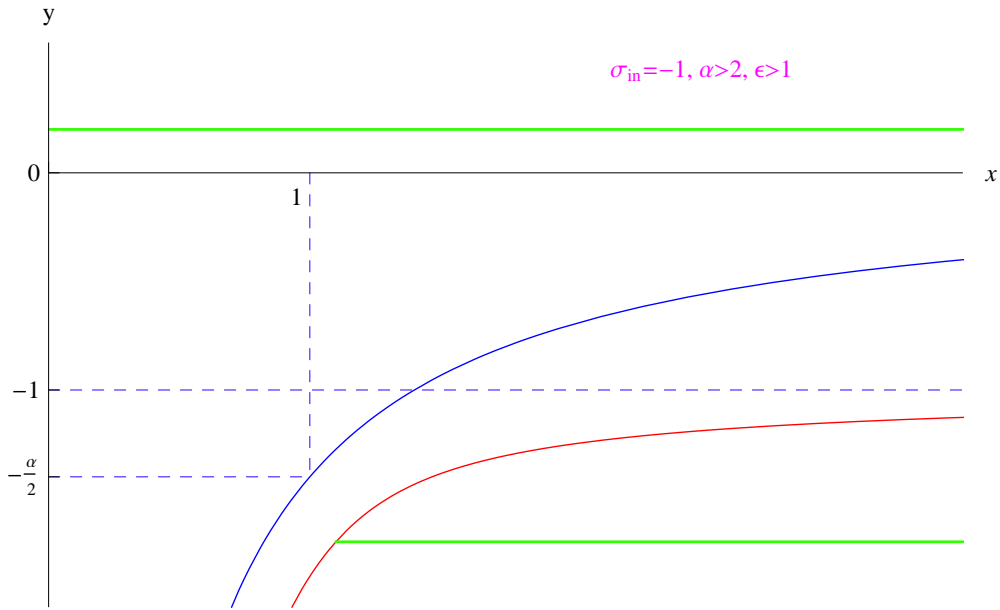}
\end{center}
\caption{The effective potential for heavy shells at $\alpha>2\epsilon$ at $\sigma_{\rm in} =-1$.}
\label{fig39}
\end{figure}

Finally, let us consider the case $\sigma_{\rm in} = - 1$, when inside the shell there is singularity at zero radius but, instead, the infinities (spatial, null and temporal). The effective potential curves look differently for $\alpha < 2 \epsilon$ (light shells) and for $\alpha > 2 \epsilon$ (heavy shells). They are shown in Figs.~\ref{fig38} and \ref{fig39}, respectively.

\begin{figure}[h]
\begin{center}
\includegraphics[angle=0,width=0.49\textwidth]{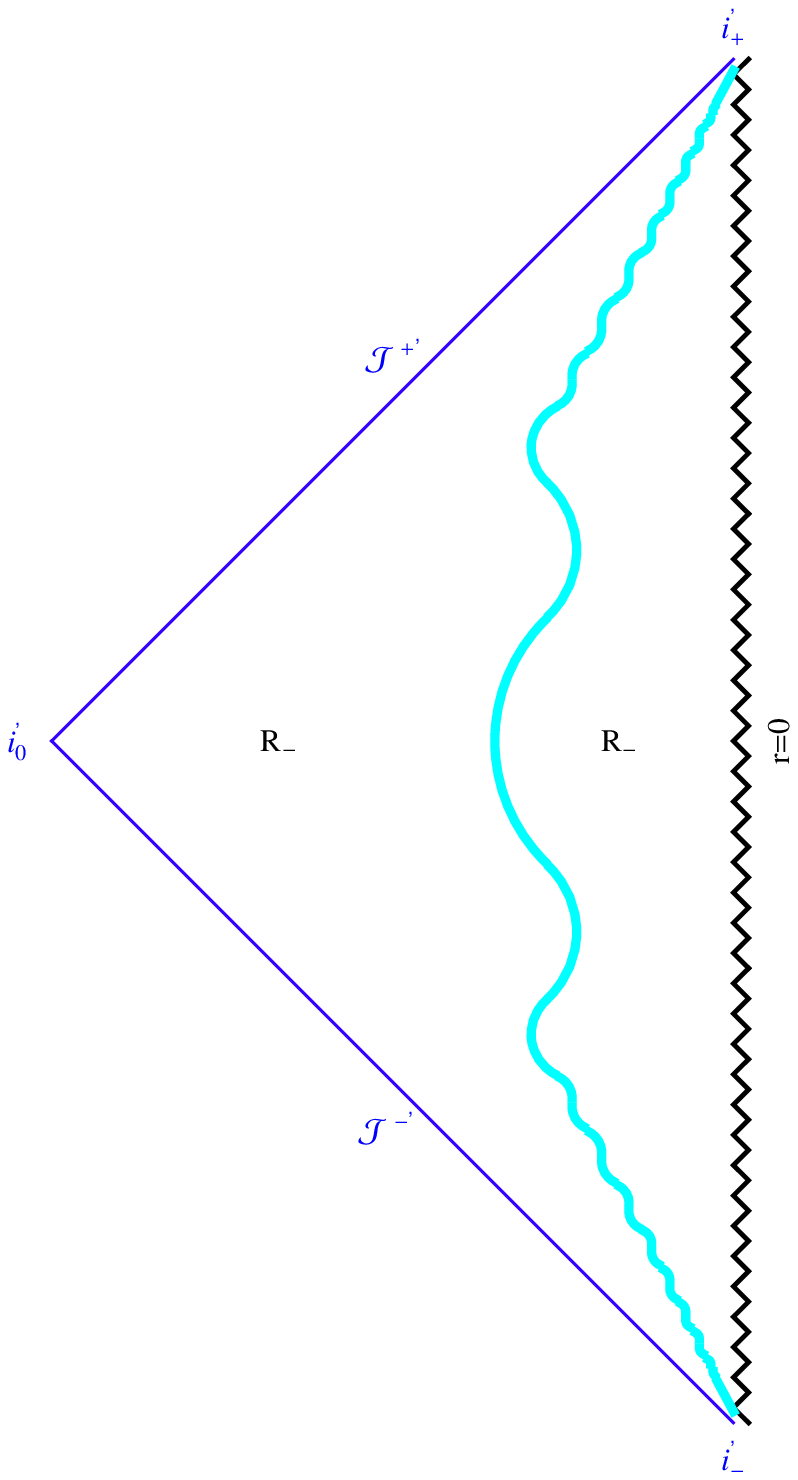}
\hfill
\includegraphics[angle=0,width=0.5\textwidth]{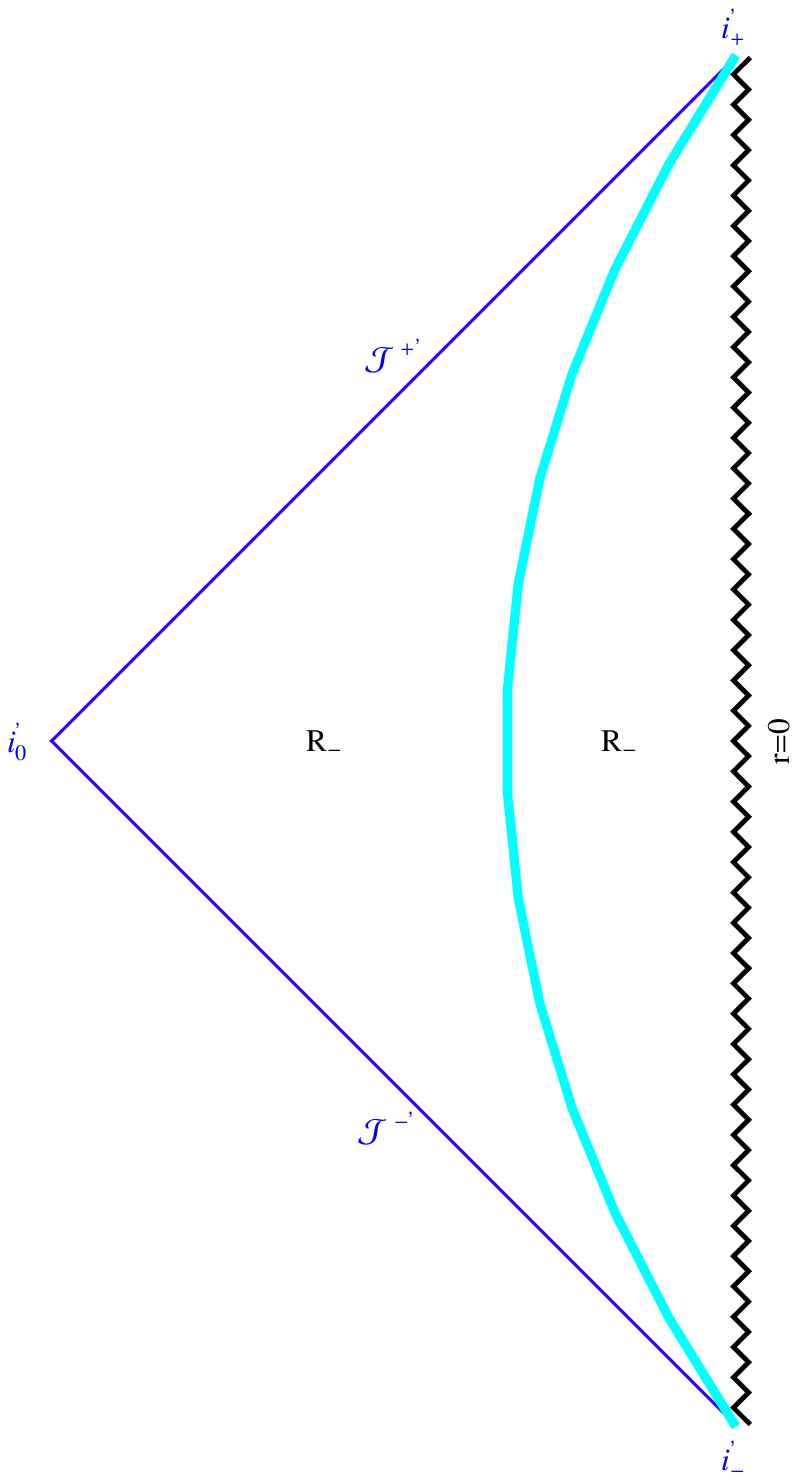}
\end{center}
\caption{The Carter-Penrose diagram at $\epsilon>1$, $\sigma_{\rm in} =-1$, $\alpha<2$ for $-1<\mu<0$ (left panel) and, respectively, for  $\alpha\gtrless2$, $\mu<-1$ (right panel).}
\label{fig40}
\end{figure}

Remember, that now the allowed trajectories (the horizontal lines $y = \mu$) lie below the potential curve. There can be either the bound motion with two turning points (for light shells), or the unbound motion with one turning point (both for light and heavy shells). The Carter-Penrose diagrams are shown in Fig.~\ref{fig40}.

\subsection{The case $\epsilon < 1$}

At last, let us consider the case $\epsilon < 1$, when the inner part of the combined manifold represents the metrics of the Reissner-Nordstrom black hole. Its structure is much more sophisticated than what we dealt with before. The mail feature is that the R-N black hole space-time contains all possible types of regions, $R_{\pm}$ and $T_{\pm}$ ones. This is the reason why the pictures for the effective potential look so differently and unusually. Fortunately, not only the authors but the readers as well, all of us have enough experience (due to reading the preceding Sections) in order to omit the details and to construct easily the Carter-Penrose diagram (= global geometry) just looking at the corresponding picture for the effective potential.

Consider, first, the case $\epsilon^2 <\alpha^2/4< \alpha^2 < 1$. Below in Fig.~\ref{fig41} is shown the effective potential $y = y_1$ together with the curves $y_2$ and $\tilde y_2 = - y_2$.
\begin{figure}[h]
\begin{center}
\includegraphics[angle=0,width=0.95\textwidth]{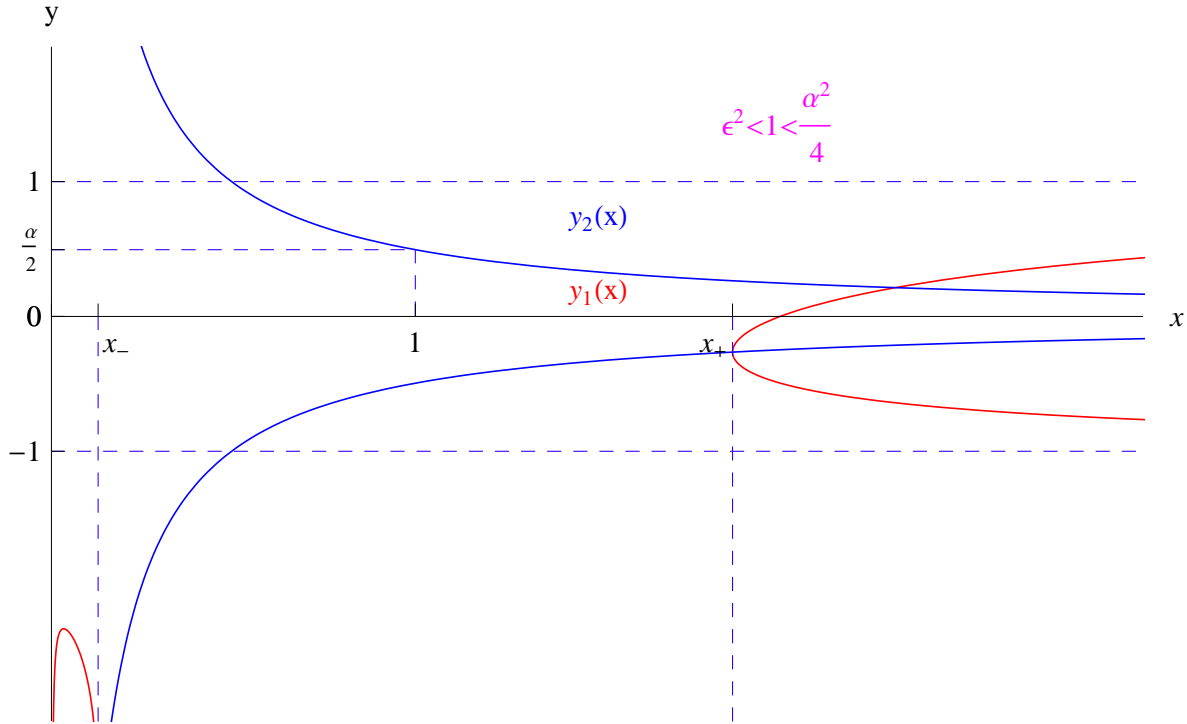}
\end{center}
\caption{The effective potential for the case $\epsilon^2 <\alpha^2/4< \alpha^2 < 1$.}
\label{fig41}
\end{figure}
\begin{figure}[h]
\begin{center}
\includegraphics[angle=0,width=0.49\textwidth]{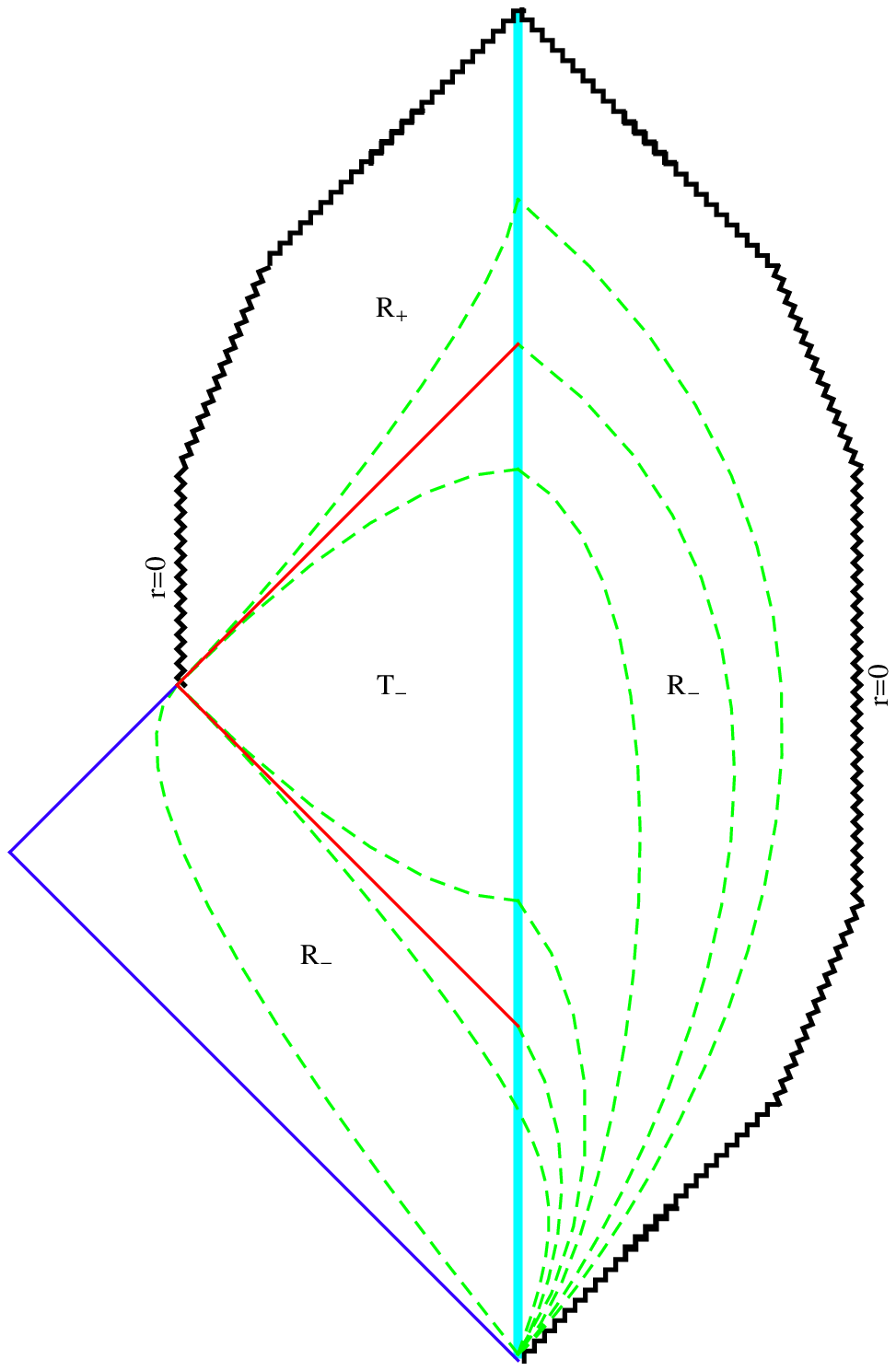}
\hfill
\includegraphics[angle=0,width=0.48\textwidth]{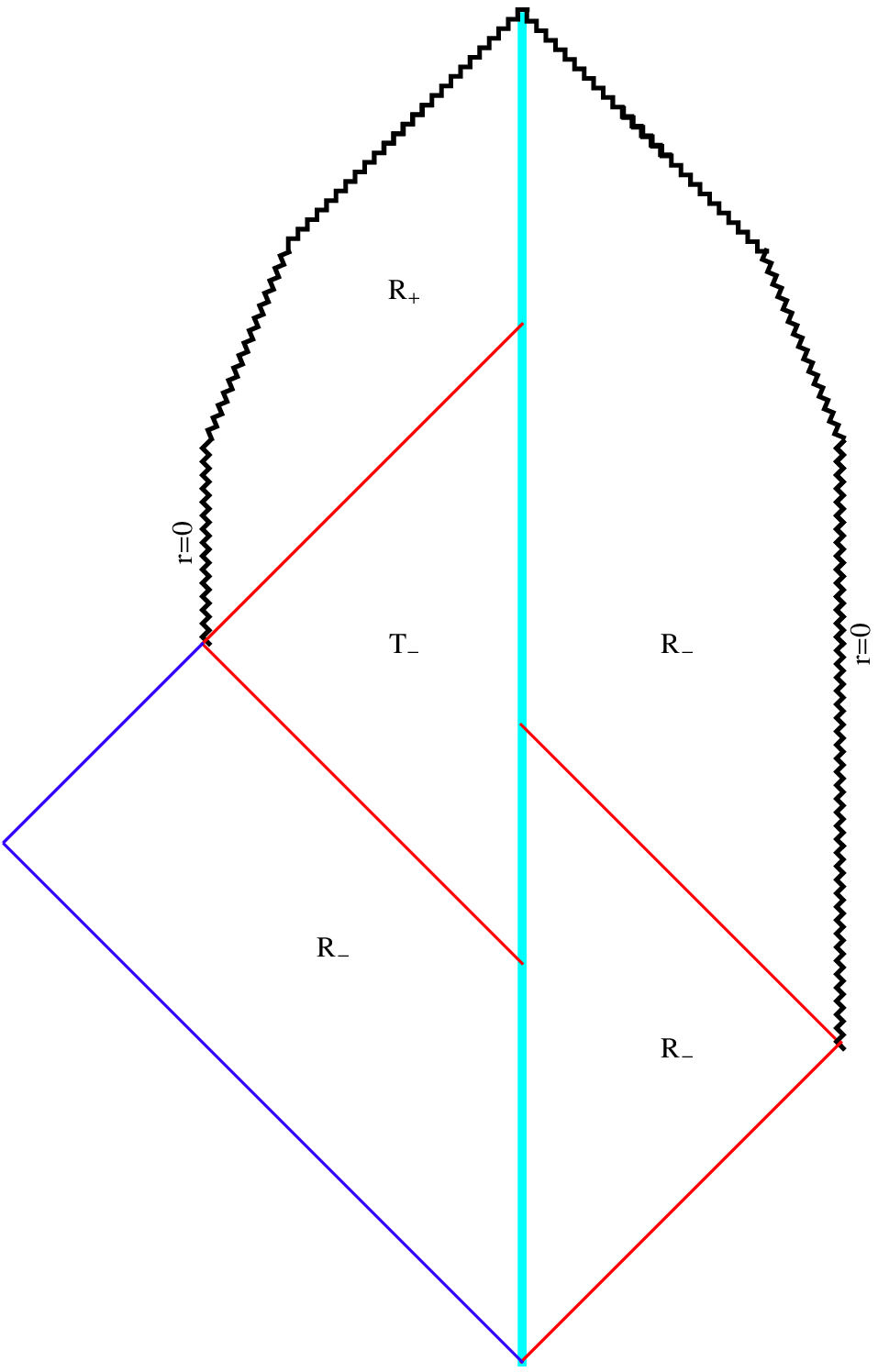}
\end{center}
\caption{The Carter-Penrose diagram at $\epsilon^2 <\alpha^2/4< \alpha^2 < 1$ for $-1/\alpha<\mu<-(1 - \epsilon^2)/\alpha<-1$ (left panel) and, respectively, for
 $-1/\alpha<\mu=-(1 - \epsilon^2)/\alpha<\mu<-1$ (right panel).}
\label{fig42}
\end{figure}
\begin{figure}[h]
\begin{center}
\includegraphics[angle=0,width=0.7\textwidth]{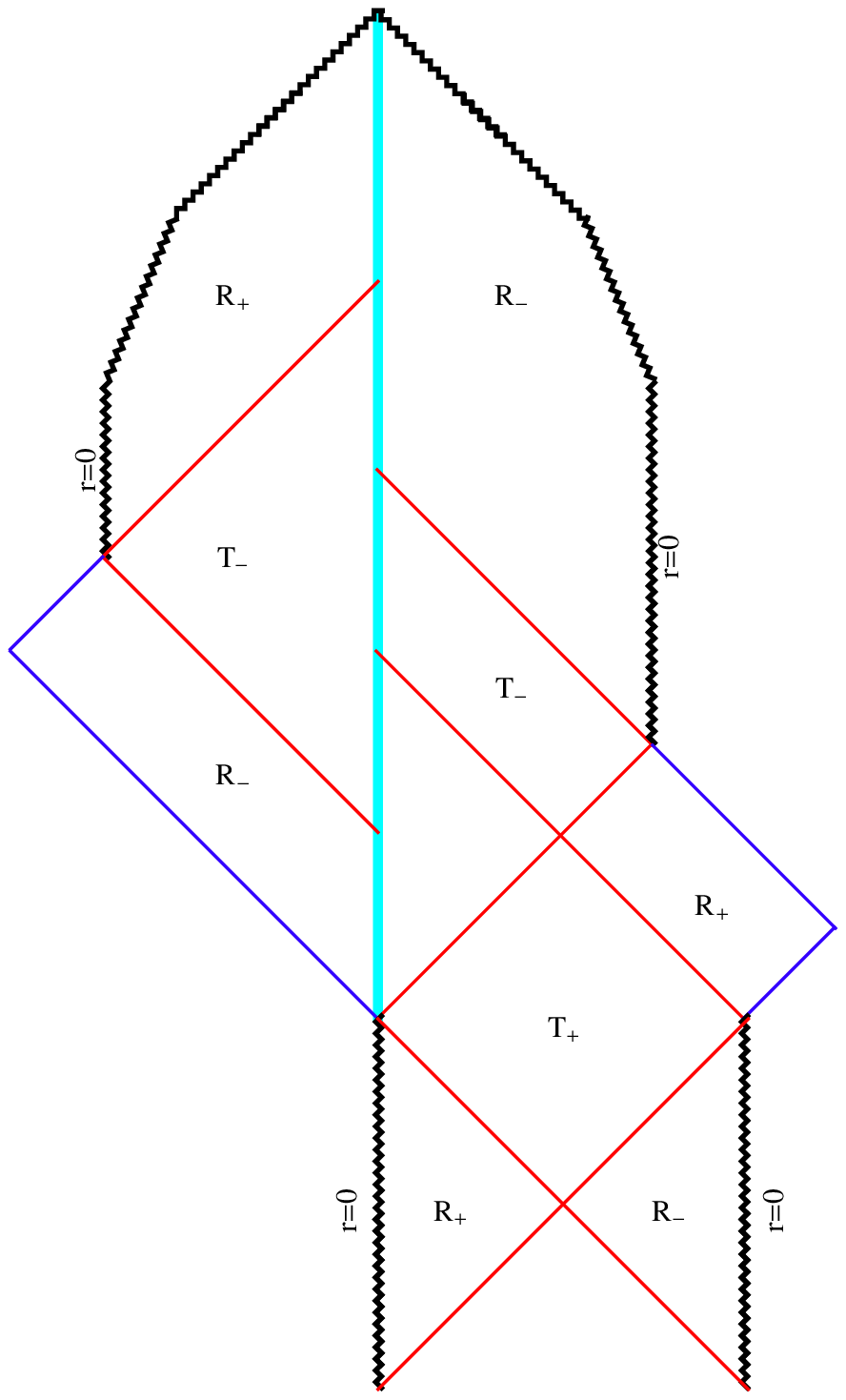}
\end{center}
\caption{The Carter-Penrose diagram at $\epsilon^2 <\alpha^2/4< \alpha^2 < 1$ for
 $-1/\alpha<-(1 - \epsilon^2)/\alpha<-1<\mu<-1$.}
\label{fig43}
\end{figure}

We enumerated the possible types of the trajectories as $\mu_1 < \mu_2 < \mu_3 < \mu_4 < \mu_5 < \mu_6$. Since we confined ourselves to only positive values of the total mass $m_{\rm out}$, there should be $1 + \mu \alpha > 0$, so $\mu > \mu_0 =1/\alpha$. As the first step, we consider the unbound motion, $\mu < - 1$, with no turning points, thus, the range for $\mu_1$ is $-1/\alpha<\mu_1<-1$. For negative values of $\mu$ the outer part of the combined manifold could be, surely, the Reissner-Nordstrom black holes, $-(1 - \epsilon^2)/\alpha < \mu_1$, as well as the extreme black hole, $\mu_1 = -(1 - \epsilon^2)/\alpha$, or the space-time with naked singularity, $\mu_1 <-(1 - \epsilon^2)/\alpha$. The Carter-Penrose diagrams are shown below  in Figs.~\ref{fig42} and ~\ref{fig43}.

\begin{figure}[h]
\begin{center}
\includegraphics[angle=0,width=0.49\textwidth]{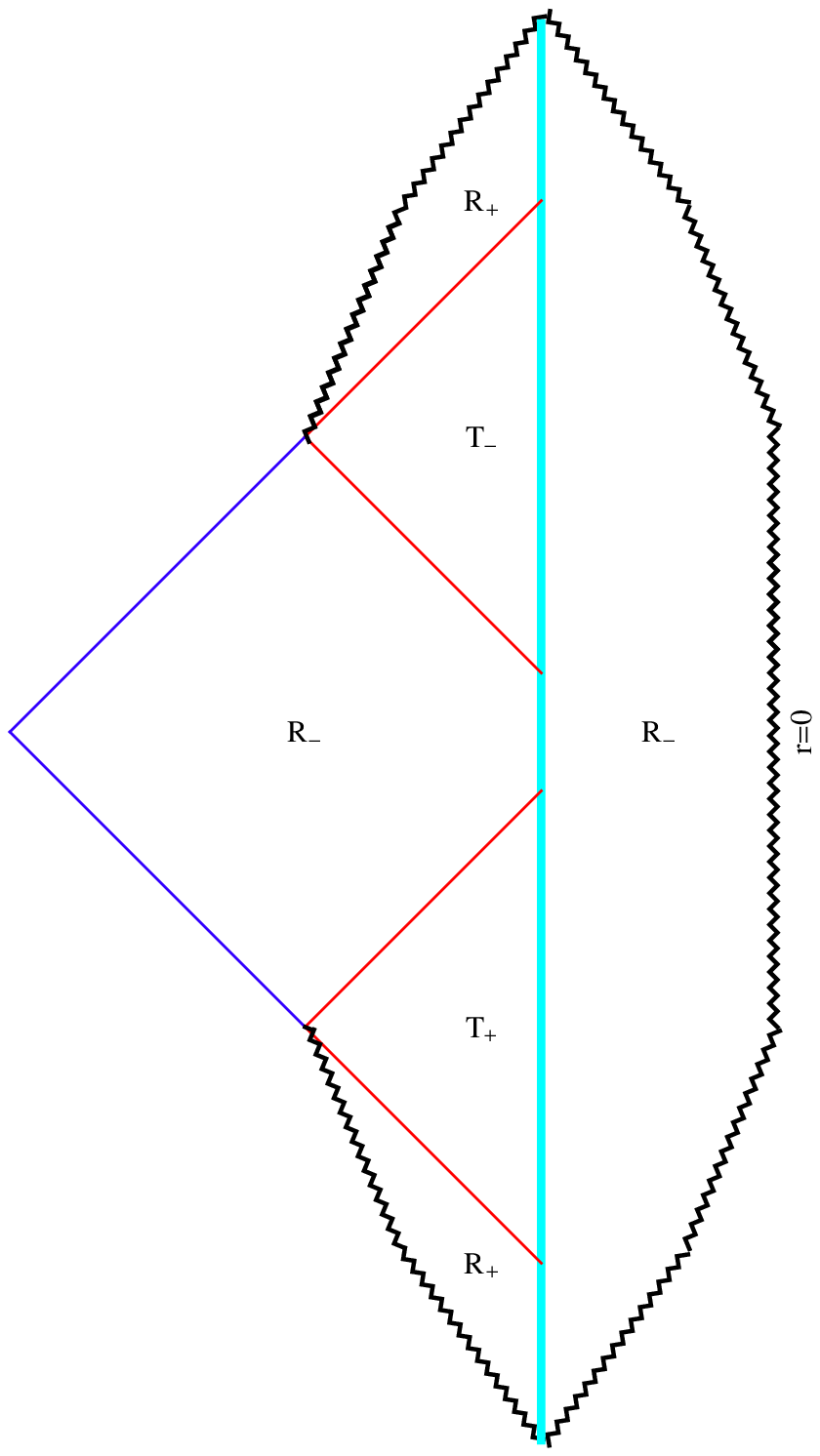}
\hfill
\includegraphics[angle=0,width=0.49\textwidth]{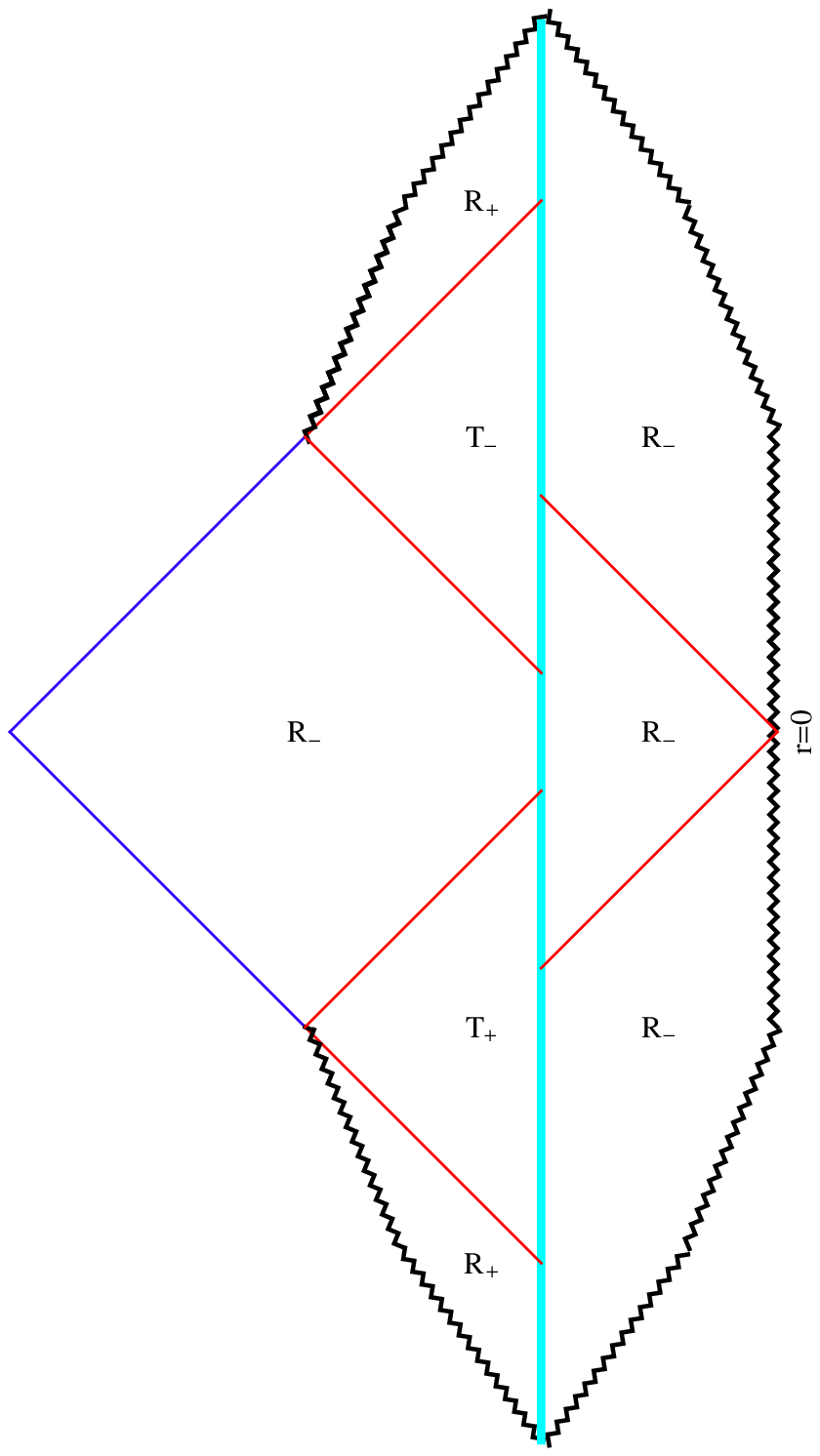}
\end{center}
\caption{The Carter-Penrose diagram at $\epsilon^2 <\alpha^2/4< \alpha^2 < 1$
for $-1<\mu<-(1 - \epsilon^2)/\alpha<\alpha(2x_+)$ (left panel) and, respectively, for
$-1<\mu+-(1 - \epsilon^2)/\alpha<\alpha/(2x_+)$ (right panel).}
\label{fig44}
\end{figure}

\begin{figure}[h]
\begin{center}
\includegraphics[angle=0,width=0.65\textwidth]{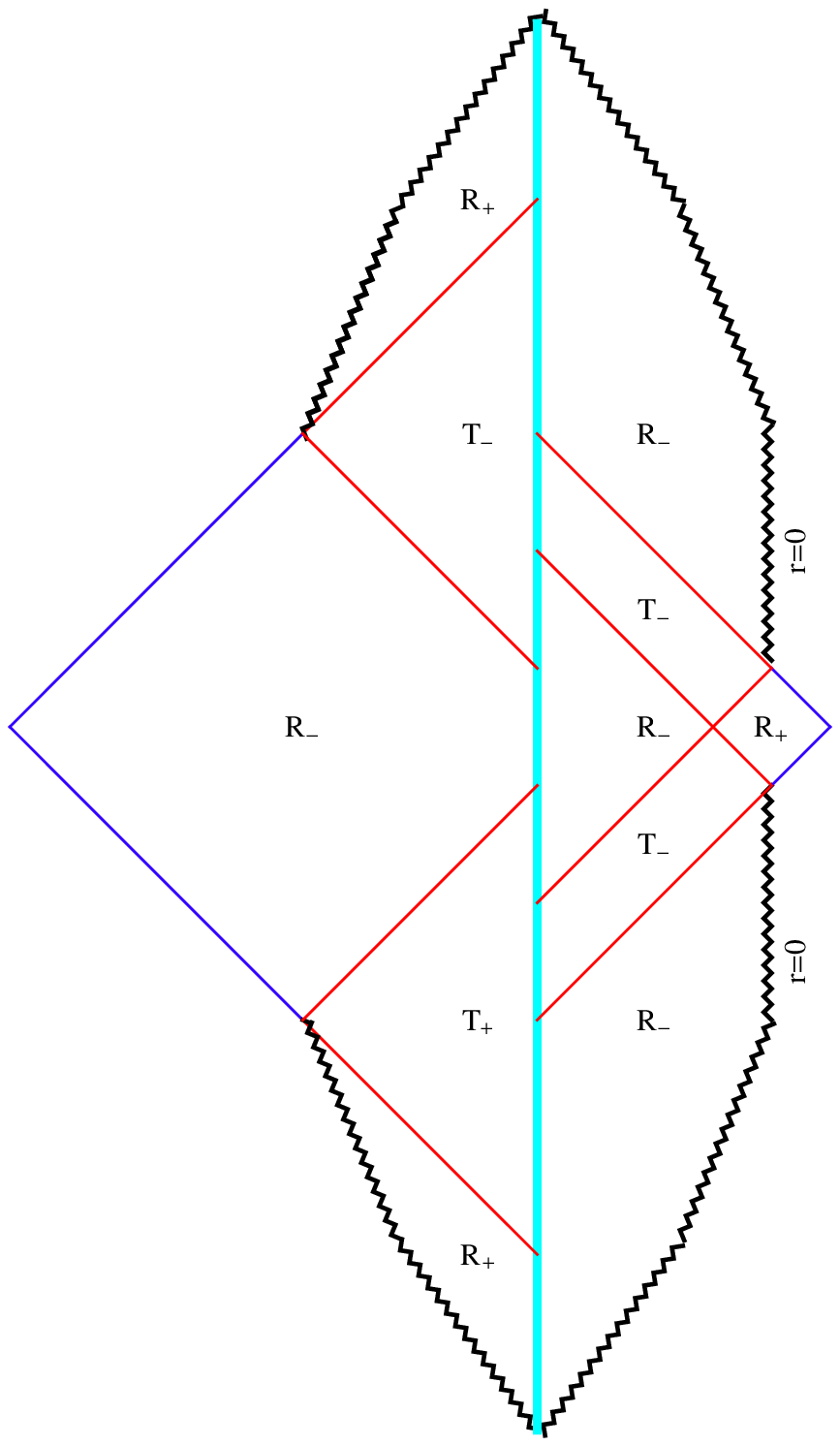}
\end{center}
\caption{The Carter-Penrose diagram at $\epsilon^2 <\alpha^2/4< \alpha^2 < 1$ for
$-1<-(1 - \epsilon^2)/\alpha<\alpha/(2x_+)$.}
\label{fig45}
\end{figure}

Next, the interval $-1 < \mu = \mu_2 < - \alpha/(2 x_+)$, where we have the bound motion that starts from the singularity at zero radius in the $R_+$-region $(\sigma_{\rm in} = + 1)$ of the inner metrics, and ends at the turning point in the $R_-$-region $(\sigma_{\rm in} = - 1)$ outside the event horizon $(x_+)$ of the inner metrics. What concerns the outer metrics, $\sigma_{\rm out} = - 1$ everywhere there. The situation is the same as for $\mu_1$: we may have all three representatives of the Reissner-Nordstrom space-time. The conformal diagrams are the following, Figs.~\ref{fig44} and ~\ref{fig45}.
\begin{figure}[h]
\begin{center}
\includegraphics[angle=0,width=0.45\textwidth]{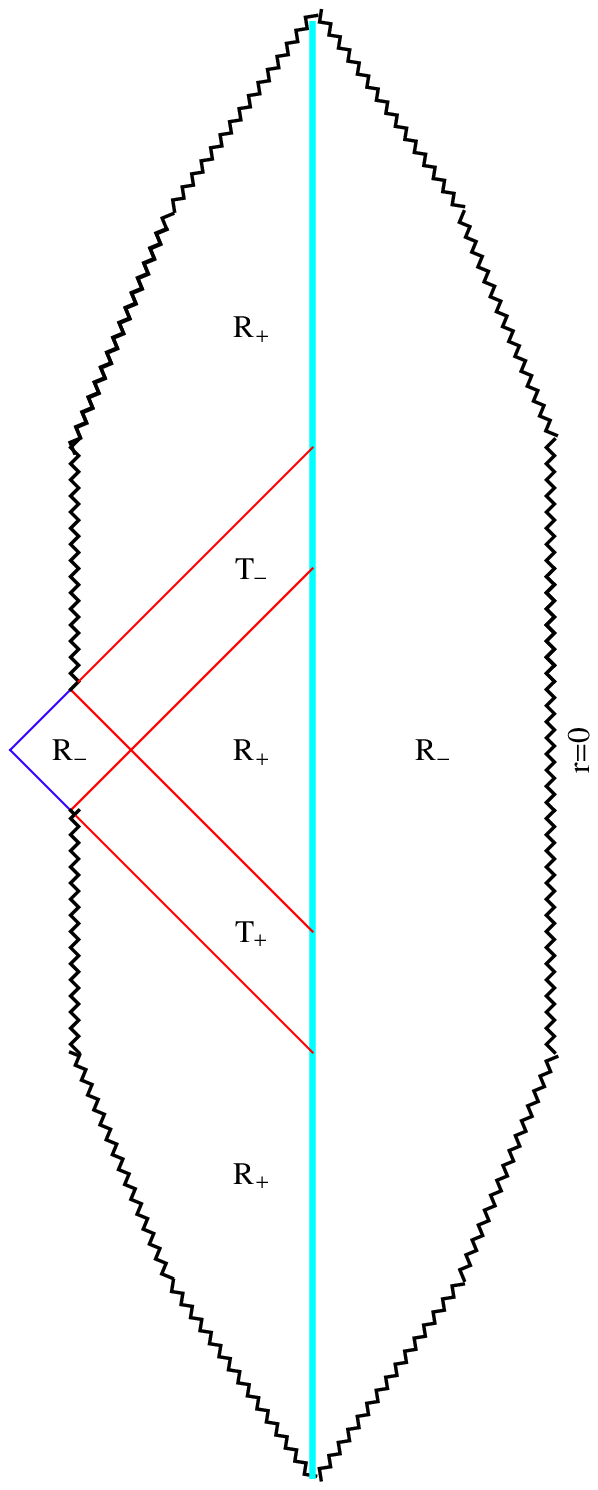}
\hfill
\includegraphics[angle=0,width=0.45\textwidth]{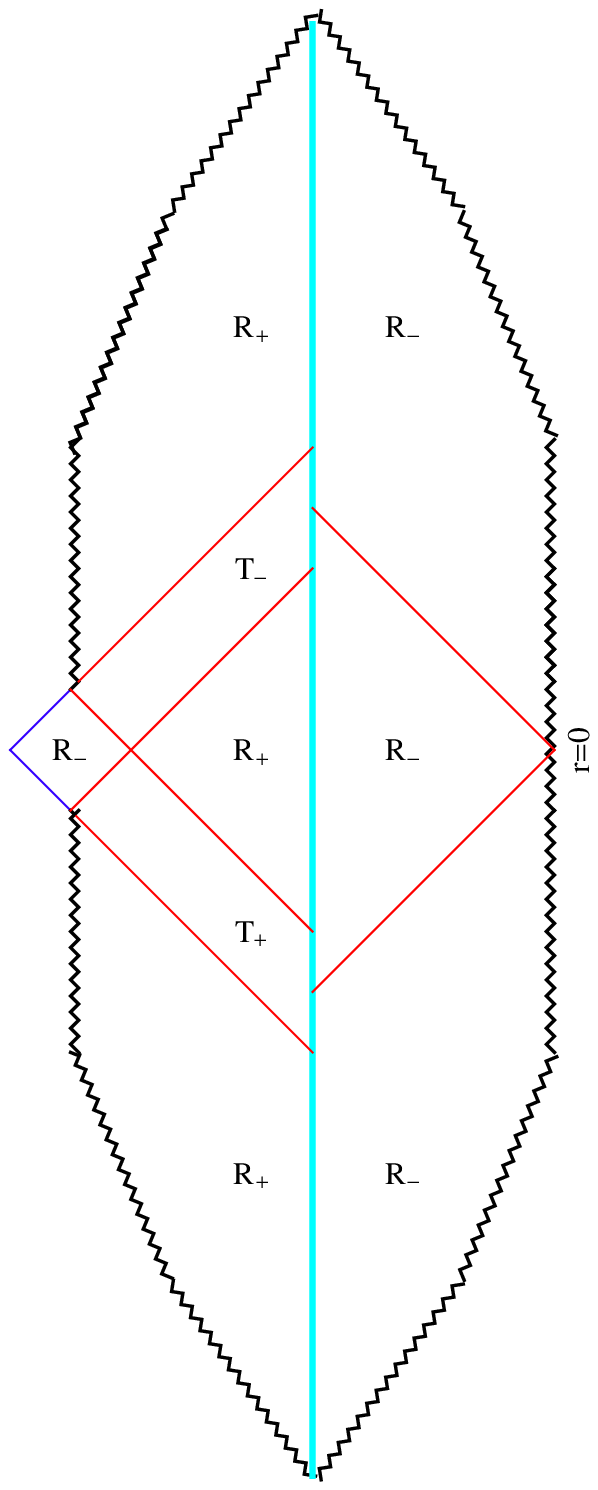}
\end{center}
\caption{The Carter-Penrose diagram at $\epsilon^2 <\alpha^2/4< \alpha^2 < 1$
for $-1<-\alpha/(2x_+)<\mu<0$ (left panel) and, respectively, for
$-1<-\alpha/(2x_+)<\mu<0$ (right panel).}
\label{fig46}
\end{figure}
\begin{figure}[h]
\begin{center}
\includegraphics[angle=0,width=0.45\textwidth]{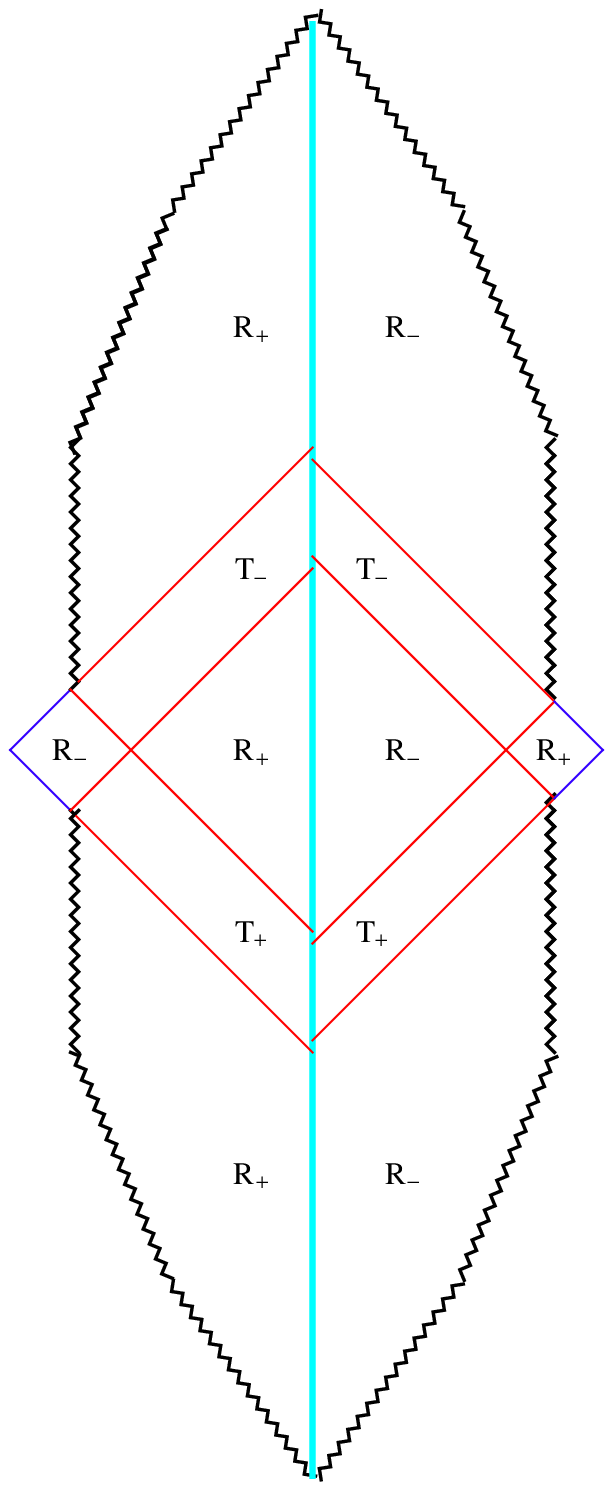}
\end{center}
\caption{The Carter-Penrose diagram at $\epsilon^2 <\alpha^2/4< \alpha^2 < 1$ for
$-1<-\alpha/(2x_+)<\mu<0$.}
\label{fig47}
\end{figure}

In the interval $- \alpha/(2 x_+)< \mu = \mu_3 < 0$, everywhere on the trajectories, $\sigma_{\rm in} = + 1$ in the inner part and $\sigma_{\rm out} = - 1$ in the outer part of the combined manifold. The motion is bound with one turning point. Below in Figs.~\ref{fig46} and ~\ref{fig47} are shown the Carter-Penrose diagrams for different types of the outer metrics.

\begin{figure}[h]
\begin{center}
\includegraphics[angle=0,width=0.46\textwidth]{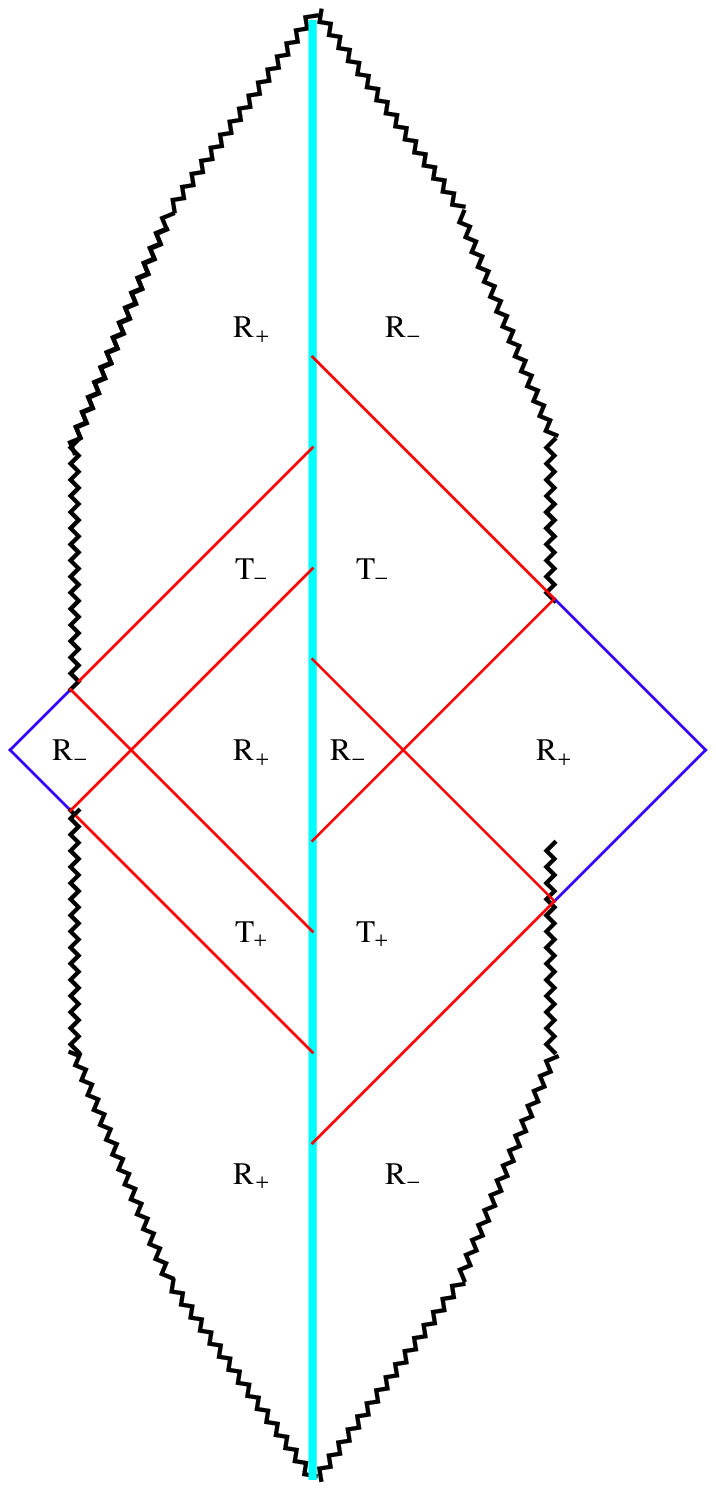}
\hfill
\includegraphics[angle=0,width=0.51\textwidth]{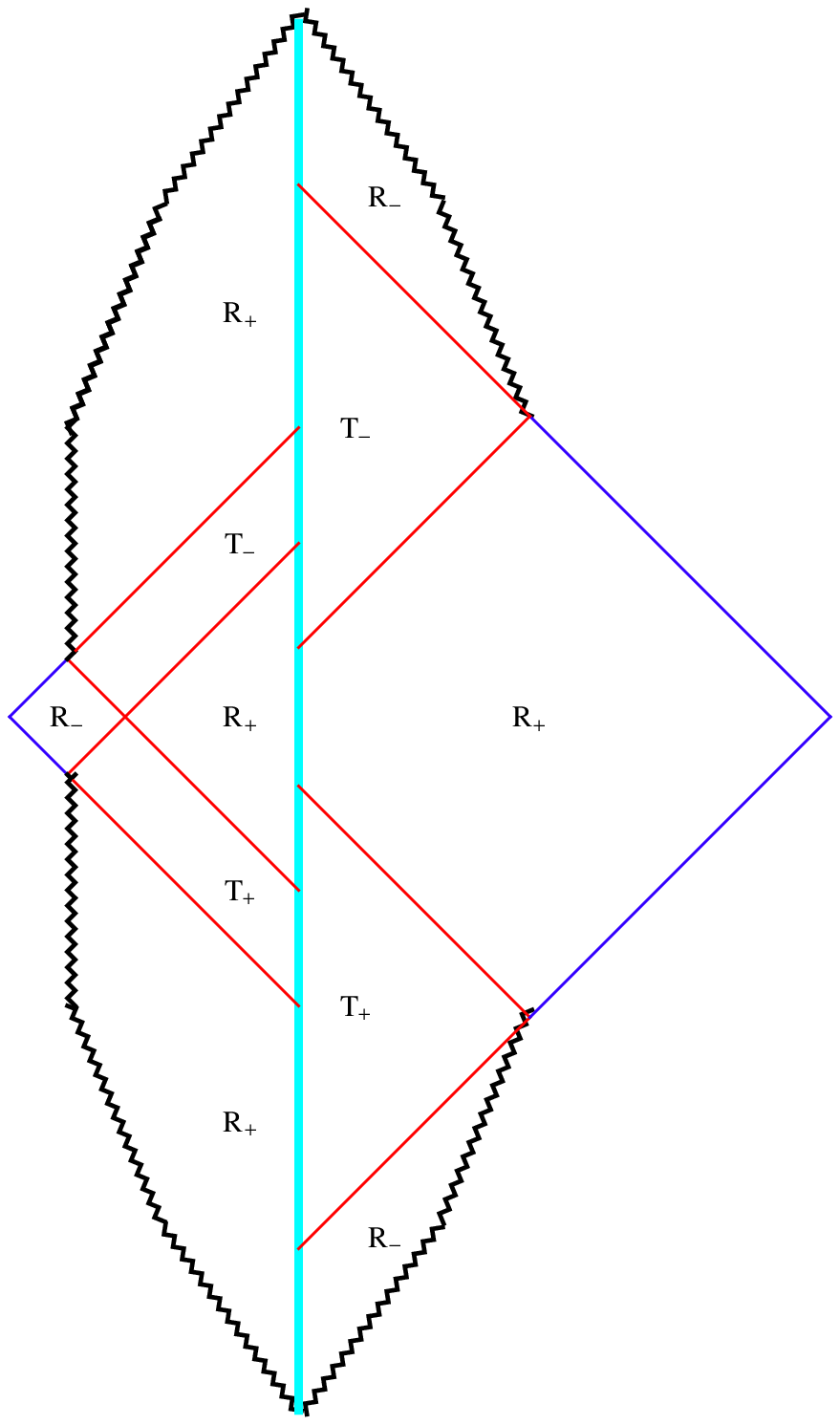}
\end{center}
\caption{The Carter-Penrose diagram at $\epsilon^2 <\alpha^2/4< \alpha^2 < 1$
for $0<\mu<\alpha/(2x_2)$ (left panel) and, respectively, for
$\alpha/(2x_2)<\mu<1$ (right panel).}
\label{fig48}
\end{figure}

\begin{figure}[h]
\begin{center}
\includegraphics[angle=0,width=0.7\textwidth]{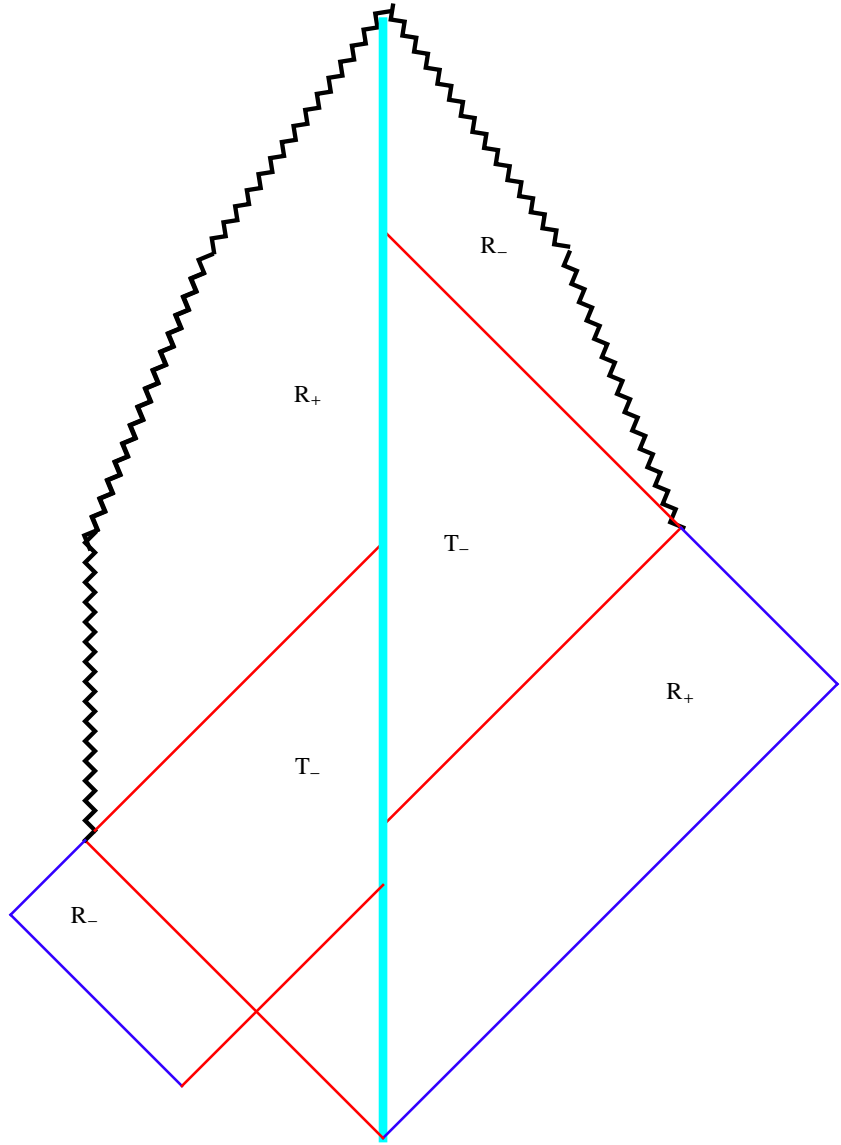}
\end{center}
\caption{The Carter-Penrose diagram at $\epsilon^2 <\alpha^2/4< \alpha^2 < 1$ for
$\mu>1$.}
\label{fig49}
\end{figure}

Let us come to the positive values of $\mu$. Now $\sigma_{\rm in} = + 1$ everywhere on the trajectories, and the metrics outside is that of the Reissner-Nordstrom black hole. There are three different intervals for $\mu$. If $0 < \mu = \mu_4 <\alpha/(2 x_2)$ (where $x_2$ is the intersection point of the right hand branch of the effective potential and the curve of changing $\sigma_{\rm out}, \; y = y_2 \; (x_2 > x_+)$), then $\sigma_{\rm out} = - 1$ everywhere on the trajectories. The Carter-Penrose diagram in this case differs from the preceding one (for the black hole outside) only by the interchange in the positions of the horizons (due to changing sign of $\mu$). The motion is bound with one turning point. If $\alpha/(2 x_2) < \mu = \mu_5 < 1$, then the turning point moves to the $R_+$-region outside the event horizon of the outer metrics. Finally, for $\mu = \mu_6 > 1$ we have the unbound motion, the shell collapses from the past temporal infinity in the $R_+$-region of the outer metrics into the $R_-$-region near singularity. Surely, there exists also the time-reversal motion. The conformal diagrams for these three intervals of $\mu$ are shown in Figs.~\ref{fig48} and ~\ref{fig49}, respectively.

\begin{figure}[h]
\begin{center}
\includegraphics[angle=0,width=0.95\textwidth]{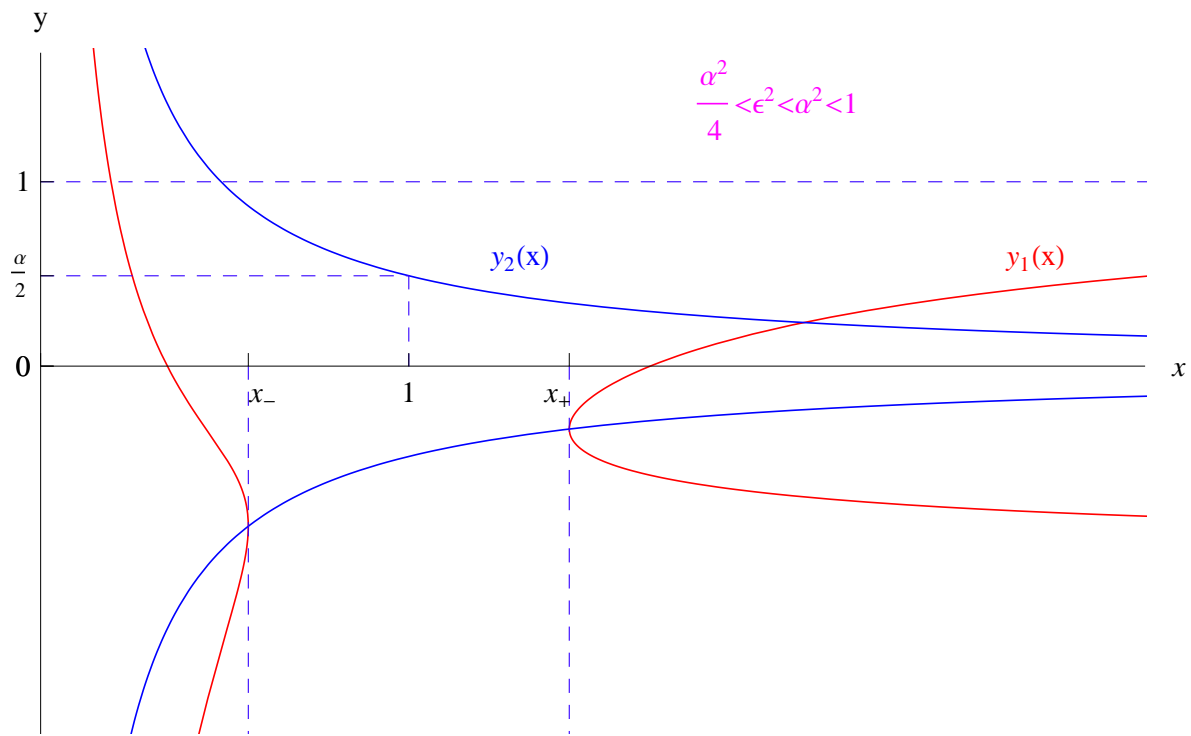}
\end{center}
\caption{The effective potential for the case  $\alpha^2/4<\epsilon^2 < \alpha^2 < 1$.}
\label{fig50}
\end{figure}

The next step is changing the relations between $\epsilon$ and $\alpha$ to the range $\alpha^2/4<\epsilon^2 < \alpha^2 < 1$. The effective potential becomes drastically different, as shown in Fig.~\ref{fig50}.

What is essential here? There appears the turning point near singularity. Consequently, all the Carter-Penrose diagrams should be replaced. One more thing: the horizontal line $\mu = \mu_0 = - 1/\alpha$, i.\,e., $m_{\rm out} = 0$, may lie lower as well as above the intersection point of the left hand branch of the effective potential with the curve $\tilde y_2$, i.\,e., $y_1 = \tilde y_2 = - y_2 = -\alpha/(2 x_-)$. So, let $\mu = \mu_1 < -\alpha/(2 x_-)$. Below in Figs.~\ref{fig51} and ~\ref{fig52} are the conformal diagrams for all three types of the outer metrics. Remember, that everywhere on the trajectories $\sigma_{\rm in} = - 1$ and $\sigma_{\rm out} = - 1$.

For higher values of $\mu$, namely, $- \alpha/(2 x_-)< \mu = \mu_2 < - 1$, we have slightly different diagrams inside the shell: now the turning point lies in the $R_+$-region near singularity, but outside the shell everything remains the same, as shown in Figs.~\ref{fig53} and ~\ref{fig54}

After shifting $\mu$ further up, into the region $- 1 < \mu = \mu_3 < - \frac{\alpha}{2 x_+}$, we get the bound motion with two turning points and with opposite signs of $\sigma_{\rm in}$ at the ends, but the same $\sigma_{\rm out} = - 1$. The Carter-Penrose diagrams are the following in Figs.~\ref{fig55} and \ref{fig56}.

For even higher, but still negative, values of $\mu$, namely, $- \alpha?(2 x_+) < \mu = \mu_4 < 0$, we have the motion of the same type as that considered above, but now both $\sigma's$ do not change their signs, $\sigma_{\rm in} = + 1, \; \sigma_{\rm out} = - 1$. See the corresponding diagrams in Figs.~\ref{fig57} and \ref{fig58}.

Let now $\mu > 0$. Outside the shell there cam be only the Reissner-Nordstrom metrics. One should consider three different intervals:

(a) $0 < \mu = \mu_5 <\alpha/(2 x_2)$ - bound motion with two turning points and $\sigma_{\rm out} = - 1$ everywhere;

(b) $\alpha/(2 x_2)< \mu = \mu_6 < 1$ - bound motion with two turning points and opposite values of $\sigma_{\rm out}$ at the ends ($\sigma_{\rm out} = - 1$ near the singularity and $\sigma_{\rm out} = + 1$ outside the event horizon);

(c) $\mu = \mu_7 > 1$ - unbound motion with one turning point and changing sign of $\sigma_{\rm out}$ ($\sigma_{\rm out} = - 1$ at the turning point). Remember, that $\sigma_{\rm in} = + 1$ everywhere.

See the corresponding diagrams in Figs.~\ref{fig59} and \ref{fig60}.

Let us now interchange $\epsilon$ and $\alpha$: $\alpha^2/4 < \alpha^2 < \epsilon^2 < 1$. There appears the intersection point, $x_1$, of the left hand branch of the effective potential with the curve $y_2$ and, therefore, the possibility for the shell to move without changing $\sigma_{\rm out} = + 1$. In what follows we will show only those new conformal diagrams when $\mu >\alpha/(2 x_1)$. Since the intersection point $y_1 = y_2$ may lie below as well as above the line $y = 1$, we get two different global geometries, shown in Fig.~\ref{fig61}.

Further increasing the values of $\alpha$ does not give us the new types of global geometries. The reader can easily find them among the already drawn diagrams.

\newpage

\newpage

\newpage
\begin{figure}[h]
\begin{center}
\includegraphics[angle=0,width=0.6\textwidth]{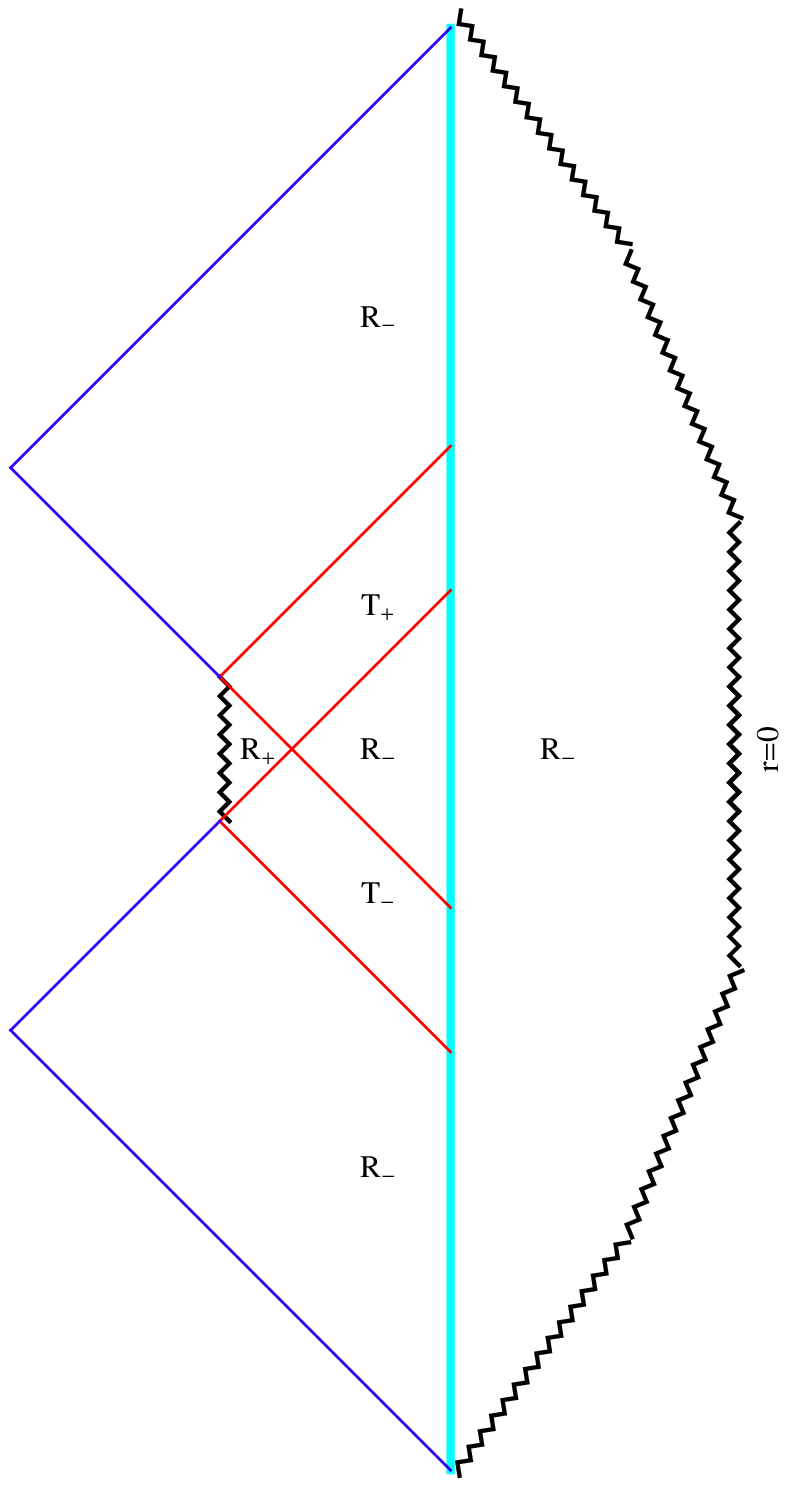}
\end{center}
\caption{The Carter-Penrose diagram at $\alpha^2/4<\epsilon^2 < \alpha^2 < 1$ for
$\mu>1$.}
\label{fig51}
\end{figure}

\newpage
\begin{figure}[h]
\begin{center}
\includegraphics[angle=0,width=0.49\textwidth]{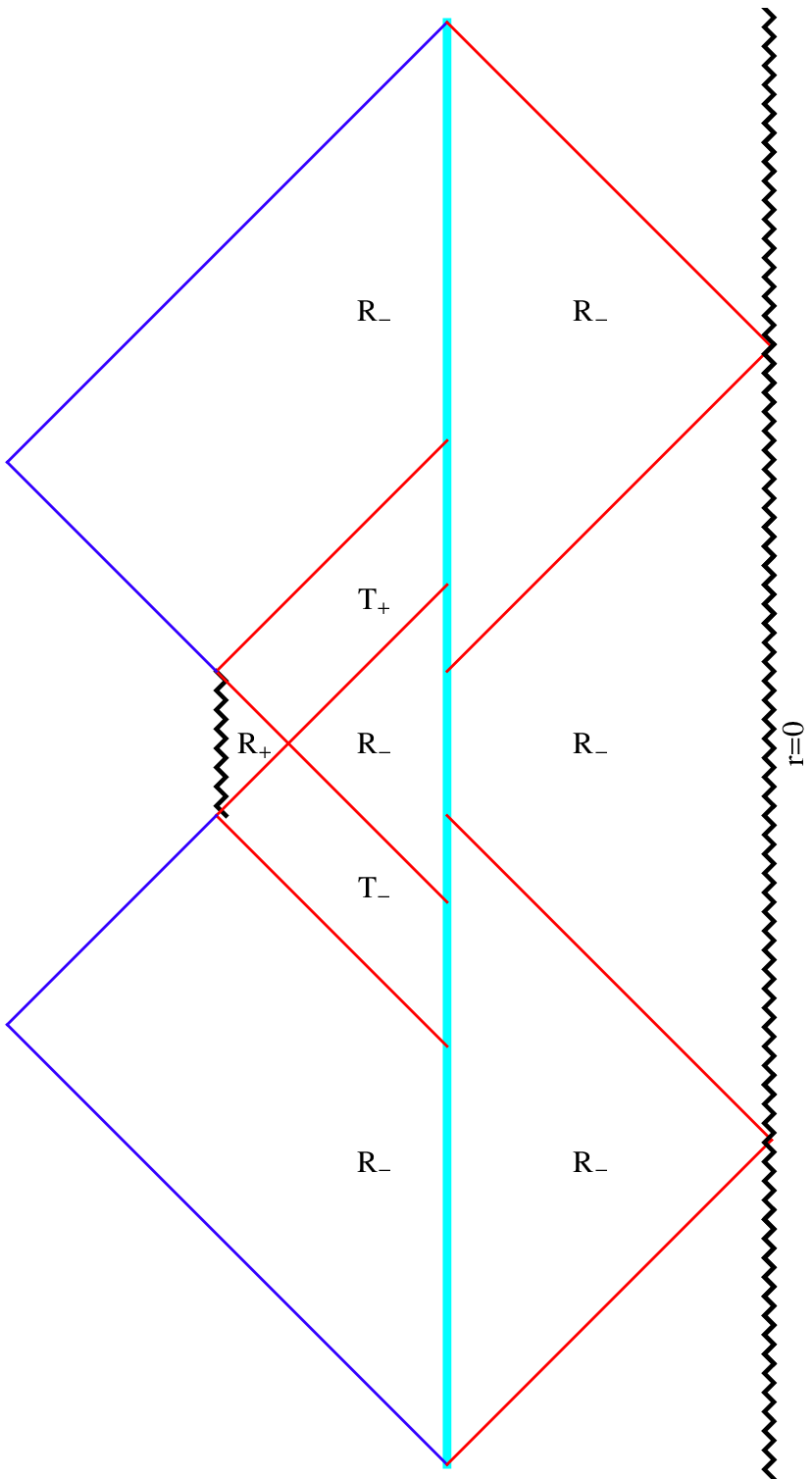}
\hfill
\includegraphics[angle=0,width=0.49\textwidth]{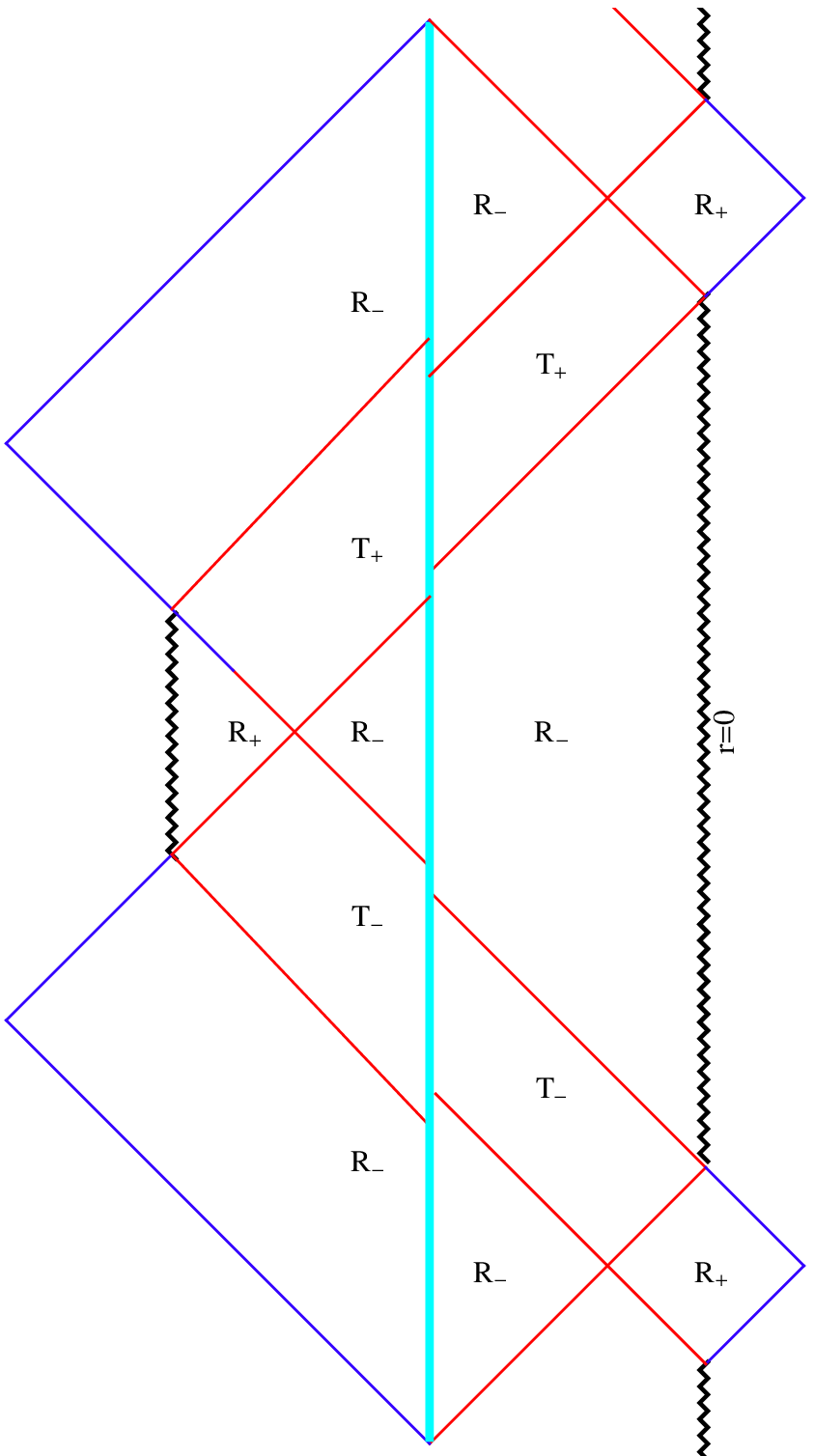}
\end{center}
\caption{The Carter-Penrose diagram at $\alpha^2/4<\epsilon^2 < \alpha^2 < 1$
for $0<\mu<\alpha/(2x_2)$ (left panel) and, respectively, for
$\alpha/(2x_2)<\mu<1$ (right panel).}
\label{fig52}
\end{figure}

\newpage
\begin{figure}[h]
\begin{center}
\includegraphics[angle=0,width=0.49\textwidth]{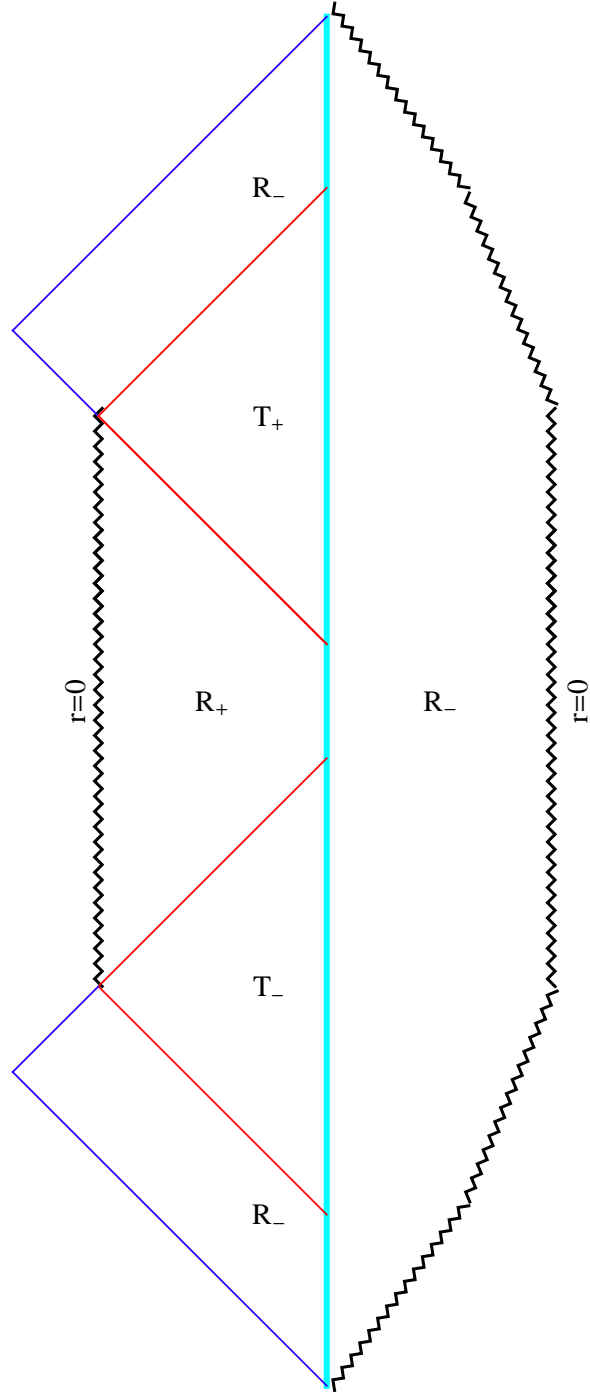}
\end{center}
\caption{The Carter-Penrose diagram $\alpha^2/4<\epsilon^2 < \alpha^2 < 1$ for
 $- \alpha/(2 x_-)< \mu = \mu_2 < - 1$.}
\label{fig53}
\end{figure}

\newpage
\begin{figure}[h]
\begin{center}
\includegraphics[angle=0,width=0.46\textwidth]{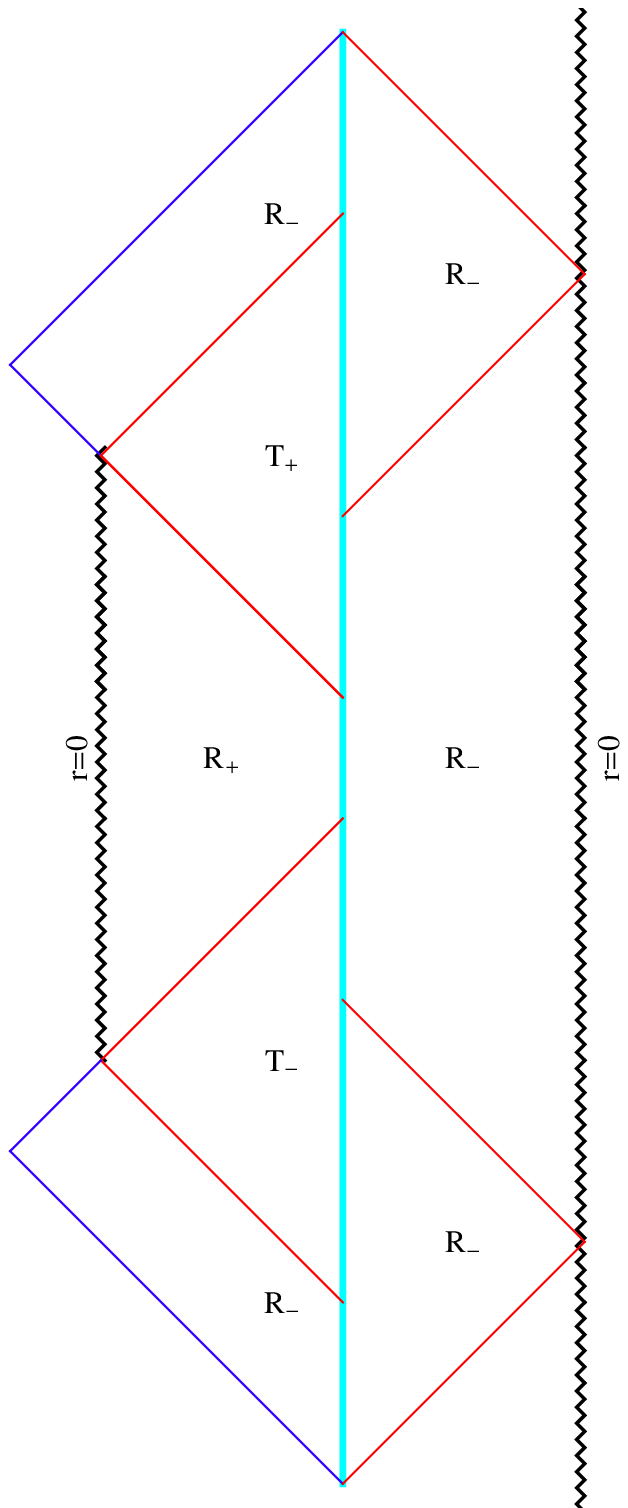}
\hfill
\includegraphics[angle=0,width=0.53\textwidth]{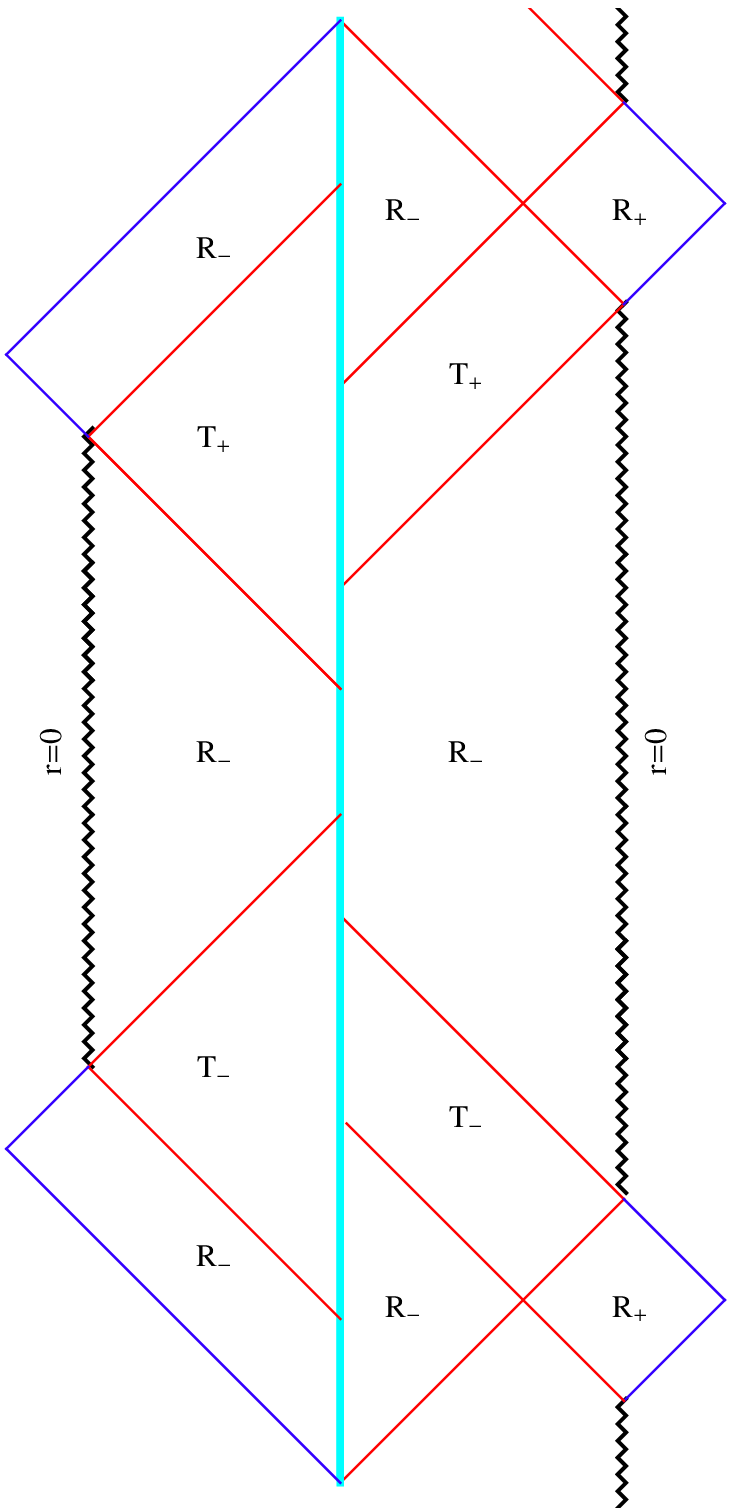}
\end{center}
\caption{The Carter-Penrose diagram at $\alpha^2/4<\epsilon^2 < \alpha^2 < 1$ for, respectively,
 $- \alpha/(2 x_-)< \mu <\mu_2 < - 1$ (left panel) and for
$- \alpha/(2 x_-)< \mu_2 < \mu < - 1$ (right panel).}
\label{fig54}
\end{figure}

\newpage
\begin{figure}[h]
\begin{center}
\includegraphics[angle=0,width=0.48\textwidth]{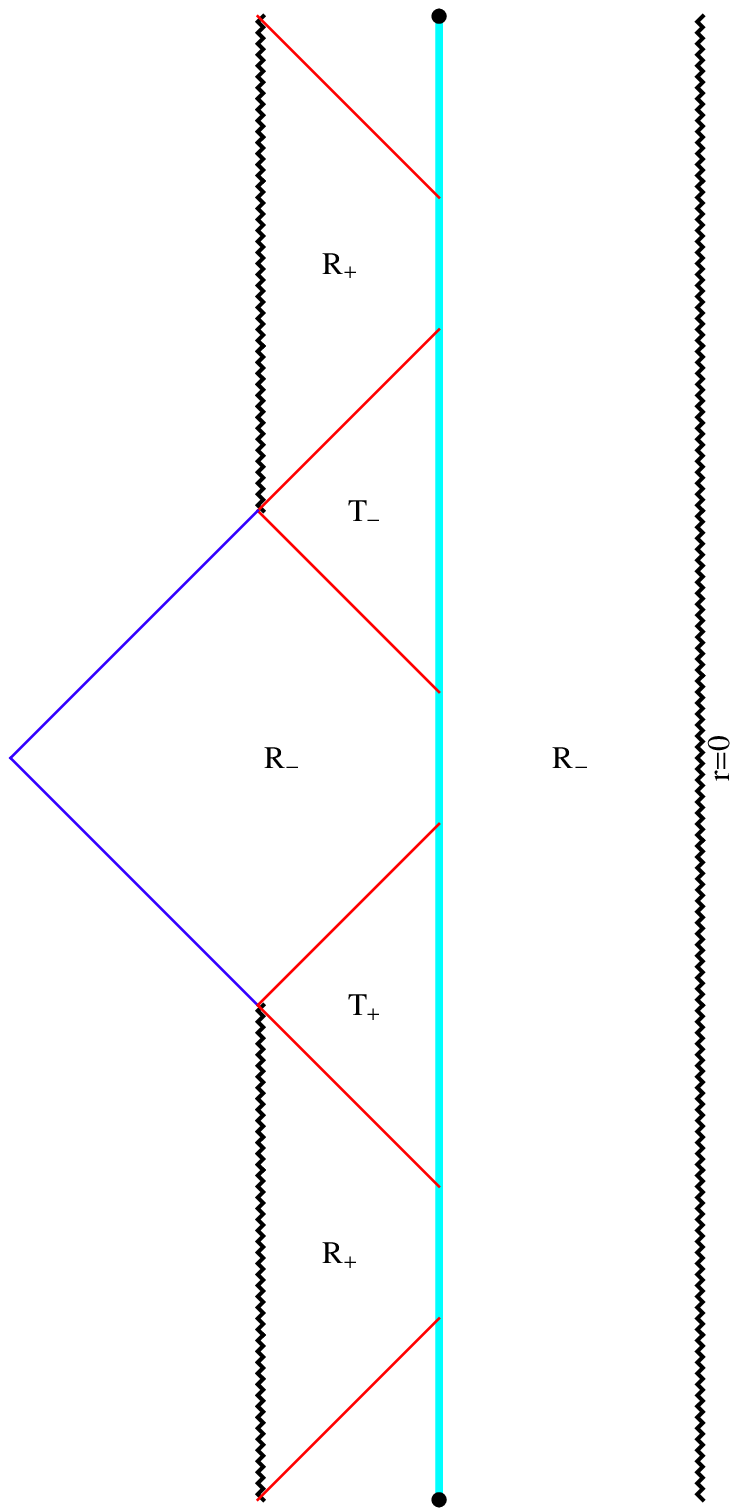}
\hfill
\includegraphics[angle=0,width=0.51\textwidth]{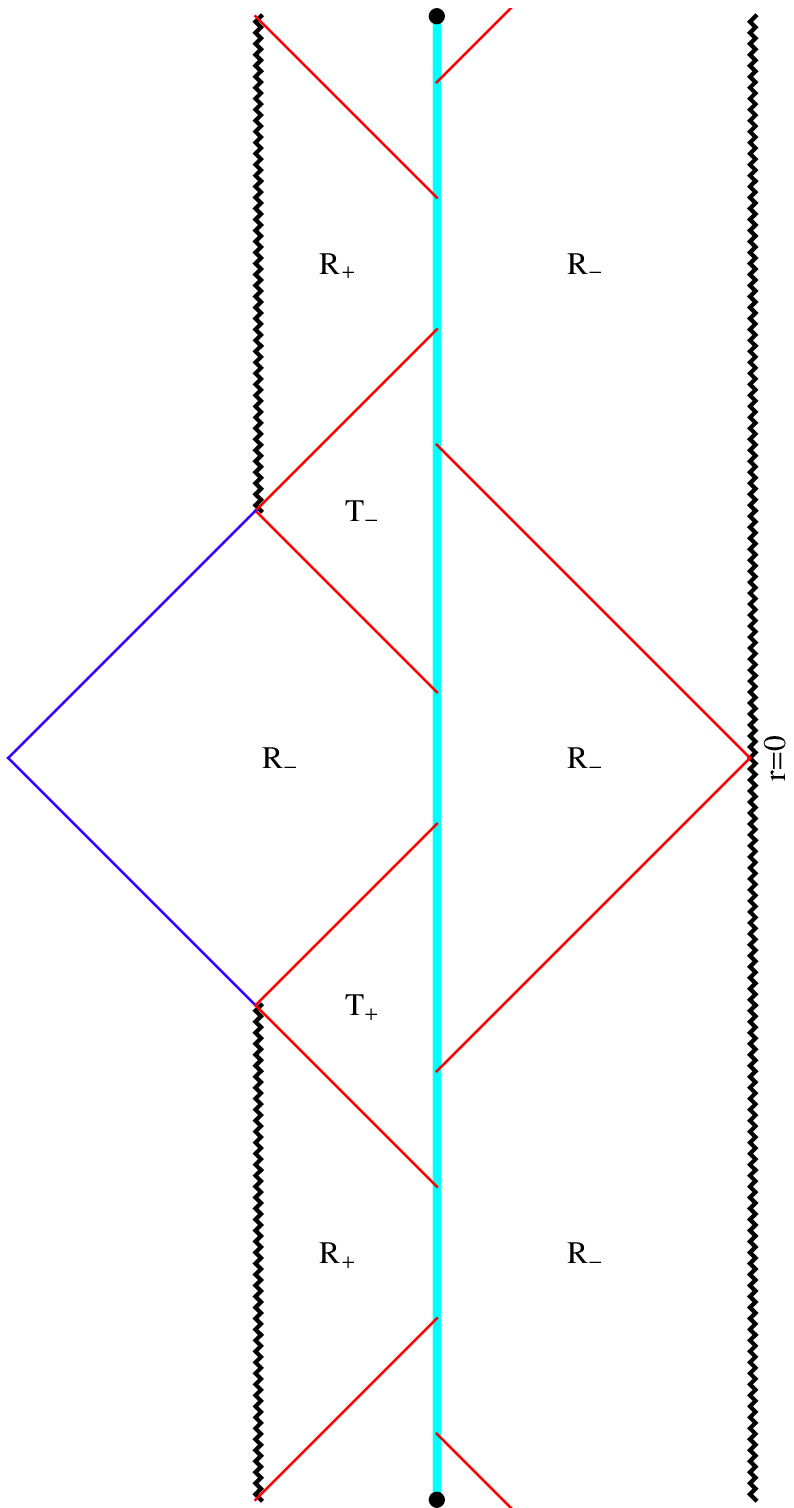}
\end{center}
\caption{The Carter-Penrose diagram at $\alpha^2/4<\epsilon^2 < \alpha^2 < 1$ for, respectively,
 $-1 \mu <-(1-\epsilon)/\alpha)<- \alpha/(2 x_-)$ (left panel) and for
 $-1 \mu=-(1-\epsilon)/\alpha)<- \alpha/(2 x_-)$ (right panel).}
\label{fig55}
\end{figure}

\newpage
\begin{figure}[h]
\begin{center}
\includegraphics[angle=0,width=0.7\textwidth]{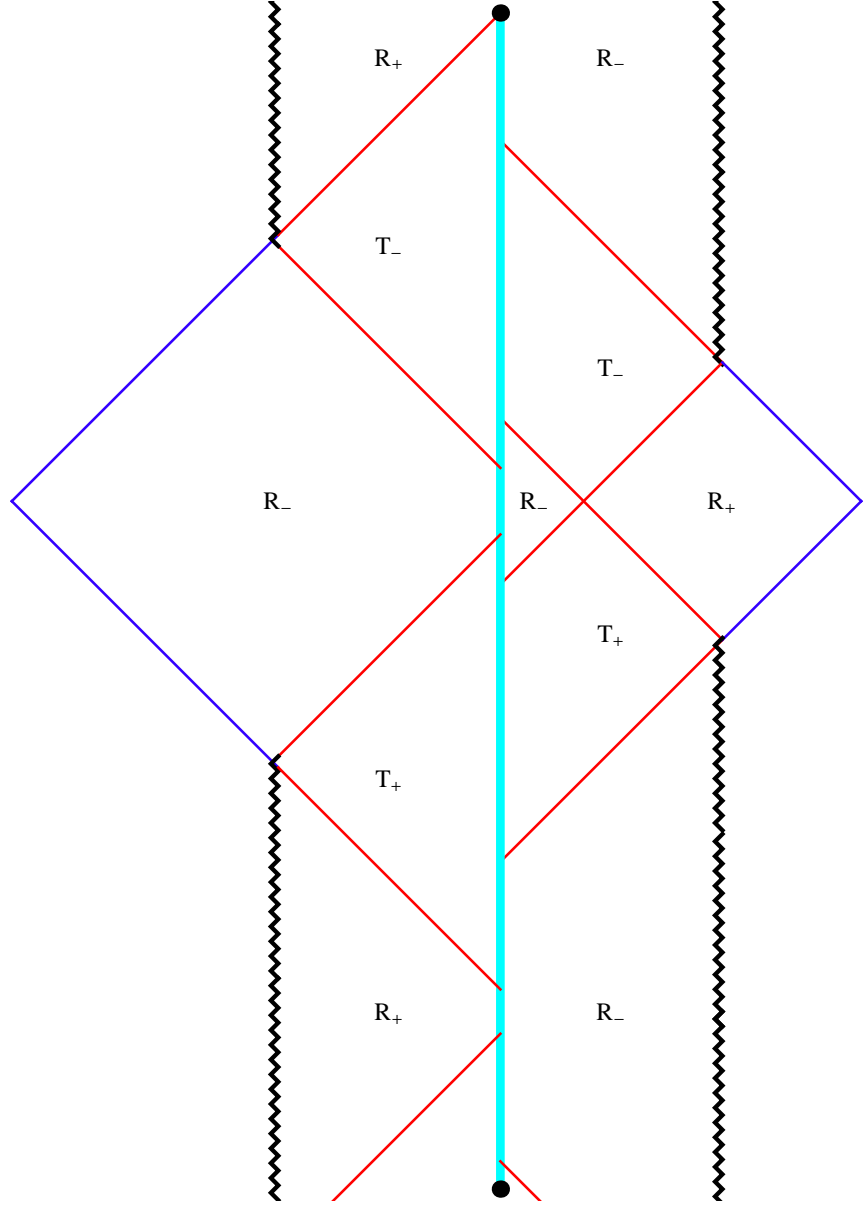}
\end{center}
\caption{The Carter-Penrose diagram at $\alpha^2/4<\epsilon^2 < \alpha^2 < 1$ for
 $-1<-(1-\epsilon)/\alpha)< \mu <- \alpha/(2 x_-)$.}
\label{fig56}
\end{figure}

\newpage
\begin{figure}[h]
\begin{center}
\includegraphics[angle=0,width=0.49\textwidth]{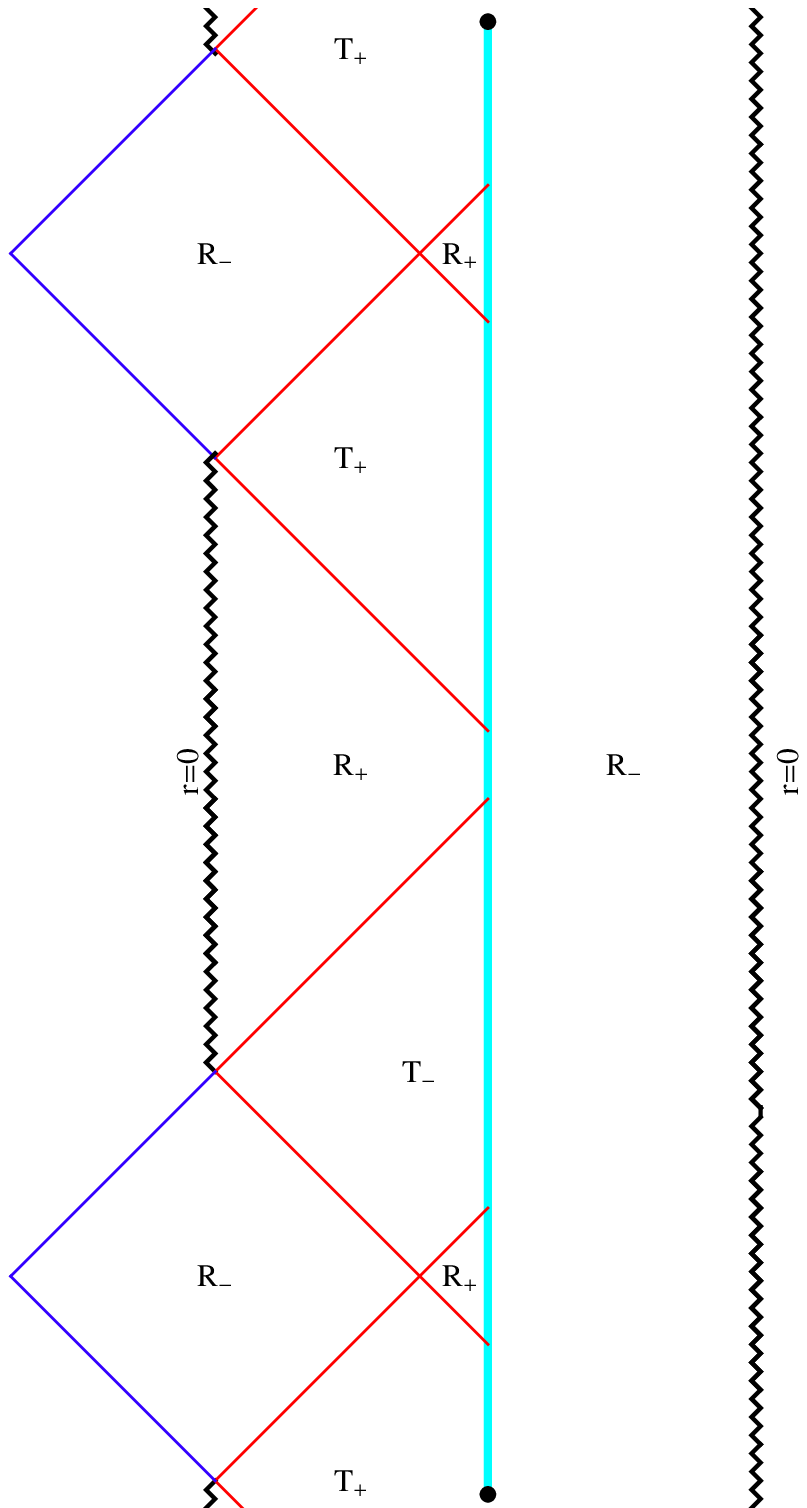}
\hfill
\includegraphics[angle=0,width=0.49\textwidth]{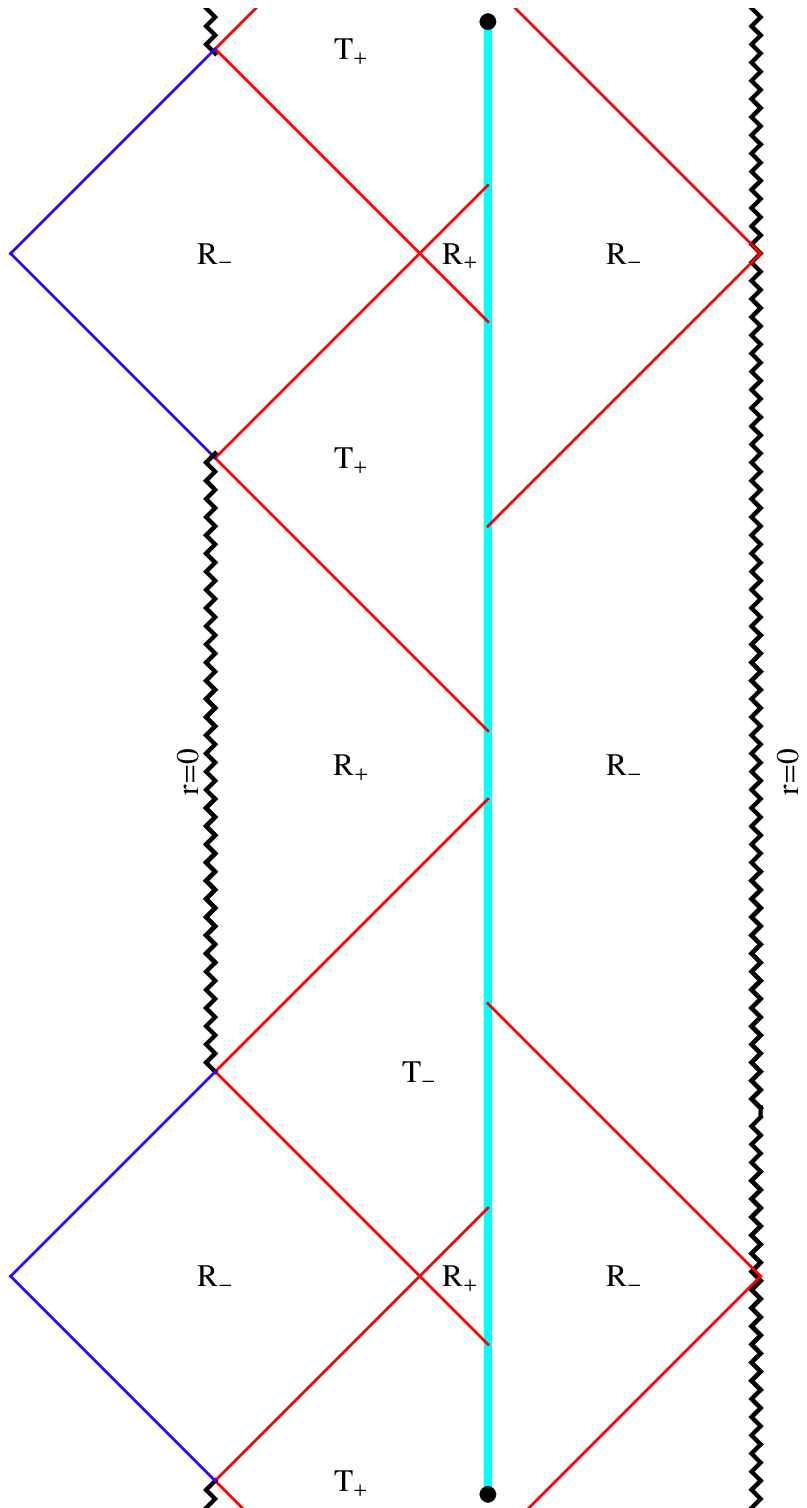}
\end{center}
\caption{The Carter-Penrose diagram at $\alpha^2/4<\epsilon^2 < \alpha^2 < 1$ for, respectively,
 $- \alpha/(2 x_-)< \mu <-(1-\epsilon)/\alpha)<0$ (left panel) and for
 $- \alpha/(2 x_-)< \mu=-(1-\epsilon)/\alpha)<0$ (right panel).}
\label{fig57}
\end{figure}

\newpage
\begin{figure}[h]
\begin{center}
\includegraphics[angle=0,width=0.7\textwidth]{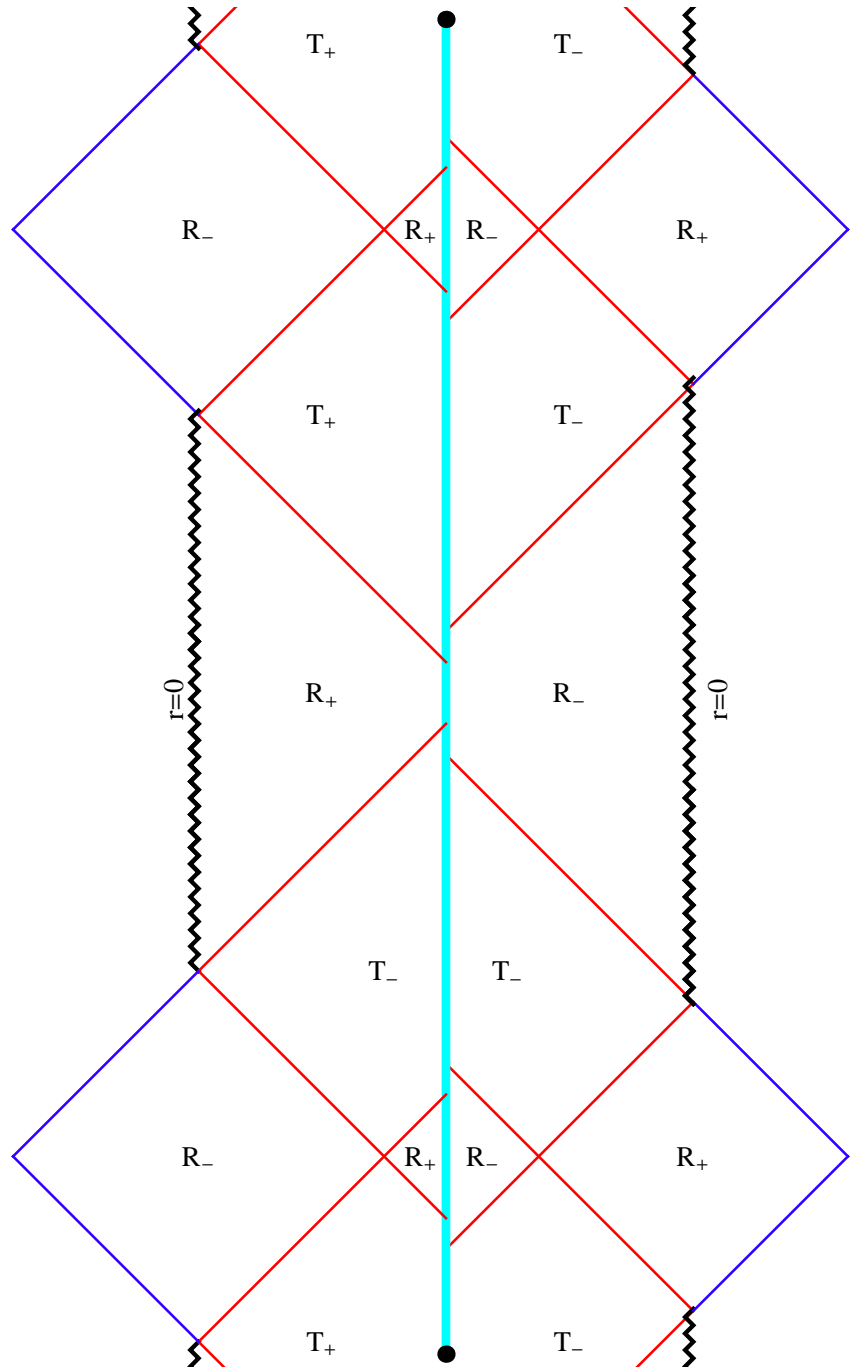}
\end{center}
\caption{The Carter-Penrose diagram at $\alpha^2/4<\epsilon^2 < \alpha^2 < 1$ for  $- \alpha/(2 x_-)< -(1-\epsilon)/\alpha)<\mu<0$.}
\label{fig58}
\end{figure}

\newpage
\begin{figure}[h]
\begin{center}
\includegraphics[angle=0,width=0.44\textwidth]{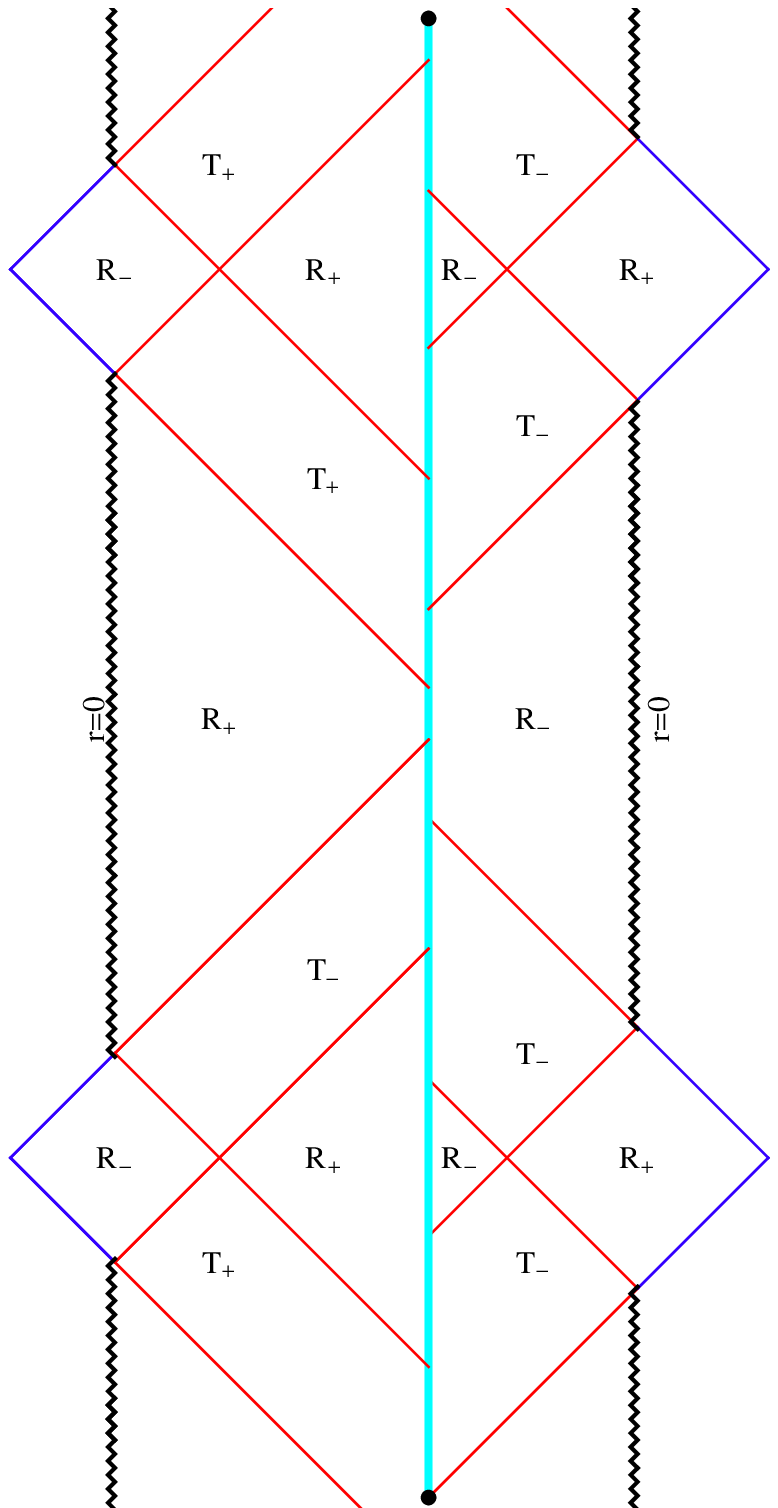}
\hfill
\includegraphics[angle=0,width=0.54\textwidth]{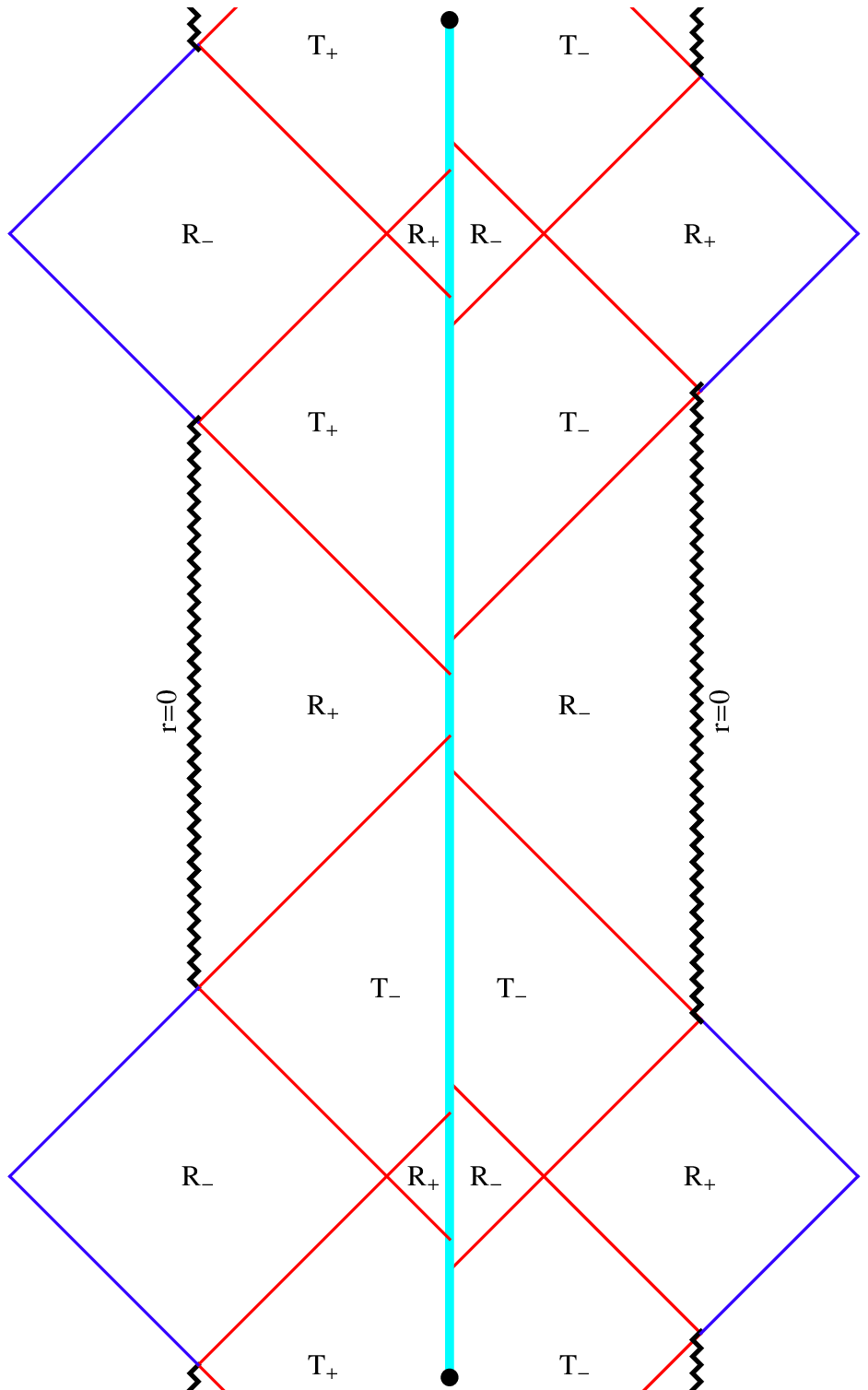}
\end{center}
\caption{The Carter-Penrose diagram at $\alpha^2/4<\epsilon^2 < \alpha^2 < 1$ for, respectively,
 $0< \mu <\alpha/(2 x_+)<1$ (left panel) and for  $\alpha/(2 x_+)<\mu <1$ (right panel).}
\label{fig59}
\end{figure}

\newpage
\begin{figure}[h]
\begin{center}
\includegraphics[angle=0,width=0.7\textwidth]{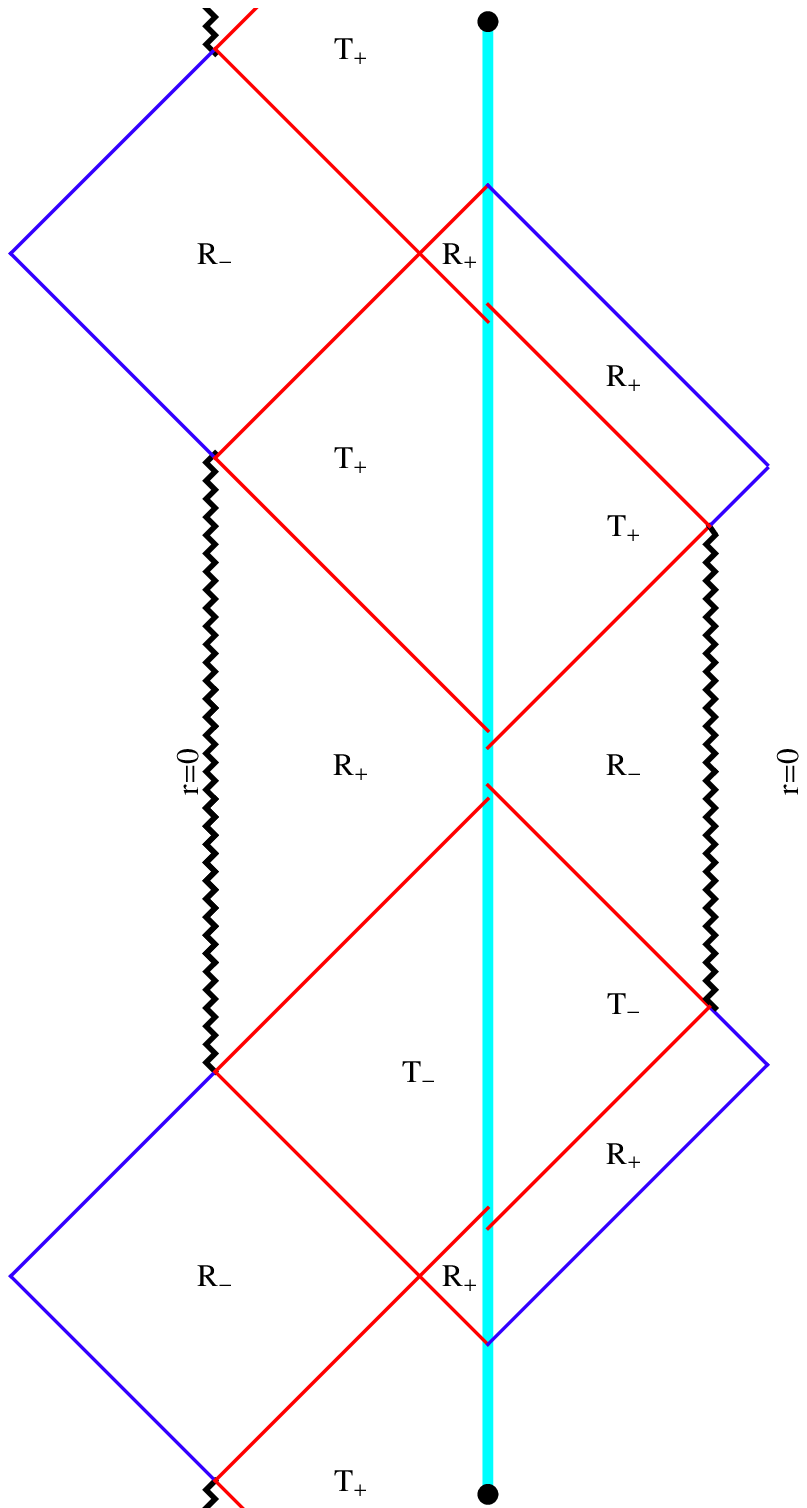}
\end{center}
\caption{The Carter-Penrose diagram at $\alpha^2/4<\epsilon^2 < \alpha^2 < 1$ for $\mu >1$.}
\label{fig60}
\end{figure}

\newpage
\begin{figure}[h]
\begin{center}
\includegraphics[angle=0,width=0.51\textwidth]{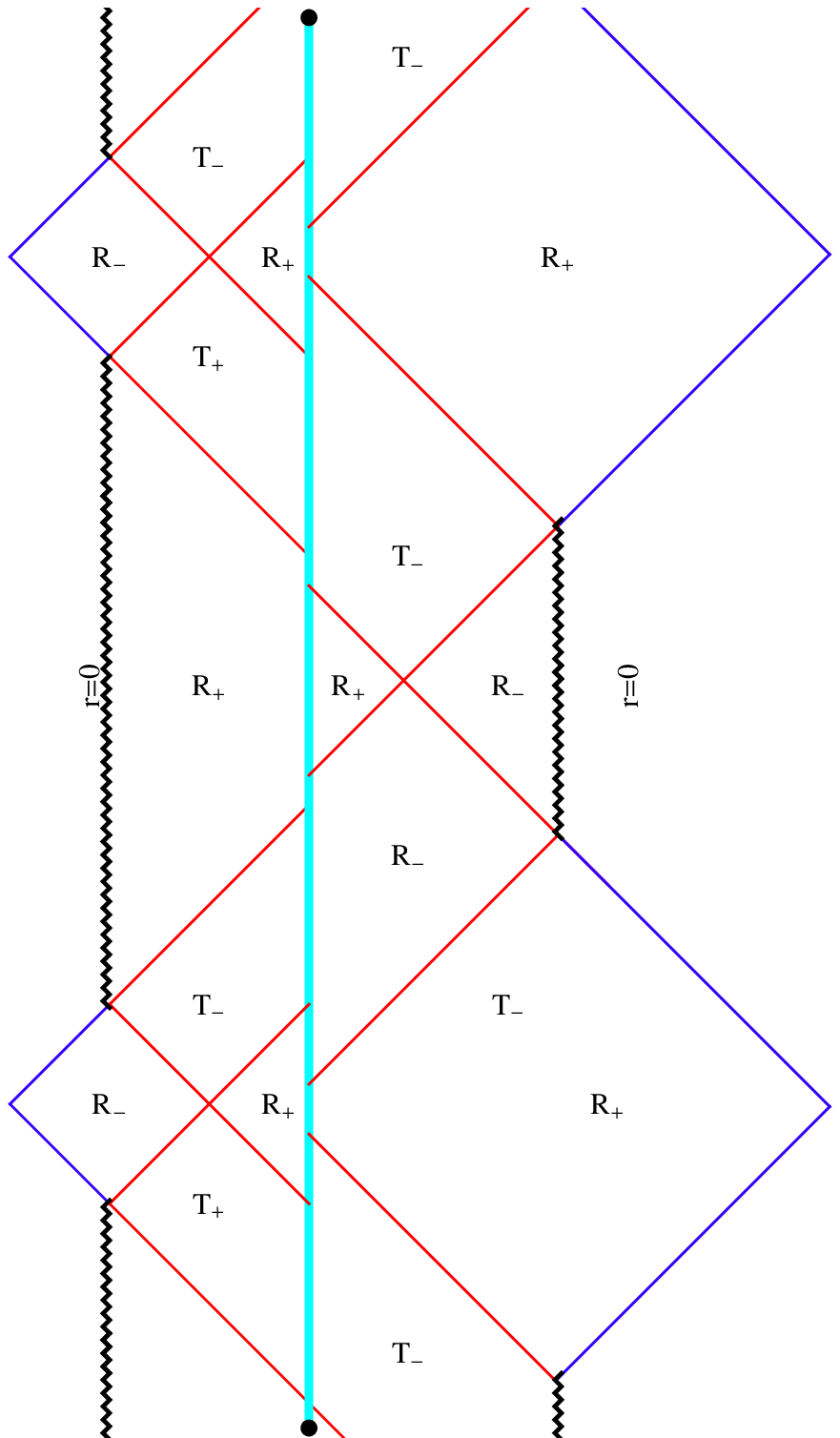}
\hfill
\includegraphics[angle=0,width=0.48\textwidth]{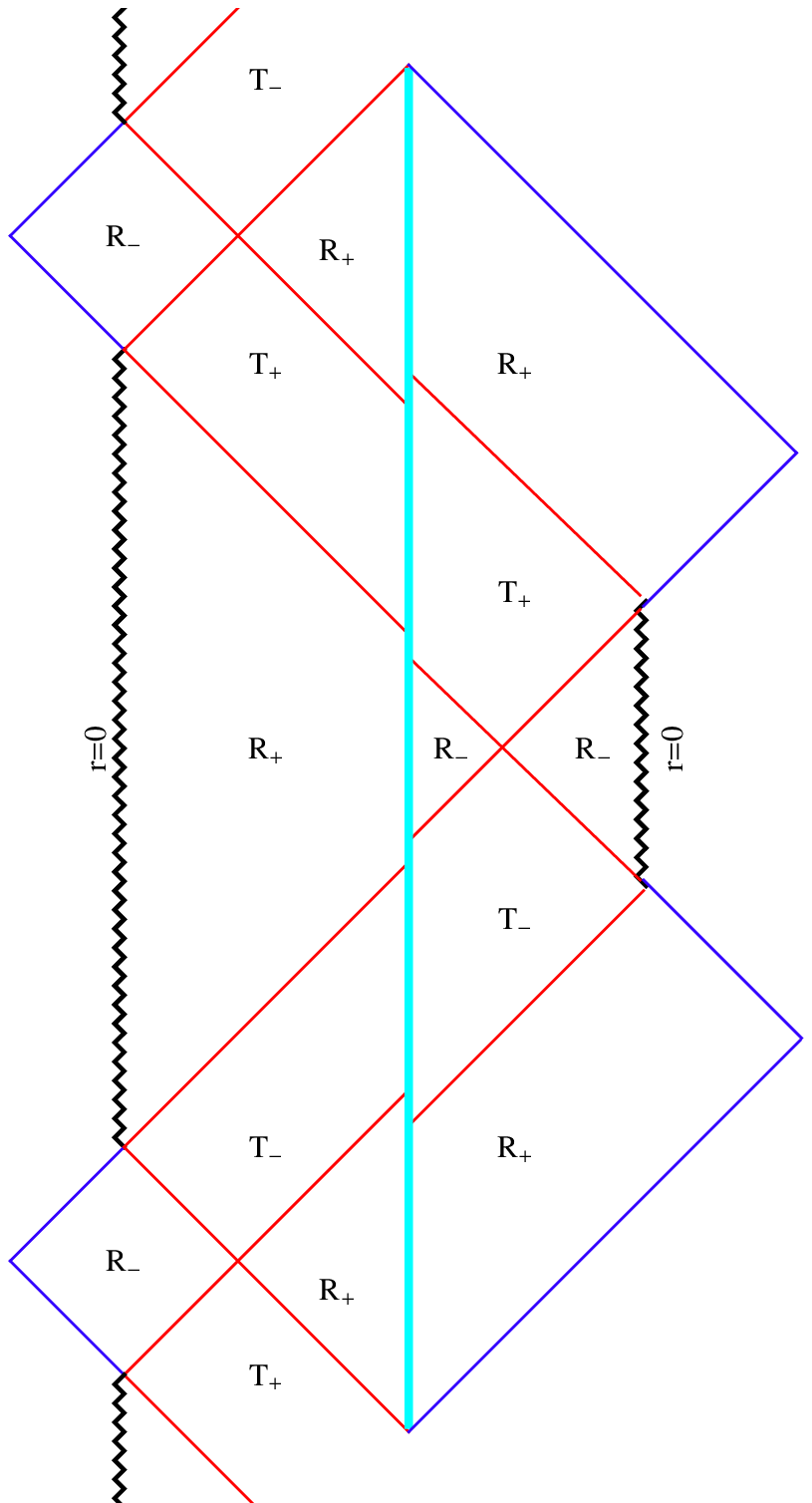}
\end{center}
\caption{The Carter-Penrose diagram at $\alpha^2/4< \alpha^2 <\epsilon^2 < 1$ for, respectively,
 $\alpha/(2 x_1)<\mu <1$ (left panel) and for  $1<\alpha/(2 x_1)<\mu$ (right panel).}
\label{fig61}
\end{figure}

\acknowledgments
This research was supported in part by the Russian Foundation for Basic Research grant RFBR~13-02-00257.


\end{document}